\documentstyle[12pt]{article}
\newlength{\pubnumber} 

\catcode`\@=11
\@addtoreset{equation}{section}

\def\section{\@startsection{section}{1}{\z@}{3.5ex plus 1ex minus .2ex}
 {2.3ex plus .2ex}{\large\bf}}
\def\subsection{\@startsection{subsection}{2}{\z@}{2.3ex plus .2ex}
 {2.3ex plus .2ex}{\bf}}
\newcommand\Appendix[1]{\def\thesection{Appendix \Alph{section}}
 \section{\label{#1}}\def\thesection{\Alph{section}}}
\begin{document}

\begin{titlepage}
\samepage{
\rightline{CERN-TH/98-100}
\rightline{\tt hep-ph/9806292}
\rightline{June 1998}
\vfill
\begin{center}
    {\Large \bf Grand Unification at Intermediate Mass Scales
       through Extra Dimensions\\}
\vfill
   {\large
      Keith R.\ Dienes$^1$\footnote{
       E-mail address: keith.dienes@cern.ch},
      Emilian Dudas$^{1,2}$\footnote{
       E-mail address:  emilian.dudas@cern.ch},
       $\,$and$\,$
        Tony Gherghetta$^1$\footnote{
       E-mail address: tony.gherghetta@cern.ch}
    \\}
\vspace{.18in}
 {\it  $^1$ CERN Theory Division, CH-1211 Geneva 23, Switzerland\\}
\vspace{.04in}
 {\it  $^2$ LPTHE, Univ.\ Paris-Sud, F-91405, Orsay Cedex, France\\}
\end{center}
\vfill
\begin{abstract}
  {\rm
   One of the drawbacks of conventional grand unification scenarios
   has been that the unification scale is too high to permit direct
   exploration.  In this paper, we show that the unification scale can
   be significantly lowered (perhaps even to the TeV scale) through
   the appearance of extra spacetime dimensions.  Such extra dimensions
   are a natural consequence of string theories with large-radius
   compactifications.  We show that extra spacetime dimensions
   naturally lead to gauge coupling unification at intermediate mass
   scales, and moreover may provide a natural mechanism for explaining
   the fermion mass hierarchy by permitting the fermion masses to evolve
   with a power-law dependence on the mass scale.  We also show that
   proton-decay constraints may be satisfied in our scenario due to
   the higher-dimensional cancellation of proton-decay amplitudes to
   all orders in perturbation theory.
   Finally, we extend these results by considering
   theories without supersymmetry;  experimental collider signatures;
   and embeddings into string theory.
   The latter also enables us to develop several novel methods of
   explaining the fermion mass hierarchy via $D$-branes.
   Our results therefore suggest a new approach towards understanding
   the physics of grand unification as well as the phenomenology of
   large-radius string compactifications.  }
\end{abstract}
\vfill }
\end{titlepage}

\setcounter{footnote}{0}

% ========================= DEFINITIONS ===================================
\newcommand{\newc}{\newcommand}

\newc{\gsim}{\lower.7ex\hbox{$\;\stackrel{\textstyle>}{\sim}\;$}}
\newc{\lsim}{\lower.7ex\hbox{$\;\stackrel{\textstyle<}{\sim}\;$}}

\newcommand{\hL}{\hat{L}}
\newcommand{\hPhi}{\hat{\Phi}}
\newcommand{\tl}{\tilde{l}}

\def\beq{\begin{equation}}
\def\eeq{\end{equation}}
\def\beqn{\begin{eqnarray}}
\def\eeqn{\end{eqnarray}}
\def\nn{{(n)}}
\def\bx{{\bf x}}
\def\by{{\bf y}}
\def\bn{{\bf n}}
\def\Fbar{{\overline{F}}}
\def\sosixteen{{$SO(16)\times SO(16)$}}
\def\e8{{$E_8\times E_8$}}
\def\V#1{{\bf V_{#1}}}
\def\half{{\textstyle{1\over 2}}}
\def\ttwo{{\vartheta_2}}
\def\tthree{{\vartheta_3}}
\def\tfour{{\vartheta_4}}
\def\ttwob{{\overline{\vartheta}_2}}
\def\tthreeb{{\overline{\vartheta}_3}}
\def\tfourb{{\overline{\vartheta}_4}}
\def\etainv{{\overline{\eta}}}
\def\Str{{{\rm Str}\,}}
\def\bone{{\bf 1}}
\def\chibar{{\overline{\chi}}}
\def\Jbar{{\overline{J}}}
\def\qbar{{\overline{q}}}
\def\calO{{\cal O}}
\def\calE{{\cal E}}
\def\calT{{\cal T}}
\def\calM{{\cal M}}
\def\calF{{\cal F}}
\def\calY{{\cal Y}}
\def\rep#1{{\bf {#1}}}
\def\ie{{\it i.e.}\/}
\def\eg{{\it e.g.}\/}
\def\eleven{{(11)}}
\def\ten{{(10)}}
\def\nine{{(9)}}
\def\Ip{{\rm I'}}
\def\oneprime{{I$'$}}
%==============================================================================
\hyphenation{su-per-sym-met-ric non-su-per-sym-met-ric}
\hyphenation{space-time-super-sym-met-ric}
\hyphenation{mod-u-lar mod-u-lar--in-var-i-ant}
%==============================================================================

%================== BLACKBOARD BOLD CHARACTERS ==============================

\def\inbar{\,\vrule height1.5ex width.4pt depth0pt}

\def\IC{\relax\hbox{$\inbar\kern-.3em{\rm C}$}}
\def\IQ{\relax\hbox{$\inbar\kern-.3em{\rm Q}$}}
\def\IR{\relax{\rm I\kern-.18em R}}
 \font\cmss=cmss10 \font\cmsss=cmss10 at 7pt
\def\IZ{\relax\ifmmode\mathchoice
 {\hbox{\cmss Z\kern-.4em Z}}{\hbox{\cmss Z\kern-.4em Z}}
 {\lower.9pt\hbox{\cmsss Z\kern-.4em Z}}
 {\lower1.2pt\hbox{\cmsss Z\kern-.4em Z}}\else{\cmss Z\kern-.4em Z}\fi}

%========================================================================
%          MACROS FOR REFERENCES
%========================================================================
\def\NPB#1#2#3{{\it Nucl.\ Phys.}\/ {\bf B#1} (19#2) #3}
\def\PLB#1#2#3{{\it Phys.\ Lett.}\/ {\bf B#1} (19#2) #3}
\def\PRD#1#2#3{{\it Phys.\ Rev.}\/ {\bf D#1} (19#2) #3}
\def\PRL#1#2#3{{\it Phys.\ Rev.\ Lett.}\/ {\bf #1} (19#2) #3}
\def\PRT#1#2#3{{\it Phys.\ Rep.}\/ {\bf#1} (19#2) #3}
\def\CMP#1#2#3{{\it Commun.\ Math.\ Phys.}\/ {\bf#1} (19#2) #3}
\def\MODA#1#2#3{{\it Mod.\ Phys.\ Lett.}\/ {\bf A#1} (19#2) #3}
\def\IJMP#1#2#3{{\it Int.\ J.\ Mod.\ Phys.}\/ {\bf A#1} (19#2) #3}
\def\NUVC#1#2#3{{\it Nuovo Cimento}\/ {\bf #1A} (#2) #3}
\def\etal{{\it et al.\/}}

% Redefine caption to put text and formulas in smaller font
\long\def\@caption#1[#2]#3{\par\addcontentsline{\csname
  ext@#1\endcsname}{#1}{\protect\numberline{\csname
  the#1\endcsname}{\ignorespaces #2}}\begingroup
    \small
    \@parboxrestore
    \@makecaption{\csname fnum@#1\endcsname}{\ignorespaces #3}\par
  \endgroup}
\catcode`@=12

\input epsf

%============================== TEXT BEGINS HERE ============================

\section{Introduction and basic idea}
\setcounter{footnote}{0}

One of the most widely investigated proposals for physics beyond
the Minimal Supersymmetric Standard Model (MSSM) is the
possible appearance of a grand unified theory (GUT).
There are several profound attractions to
the idea of grand unification.
Perhaps the most obvious is that GUT's have the potential to unify
the diverse set of particle representations and parameters found in
the MSSM into a single, comprehensive, and hopefully predictive framework.
For example, through the GUT symmetry one might hope to explain
the quantum numbers of the fermion spectrum, or even the origins of
fermion mass.
Moreover, by unifying all $U(1)$ generators within a non-abelian theory,
GUT's would also provide an explanation for the quantization of electric
charge.  
Furthermore, because they generally lead to baryon-number
violation,
GUT's have the potential to explain the cosmological baryon/anti-baryon
asymmetry.  By combining GUT's with supersymmetry in the context of
SUSY GUT's, it might then be possible to realize
the attractive features of GUT's simultaneously with those of supersymmetry
in a single theory.
Indeed, there is even a bit of ``experimental'' evidence for the idea of
grand unification, for the three MSSM gauge couplings
appear to unify
when extrapolated towards higher energies
within the framework of the MSSM, with a unification scale
$M_{\rm GUT} \approx 2\times 10^{16}$ GeV.
The phenomenon of gauge coupling unification thus sets
the natural energy scale for grand unification.

Unfortunately, because this energy scale is so high,
it proves rather difficult to probe the physics of
grand unification directly at low energies.
There are, of course, numerous {\it indirect}\/ probes of such
high-scale GUT physics, most notably through rare decays such
as proton decay.
However, all of these tests are ultimately limited by the
remoteness of the GUT scale.

In this paper, we shall investigate whether it is
possible to bring grand unification down towards more
accessible energy scales by  {\it lowering the unification scale itself}.
Of course,
it hardly seems possible to consider a unification
of the MSSM gauge groups at any lower scale $M\ll M_{\rm GUT}$
at which their couplings have not yet unified.
Therefore, in order to sensibly contemplate the possibility
of an intermediate-scale GUT, we are forced to consider
whether it is possible to achieve gauge coupling unification
at scales $M\ll M_{\rm GUT}$.
It is immediately clear that adding arbitrary extra matter states
to the MSSM cannot achieve the desired effect, for in general
such extra matter not only tends to raise (rather than lower)
the unification scale, but also tends to drive the theory towards
strong coupling.
What we require, therefore, is a different mechanism.

The underlying reason that gauge coupling unification
is delayed until such a high energy scale is essentially
that the one-loop MSSM gauge couplings run only logarithmically
with energy scale $\mu$ (or linearly versus $\log\,\mu$).
Thus, given the different values of these couplings at the
weak scale, one must extrapolate upwards over many orders of magnitude
in energy before they have a chance of unifying.
Clearly, if there were a way to change the running of
the gauge couplings so that they ran more quickly (\eg, exponentially
rather than linearly), we would have a chance to achieve a more
rapid unification.

What physical effect could cause the running of gauge
couplings to be exponential rather than linear?
Remarkably, there does exist a simple way in which such an exponential
running can arise:  the appearance of extra spacetime dimensions.
Since extra spacetime dimensions are
naturally predicted in string theory (both through
the need to compactify as well as through various non-perturbative
effects), we expect that such a scenario
might find a natural home within the context of string theory.
However, as we shall see, such a scenario
can be discussed in purely field-theoretic
terms.

Of course, there is one subtle complication with this na\"\i ve picture:
in field theory, extra spacetime dimensions lead
to a loss of renormalizability.  Thus, strictly speaking,
quantities such as gauge couplings do not ``run'' in the usual
sense.  However, our basic idea is nevertheless correct, and as we shall see,
the ``exponential running'' is more correctly described as an
exponential dependence on the cutoff pertaining to high-scale physics
(such as the appearance of a fundamental string theory) at energy
scales beyond those we shall be considering.  Moreover, as we shall
demonstrate, even though our theory is non-renormalizable,
there exists a {\it renormalizable}\/ theory which, for our purposes,
is essentially equivalent to our non-renormalizable theory.
Thus, our basic intuitive idea of exponential ``running'' remains
intact.

Given this motivation, in this paper we shall undertake
a general analysis of the effects of extra spacetime dimensions
on the MSSM.
We shall begin, in Sect.~2, with a discussion
of various issues that arise when attempting to extrapolate
the MSSM to higher dimensions.
Then, in Sect.~3,
we shall discuss the effects of extra spacetime dimensions
on ordinary gauge coupling unification.  Quite remarkably, we shall
find that within the MSSM, the appearance of {\it any}\/ number of
extra spacetime dimensions at {\it any}\/ intermediate scale
always preserves gauge coupling unification --- indeed, we shall
find that the effect
of the extra dimensions is simply to shift the unification
scale downwards towards lower energies, as desired.
Thus, within the MSSM,
we find that the appearance of extra spacetime dimensions naturally leads
to an intermediate-scale grand unified theory.

In Sect.~4 we shall then consider the effects of extra spacetime dimensions
of the proton-lifetime problem, and propose that
to all orders in perturbation theory,
proton-decay amplitudes
are exactly cancelled as the result of new Kaluza-Klein selection
rules corresponding to the extra spacetime dimensions.
This is therefore an intrinsically {\it higher-dimensional}\/
solution to the proton-decay problem.

In Sect.~5,
we shall then turn our attention
to the effects of extra spacetime dimensions on the evolution of
the fermion Yukawa couplings.  We shall find that extra
dimensions cause the Yukawa couplings to
run exponentially as well, and thereby show that this exponential running
for the Yukawa couplings has the potential to explain
the fermion mass hierarchy.

In Sect.~6, we will then consider the effects of extra dimensions
on the {non-supersymmetric}\/ Standard Model, and show that once again
it is possible to obtain gauge coupling unification at relatively low
energy scales, even without supersymmetry.
Thus, if the unification scale is sufficiently low, we can actually
avoid the {\it gauge}\/ hierarchy problem.

In Sect.~7, we will then discuss how our scenario can be embedded
into string theory, and make some general remarks
concerning the manner in which the appearance of extra spacetime dimensions
and intermediate-scale grand unification
can be incorporated and interpreted in terms
of the mass scales expected within string theory.
We shall also consider various non-perturbative
$D$-brane realizations of our scenario.
Then, in Sect.~8, we shall revisit the fermion mass hierarchy problem
using some of these non-perturbative $D$-brane insights,
and we shall propose several new $D$-brane methods for addressing
the fermion mass hierarchy problem which do not rely on the {\it ad hoc}\/
introduction of extra low-energy matter states or flavor-dependent
couplings.
The mechanisms we propose in this section might therefore be of general
use for the phenomenology of Type~I model-building.

In Sect~9, we shall discuss how our our work relates to prior
work in the literature, and in Sect.~10 we shall discuss
some of the experimental consequences of our scenario.
As we shall see, our scenario can be expected to lead to numerous
exciting collider signals;  it can also have important cosmological
implications.
We will then conclude in Sect.~11 with a summary of our results,
and as well as future prospects and ramifications.
Three Appendices contain
ancillary calculations
which justify the basic approach that we shall be following in this paper.
Note that an abbreviated discussion of some of the ideas in this paper
can also be found in Ref.~\cite{shortpaper}.

%========================================================================
%  \vfill\eject
\section{Preliminaries}
\setcounter{footnote}{0}

How can we incorporate extra spacetime dimensions
into a field-theoretic analysis of the MSSM?
In this section, we shall provide a discussion of some
of the issues that arise, including the appearance and proper treatment
of Kaluza-Klein modes as well as the resulting lack
of renormalizability that afflicts higher-dimensional field theories.
Throughout, we shall focus on taking a ``bottom-up'' approach,
and seek the ``minimal'' scenarios that consistently
embed the MSSM into higher dimensions.
Thus, our approach will necessarily be part of any larger
structure that may ultimately be derived
from a more complete high-energy theory such as string theory.

\subsection{Incorporating extra dimensions into the MSSM}

Given that the observed low-energy world consists of only four flat
dimensions,
the only rigorous way in which to discuss the appearance of
extra spacetime dimensions is to assume that they are compactified.
For this purpose, we shall begin by assuming that they are simply
compactified on a circle of a certain
fixed radius $R$, where $R^{-1}$ exceeds presently observable
energy scales.  Thus $\mu_0\equiv R^{-1}$ sets the mass scale at which the
extra
dimensions become significant.

The appearance
of extra dimensions of radius $R$ implies that a given
complex quantum field $\Phi(x)$ now depends on not only
the usual four-dimensional
spacetime coordinates $\bx\equiv (x_0,x_1,x_2,x_3)$,
but also the additional spacetime coordinates
$\by \equiv (y_1,y_2,...,y_\delta)$
where $\delta\equiv D-4$ is the
number of additional dimensions.
We shall denote these coordinates collectively as
$x=({\bf x},{\bf y})$.
Demanding periodicity of $\Phi(x)$ under
\beq
         y_i\to y_i+2\pi R
\label{circleidentify}
\eeq
then implies that $\Phi(x)$ takes the form
\beq
   \Phi(x) ~=~
      \sum_{n_1= -\infty}^\infty
      \sum_{n_2= -\infty}^\infty
      \cdots \sum_{n_\delta= -\infty}^\infty
         \Phi^{(\bn)}(\bx)
         \,\exp( i {\bf n}\cdot {\bf y}/R)~
\label{KKexpansion}
\eeq
where $\bn = (n_1,n_2,...,n_\delta)$, with $n_i\in \IZ$.
The ``four-dimensional'' fields $\Phi^{(\bn)}(\bx)$
are the so-called {\it Kaluza-Klein modes}\/,
and $n_i$ are the corresponding Kaluza-Klein
excitation numbers (with $p_i\equiv n_i/R$ serving
as the Kaluza-Klein momenta).
In general, the mass of each Kaluza-Klein mode
is given by
\beq
         m^2_n ~\equiv~ m^2_0~+~ \,{\bn \cdot \bn \over R^2}~
\label{KKmasses}
\eeq
where $m_0$ is the mass of the zero-mode.
At energies far below $R^{-1}$, we expect our extra dimensions to
be unobservable.  However, in this limit, none of the non-zero Kaluza-Klein
modes can be excited, and we will observe only the zero-mode
field $\Phi^{({\bf 0})}({\bf x})$.
Thus, the zero-mode field corresponds to the
usual four-dimensional state, and the appearance of the extra
spacetime dimensions is felt through the appearance of an
infinite tower of associated Kaluza-Klein states of increasing mass.

How then do we incorporate extra spacetime dimensions into the MSSM?
It might initially seem
that for every MSSM state
of mass $m_0$,
there will exist a corresponding infinite
tower of Kaluza-Klein states with masses given by (\ref{KKmasses}),
each of which exactly mirrors the zero-mode MSSM ground state.
Since $R^{-1}$ is presumed to exceed presently observable energy
scales, we are free to neglect $m_0$ in (\ref{KKmasses}).

However, it turns out that
not every MSSM state can have Kaluza-Klein excitations
that exactly mirror the MSSM state itself.
This complication arises because it is necessary for
Kaluza-Klein excitations to fall into representations
that permit suitable Kaluza-Klein mass terms to be formed.
This issue is particularly important for us because
a chiral MSSM state (such as a quark or lepton)
by itself cannot be given a Kaluza-Klein mass.
Thus, we cannot have an infinite tower of chiral
Kaluza-Klein excitations.
Instead, we have two choices:  either
a given chiral MSSM representation will {\it not}\/ have
Kaluza-Klein excitations, or it will, in which case
the corresponding Kaluza-Klein excitation
will consist of not only the original MSSM representation,
but also its chiral-conjugate mirror.
Both cases are also consistent with the situation in string theory.

In this paper we shall consider a variety of different cases.
Specifically, we shall let the variable $\eta$ denote the
number of generations of MSSM chiral fermions that we shall permit
to have Kaluza-Klein excitations.
Of course, the simplest scenario is one for
which $\eta=0$ --- \ie, no chiral MSSM fermions having Kaluza-Klein
excitations.
Thus, in this scenario,
only the {\it non-chiral}\/ MSSM states
(namely the gauge bosons and Higgs fields) will have
Kaluza-Klein excitations, while
the quark and lepton representations
will be assumed not to have Kaluza-Klein excitations.
We shall refer to this as the ``minimal scenario'', and as we shall
see it has a number of special properties.
However, in the expectation that string theory will ultimately give
rise to various mixtures of configurations,
in this paper we shall also consider the cases
with $\eta=1,2,3$.

It is important to consider the exact form of the Kaluza-Klein
excitations that the Higgs fields, gauge bosons, and chiral fermions will have.
In the case of the Higgs fields, for each set of non-zero
Kaluza-Klein momenta $\lbrace n_i\rbrace$ the corresponding
Kaluza-Klein state will be a chiral $N=1$ multiplet
${\cal H}^\nn \equiv (H^\nn,\psi_H^\nn)$.
Note that the structure of these chiral multiplets
does not depend on whether they are massless or massive.
Since the MSSM contains
two separate Higgs fields, we thus find that
at each Kaluza-Klein mass level
there will be two massive chiral $N=1$ supermultiplets ${\cal H}_{1,2}$.
As we shall see, it will prove convenient
to combine these two $N=1$ supermultiplets to form a
single $N=2$ hypermultiplet $H$:
\beq
  H^\nn~=~\pmatrix{
   H^\nn_1 & H^\nn_2 \cr
   \psi^\nn_{H_1} & \psi^\nn_{H_2} \cr}~
\label{Thyp}
\eeq
where we have suppressed gauge and Lorentz indices.

A similar situation exists for the gauge bosons.
An ordinary massless gauge boson is an $N=1$  vector supermultiplet.
However, a massive gauge boson is represented by a massive $N=1$
vector supermultiplet, which is equivalent to
an $N=1$ massless vector supermultiplet ${\cal A}\equiv(A,\lambda)$
plus an additional $N=1$ chiral supermultiplet ${\cal A}'\equiv(\phi,\psi)$.
Together, these form a massive $N=2$ vector supermultiplet:
\beq
  V^\nn~=~\pmatrix{
     A^\nn & \phi^\nn\cr
    \lambda^\nn & \psi^\nn \cr}~.
\label{Tvec}
\eeq
One of the
real scalar fields in the chiral supermultiplet ${\cal A}'$ becomes the
longitudinal
component of the massive gauge boson, while the other real scalar field
and the Weyl fermion remain in the spectrum at the massive level.
Thus, once again, we see that our Kaluza-Klein towers of states are effectively
$N=2$ supersymmetric.\footnote{
   Strictly speaking, the extra Kaluza-Klein towers of states will
   effectively be $N=2$ supersymmetric only for $\delta=1$ or $2$.
   For higher values of $\delta$, the situation can be more complicated ---
   \eg, for $\delta=6$ we na\"\i vely expect our Kaluza-Klein towers
   of states to be $N=4$ supersymmetric.
   In general, the enhanced supersymmetry for the excited Kaluza-Klein
   arises because the minimum number of supersymmetries
   in higher spacetime dimensions (as counted in terms of four-dimensional
   gravitino spinors) grows with the spacetime dimension.
   Thus, since the higher number of supersymmetries must
    be always restored in the limit $R\to\infty$,
   we see that our Kaluza-Klein towers must exhibit a higher number
   of supersymmetries than the ground states, even at finite $R$.
    However, by making suitable
   choices of orbifolds (as we shall see will be necessary in any case),
   it is always possible to project the relevant Kaluza-Klein towers down
   to representations of $N=2$ supersymmetry, even if $\delta >2$.
   Hence, without loss of
   generality, we shall consider $N=2$ supersymmetric Kaluza-Klein
   towers for arbitrary values of $\delta$.

     In this context, we also remark that (\ref{Thyp}) is not the
     only way in which we might have constructed an $N=2$ supermultiplet
     from the MSSM Higgs fields.  Rather than combining the two MSSM Higgs
     fields into a single $N=2$ multiplet, another possibility would
     have been to augment each Higgs field separately into its own
     $N=2$ multiplet.
     This would therefore have required the introduction of even more fields.
     We shall adopt the ``minimal'' approach of (\ref{Thyp}) in this
     section, and defer a discussion of the remaining possibilities
     to Sect.~7.
   }

Finally, in the cases $\eta \geq 1$ for which we allow certain
chiral fermions to have Kaluza-Klein excitations, these excitations
will have the form
\beq
            F^\nn ~=~ \pmatrix{
             \phi_{1}^\nn & \phi_2^\nn \cr
             \psi_{1}^\nn & \psi_2^\nn \cr}~
\label{Fmultiplet}
\eeq
where ${\cal F}_1^\nn\equiv (\phi_1^\nn,\psi_1^\nn)$ are the Kaluza-Klein
excitations of the original fermion field ${\cal F}$, and where
${\cal F}_2^\nn\equiv (\phi_2^\nn,\psi_2^\nn)$
are the Kaluza-Klein excitations of the mirror fermions.
Together, (\ref{Fmultiplet}) forms an $N=2$ hypermultiplet
(like the Higgs field).

For the purposes of this paper, it will not often be necessary to
use $N=2$ language to describe these towers of Kaluza-Klein states.
But the main point is that we shall give Kaluza-Klein excitations
to the gauge bosons, to the Higgs fields, and to only $\eta$
generations of the MSSM fermions, where $\eta=0,1,2,3$.
Thus, the effects of the extra dimensions are restricted
to only the corresponding subset of the MSSM.

At first glance, this may seem to be an inconsistent
situation because of two worries.
First, how can we have Kaluza-Klein
towers of gauge-boson states that fall into $N=2$ representations
when we know that their zero-modes (their corresponding
observable states at low energies) are only $N=1$ supersymmetric?
How is it possible to ``decouple'' the gauge-boson zero-modes from the
excited states in this way?
Second, we may also ask
how can we have Kaluza-Klein excitations for
some fields, while forbidding them for other fields.
How can this be reconciled with the presence of additional
spacetime dimensions, which are presumed to apply to
the entire theory at once?

To understand both of these points,
let us consider the case of a single additional
dimension (\ie, $\delta=1$) for simplicity, so that
our spacetime coordinates are $x\equiv ({\bf x},y)$
where $\bx\equiv (x_0,x_1,x_2,x_3)$.
Note that we can recast (\ref{KKexpansion})
into the form $\Phi(x)=\Phi_+(x) + i\Phi_-(x)$, where
\beqn
         \Phi_+(x) &=& \sum_{n=0}^\infty
        \,\lbrack \Phi^{(n)} ({\bf x})
                 + \Phi^{(-n)} ({\bf x})\rbrack\,
                      \cos(n y/R)\nonumber\\
         \Phi_-(x) &=& \sum_{n=1}^\infty
        \,\lbrack \Phi^{(n)} ({\bf x})
                 - \Phi^{(-n)} ({\bf x})\rbrack\,
                      \sin(n y/R)~.
\label{KKcossin}
\eeqn
Note that $\Phi_{\pm}(x)$ are generally complex fields.
Of course, if $\Phi(x)$ were a real field,
then only $\Phi_+$ would be non-zero.
However, even if  $\Phi(x)$ is complex, it is possible to
distinguish between $\Phi_+$ and $\Phi_-$ through their
properties under the $\IZ_2$ transformation
\beq
          y\to -y~.
\label{Z2relation}
\eeq
Specifically, we have
\beqn
         \Phi_+ ({\bf x},-y) &=& +\,\Phi_+({\bf x},y)\nonumber\\
         \Phi_- ({\bf x},-y) &=& -\,\Phi_-({\bf x},y)~.
\eeqn
What is particularly useful about the decomposition (\ref{KKcossin})
is that $\Phi_-(x)$ lacks a zero-mode.
Thus, even though our Kaluza-Klein
tower of states for the gauge bosons are
$N=2$ supersymmetric, as in (\ref{Tvec}),
we can ensure that their corresponding zero-mode is only
$N=1$ supersymmetric (as appropriate for the MSSM)
by additionally demanding that $A$ and $\lambda$ transform as
even functions under (\ref{Z2relation}), while $\phi$ and $\psi$
transform as odd functions.

What are the implications of making such additional requirements?
By demanding that our wavefunctions exhibit certain
symmetry properties under the transformation (\ref{Z2relation}),
we are not, strictly speaking, compactifying on a circle.
Instead, we are implicitly making
the additional $\IZ_2$ identification $y\approx -y$.
Such an identification changes the circle into a so-called
$\IZ_2$ {\it orbifold}\/,
so what we are really doing is compactifying on a $\IZ_2$ orbifold
rather than on a circle.
The fact that we are compactifying on an orbifold is what
allows us to demand specific symmetry properties under
the orbifold relation (\ref{Z2relation}).
Such orbifold choices are completely natural from the point
of view of string theory, and are therefore completely consistent
with an ultimate embedding of our scenario into string theory.\footnote{
  In this regard, it is also important to note that in this paper
  we are considering only the Kaluza-Klein {\it momentum}\/ states
  for which $m_n\sim n/R$.  These are the states which are appropriate
  for a field-theoretic treatment of extra spacetime dimensions.
  In string theory, however, there are also Kaluza-Klein {\it winding-mode}\/
  states for which $m_w\sim w M^2_{\rm string}R$, where $w$ is the Kaluza-Klein
  winding number and $M_{\rm string}$ is the string scale.
  This will cause no inconsistency for us
  because we will ultimately take $R^{-1}\ll M_{\rm string}$
  in our scenario.
  Thus, winding-mode states will play no role in the
  field-theoretic limit.}

Let us now consider our second question:  in the scenarios
with $\eta <3$, how can we ensure
that $3-\eta$ generations of MSSM
fermions lack Kaluza-Klein excitations altogether?
Once again, it is the fact that we are compactifying
on an orbifold which provides the explanation.
For such an orbifold compactification, we see that there
are two special points, $y^{(A)}=0$ and $y^{(B)}=\pi R$, which
are invariant under the orbifold relation (\ref{Z2relation})
in conjunction with the circle relation (\ref{circleidentify}).
Such special points are called {\it fixed points}\/ of the orbifold.
The existence of such fixed points implies that rather
than having a mode expansion of the form (\ref{KKexpansion}),
a perfectly consistent alternative mode-expansion
would be
\beq
     \Phi(x) ~=~ \Phi^{(A)}({\bf x}) \,\delta(y) ~+~
     \Phi^{(B)}({\bf x}) \,\delta(y-\pi R)~.
\eeq
Generalizations to higher spacetime dimensions are obvious.
Note that such mode-expansions also respect the symmetries of the
orbifold.
However, such mode-expansions do not
give rise to infinite Kaluza-Klein towers because
such states exist only at the fixed points of the orbifold.
Thus, it is possible to ensure that a given MSSM fermion
will have no Kaluza-Klein excitations by requiring that it
be situated only at the fixed points of the orbifold.
Once again, this is completely natural from the point of
view of string theory.  In string theory, the act of
``twisting'' by the orbifold element (\ref{Z2relation})
naturally gives rise to so-called ``twisted string sectors'',
and the physical states that arise in such twisted
sectors will precisely be of the ``fixed point'' variety.
Note that it will not matter whether these fermions are located
at $y=0$ or $y=\pi R$.

Strictly speaking, we remark that this orbifold ``fixed point''
mechanism is valid only for {\it closed}\/ string
theories (such as the heterotic string).
For open string theories, by contrast, an analogous mechanism
will involve $D$-branes and will be discussed in Sect.~7.

Thus, putting the pieces together, we see that
our higher-dimensional MSSM must be described in
terms of a spacetime consisting of four flat dimensions
along with a certain number of extra dimensions compactified on 
orbifolds of radius $R$.  At the massless (zero-mode) level,
our particle content will consist of the
full MSSM.  At the higher Kaluza-Klein levels, however, we will
have infinite towers of Kaluza-Klein states
associated with the Higgs field, the gauge-boson states,
and $\eta$ generations of the chiral MSSM fermions, where
$\eta=0,1,2,3$..
The remaining $3-\eta$ generations of chiral fermions
will not have Kaluza-Klein states, and will instead
be restricted to the fixed points of the orbifold.

Let us now consider how this system behaves at different energy scales.
At energy scales much smaller than $\mu_0\equiv R^{-1}$,
the energy of the system is less than the mass
of the lowest Kaluza-Klein excitations, and the existence of the
Kaluza-Klein states (and indeed that of the extra dimensions) can be ignored.
Thus, in this limit, our theory reduces to the usual four-dimensional
MSSM.
For $\mu\gg\mu_0$, by contrast, excitations of many Kaluza-Klein modes
become possible, and the contributions of these Kaluza-Klein states
must be included in all physical calculations.
For example, these contributions must be included in the
running of gauge couplings, and they tend to accelerate this running,
ultimately changing the scale-dependence of the gauge couplings
from logarithmic to power-law as a function of $\mu$.
This reflects the fact that beyond the scale $R^{-1}$, a certain
subset of the MSSM
is essentially higher-dimensional, and the effective radius $R$ of
these extra dimensions appears to be infinite relative to
the energy scale $\mu$.
Indeed, in this limit, the new spacetime dimensions that appear
are effectively flat.
Thus, we see that
the effect of the extra Kaluza-Klein excitations is
essentially to make the spacetime appear
to be $D$-dimensional rather than four-dimensional
for the appropriate subset of the MSSM.

At certain points in this paper, we shall need to make recourse to
a slightly modified description of the extra dimensions.
Strictly speaking,
we know that our theory contains infinite towers of Kaluza-Klein states.
However, it is clear that only the lowest-lying Kaluza-Klein states
can possibly play an important role in the physics because
the contributions of the very heavy Kaluza-Klein states are
suppressed by their large masses.
Thus, in some cases it will prove useful to
retain only a  {\it finite}\/ number of
low-lying Kaluza-Klein states in the theory.
The point at which the Kaluza-Klein towers are truncated will ultimately depend
on the energy scale at which we wish our theory to apply, but for our
purposes it will always be possible to choose such a fixed truncation point.
We shall refer to this as the ``truncated'' Kaluza-Klein description
of the extra spacetime dimensions.
As we shall see in Sect.~3, this truncated description will prove very useful.

The appearance of extra spacetime dimensions has an important
effect on the gauge couplings of the theory.
In four spacetime dimensions, the gauge couplings $g_i$ are quantities
of zero mass dimension (\ie, pure numbers).
In $D$ spacetime dimensions, however, the
gauge couplings $\tilde g_i$ accrue a classical mass dimension
\beq
            \lbrack \tilde g_i \rbrack ~=~ 2-{D\over 2}~~~~\Longrightarrow~~~~
            \lbrack \tilde \alpha_i^{-1}\rbrack ~=~ D-4 ~=~\delta~
\label{classdims}
\eeq
where $\delta \equiv D-4$.
It is therefore important to understand the connection between
the higher- and lower-dimensional couplings.
However, since our extra spacetime dimensions have a fixed radius
$R$,
we can follow the standard compactification procedure to find
that the four- and $D$-dimensional gauge couplings are related
to each other via
\beq
          \alpha_i~=~ R^{-\delta}\, \tilde\alpha_i~.
\label{Irelation}
\eeq

\subsection{Renormalizability and the interpretation of cutoffs}

Finally, we conclude this section with a few important comments regarding the
renormalizability of these higher-dimensional theories.

As is well-known, higher-dimensional field theories are non-renormalizable
because of their enhanced divergence structure.
This non-renormalizability stems from the
presence of infinite towers of non-chiral Kaluza-Klein states which circulate
in
the loops of all quantum-mechanical processes.
Even if we choose to ignore these Kaluza-Klein states by treating
the non-chiral sector of the MSSM as being in $D$ flat spacetime dimensions,
the resulting description is
still non-renormalizable because of the
need to integrate over $D$ dimensions' worth of uncompactified loop
momenta in such sectors.

Given the non-renormalizable nature of such higher-dimensional field theories,
it therefore makes no sense to talk of a ``running'' of gauge couplings
as a function of a floating energy scale $\mu$.
Instead, for a non-renormalizable theory, we must introduce
an explicit {\it cutoff}\/ parameter $\Lambda$.  Consequently,
strictly speaking,
the values of physical parameters such as gauge couplings do not ``run''  ---
they instead receive {\it finite quantum corrections}\/ whose magnitudes depend
explicitly on the value of this cutoff parameter.
Therefore, in the language appropriate to a non-renormalizable field theory,
we do not seek to calculate the ``running''
of the gauge or Yukawa couplings;
we instead seek to calculate the one-loop-corrected values of
the gauge couplings $\alpha_i(\Lambda)$ as functions of the value
of this cutoff parameter $\Lambda$.

In many cases, this mathematical dependence on the cutoff is identical
to the scale-dependence that we would have na\"\i vely calculated
if the theory had been renormalizable.
Therefore, we will occasionally continue to use words such as ``running'',
even in our non-renormalizable context, to describe
the dependence on the cutoff.

There is, however, one profound distinction that arises due to the
fact that our theory is non-renormalizable.
Since our theory is non-renormalizable, it can only be viewed as an
 {\it effective}\/ theory, valid up to some even higher mass scale $M$.
Thus, throughout this paper, our higher-dimensional theory will
be interpreted in precisely this way, as an effective theory requiring
the emergence of an even more fundamental theory (such as a string theory)
at an even higher energy scale.
Indeed, given our results, we shall see that this interpretation
will be particularly natural.

In itself, this is not a problem.
However, this then broaches the question:
just how can we interpret the cutoff parameter $\Lambda$ which will appear
throughout our calculations?
It is important to resolve this issue
because such a cutoff $\Lambda$ is not a physical parameter with
intrinsic meaning.  Indeed,  such a cutoff
 ultimately depends on the form of the
regulator in which it is presumed to appear and
thereby on the normalization that is used for defining $\Lambda$
within this regulator.

It might seem natural, of course,
to associate the cutoff parameter
$\Lambda$ with the physical mass scale $M$ at which we presume new physics to
appear
 {\it beyond}\/ our higher-dimensional non-renormalizable theory.
This would seem to make sense because, as we stated above,
we must ultimately assume that our higher-dimensional non-renormalizable theory
is
only an {\it effective}\/ description of physics for energy scales
below some new fundamental mass scale $M$.
In general, such an association works well.
However, this is not always the case.
For example, it has been pointed out~\cite{burglon} that
in certain situations one may obtain misleading results,
essentially due to poor choices of cutoff and regulator variables.
In such a case, the value of the cutoff $\Lambda$ gives no information
about the underlying mass scale $M$.
Moreover, it is not even straightforward to recognize the situations
in which such misleading results will arise.

Fortunately, in our case we will be able to completely sidestep
all of these issues by making a crucial observation.
As we have stated, the lack of renormalizability can be
attributed to the fact that our towers of Kaluza-Klein states are
 {\it infinite}\/.  However, for {\it calculational}\/ purposes
it is often unnecessary to include
 {\it all}\/ of the Kaluza-Klein states --- indeed, one may
often truncate the tower at a suitable energy level without seriously altering
the
results of a given calculation.
However, because this tower of states is truncated,
this description of the physics has the potential
to give rise to a {\it completely renormalizable field theory}\/.
This issue will be discussed in more detail in Appendix~B.
In such cases, we then have a remarkable situation:
Although our full underlying theory is
non-renormalizable, there will exist a fully renormalizable field
theory which gives essentially the same results for certain calculations.
This in turn implies that any ambiguities or uncertainties that
might arise in relating $\Lambda$ to $M$ in the full theory
can be completely resolved by making recourse to the fully renormalizable
approximation.
Therefore, by making recourse to our renormalizable approximate description
of the physics,
we shall be able to formulate a clear relation
between the cutoff parameter $\Lambda$ and the corresponding
physical mass scale $M$.
This will thereby enable us to interpret our cutoff $\Lambda$,
and likewise determine our new fundamental mass scale $M$,
without ambiguity.

%========================================================================
%  \vfill\eject
\section{Extra dimensions and gauge coupling unification}
\setcounter{footnote}{0}

Let us now begin by considering how extra spacetime dimensions
affect the Standard Model gauge couplings and their unification.

In ordinary four-dimensional field theory, gauge couplings $g_i$ are
dimensionless quantities and their evolution as a function of the
mass scale $\mu$ is given by the
usual one-loop renormalization group equation (RGE)
\beq
          {d\over d \ln \mu} \, \alpha^{-1}_i(\mu) ~=~ -{ b_i \over 2\pi}~
\label{diffeqfour}
\eeq
for which the solution is given by
\beq
         \alpha_i^{-1}(\mu) ~=~
         \alpha_i^{-1}(M_Z) ~-~ {b_i\over 2\pi}\,\ln {\mu\over M_Z}~.
\label{oneloopRGEfour}
\eeq
This is the usual logarithmic running of the gauge couplings.
Here $\alpha_i\equiv g_i^2/4\pi$, the $b_i$ are
the MSSM one-loop beta-function coefficients
\beq
        (b_Y,b_2,b_3)~=~ (11,1,-3)~,
\label{bdef}
\eeq
and we have taken the $Z$-mass $M_Z\equiv 91.17$ GeV as
an arbitrary low-energy reference scale.
At this scale (and within the $\overline{\rm MS}$ renormalization group
scheme),
the gauge couplings are given by
\beqn
        \alpha_Y^{-1}(M_Z)|_{\overline{\rm MS}} &\equiv&  98.29 \pm 0.13
                    \nonumber\\
        \alpha_2^{-1}(M_Z)|_{\overline{\rm MS}} &\equiv&  29.61 \pm 0.13
                    \nonumber\\
        \alpha_3^{-1}(M_Z)|_{\overline{\rm MS}} &\equiv&  8.5 \pm 0.5 ~
\label{lowenergycouplingsa}
\eeqn
and we shall henceforth define $\alpha_1\equiv (5/3)\alpha_Y$ and
$b_1\equiv (3/5) b_Y$.
As is well-known, an extrapolation of these low-energy couplings
according to (\ref{oneloopRGEfour})
leads to the celebrated unification relation
\beq
          \alpha_1(M_{\rm GUT}) ~=~ \alpha_2(M_{\rm GUT})
               ~=~ \alpha_3(M_{\rm GUT}) ~\approx ~ {1\over 24}
\eeq
at the unification scale
\beq
     M_{\rm GUT} ~\approx~ 2 \,\times\, 10^{16} ~{\rm GeV}~.
\label{MGUT}
\eeq

Let us now consider how this picture is modified in the presence of
extra spacetime dimensions.
Recall that, as discussed in
Sect.~2,
we shall take our  spacetime to consist
of four flat spacetime dimensions
and $\delta$ additional dimensions of radius $R$.
This results in the appearance of infinite towers of Kaluza-Klein states
of masses $m_n\approx n\mu_0$, $n\in\IZ^+$, where $\mu_0\equiv R^{-1}$.
Recall that, in general, these Kaluza-Klein states will be assumed to appear
for the gauge bosons, the Higgs fields, and $\eta$ generations of
chiral fermions of the MSSM, where we shall consider the cases
$\eta=0,1,2,3$.
By contrast, the remaining $3-\eta$ generations of MSSM
quarks and leptons will be assumed not to
have Kaluza-Klein excitations.

Below $\mu_0$, we may ignore the effects of the Kaluza-Klein
states, and simply treat our higher-dimensional theory as
equivalent to the ordinary four-dimensional MSSM.
We shall show in Appendix~A that this is an excellent approximation.
Thus, for $\mu\leq \mu_0$, we may assume that the gauge couplings run
in the usual fashion described above.

Above $\mu_0$, however, we must
take into account
the full spectrum of Kaluza-Klein states.
Because these towers of Kaluza-Klein
states are infinite, our higher-dimensional theory
is non-renormalizable.
Thus, for $\mu\geq \mu_0$,
we cannot talk of the ``running'' of these gauge
couplings.
Rather, as we discussed in Sect.~2, the gauge couplings instead receive
finite corrections which depend on the appropriate cutoffs in the theory.
In the present case, we have both an infrared cutoff $\mu_0$ (below which
the physics is effectively described by the usual four-dimensional MSSM
with no Kaluza-Klein states), and an ultraviolet cutoff $\Lambda$
(which marks the scale at which some new physics
beyond our higher-dimensional theory is presumed to emerge).

The corresponding corrections to these gauge couplings
can then be calculated
in the usual manner by
evaluating the same one-loop diagrams (particularly the
wavefunction vacuum polarization diagram) that renormalize
the usual gauge couplings in four dimensions.
Denoting the value of this diagram as
$\Pi_{\mu\nu}(k)$ where $k$ is the overall momentum flowing
through this diagram, and using gauge invariance to write
\beq
    \Pi_{\mu\nu}(k) ~=~ (k_\mu k_\nu - g_{\mu\nu} k^2) \,\Pi(k^2)~,
\label{gauginv}
\eeq
we then find the approximate one-loop result
\beq
      \Pi(0) ~=~  {g_i^2 \over 8\pi^2}\, \left\lbrack
          (b_i-\tilde b_i) \ln {\Lambda\over \mu_0} ~+~
            \tilde b_i\,  \mu_0^{-\delta} \,{X_\delta \over \delta}\,
          \left(\Lambda^\delta  - \mu_0^\delta\right)\right\rbrack~.
\label{pizero}
\eeq
In this expression,
$g_i$ is the original, uncorrected
gauge coupling,
$\delta\equiv D-4$, and
the numerical coefficient $X_\delta$ will be discussed below.
The beta-function coefficients $b_i$ are those of the usual
MSSM given in (\ref{bdef}), which correspond to the zero-mode states,
while the new beta-function coefficients $\tilde b_i$
are given by
\beq
      (\tilde b_1,\tilde b_2,\tilde b_3) ~=~ (3/5, -3,-6)~
           + ~\eta\,(4,4,4)~.
\label{btilde}
\eeq
These beta-function coefficients correspond to the contributions
of the appropriate Kaluza-Klein states at each
massive Kaluza-Klein excitation level.
As discussed in Sect.~2, at each mass level these Kaluza-Klein states
consist of two Higgs chiral $N=1$ supermultiplets as well as
the massive gauge-boson multiplets, each of which consists of
one vector and chiral $N=1$ supermultiplet.
They also include $\eta$ generations of chiral MSSM fermions,
along with their appropriate mirrors.

It is easy to interpret the form of the one-loop correction (\ref{pizero}).
The second term reflects the contributions
from the infinite towers of Kaluza-Klein states corresponding
to the MSSM states that feel the extra dimensions,
with beta-function coefficients $\tilde b_i$.
If the states in the excited levels of these Kaluza-Klein towers had matched
the zero-mode states in our theory, this term (which effectively combines
both the zero-modes and all the excited modes) would have been sufficient.
However, as we have seen, the states at the zero-mode level of our theory
actually differ from those at the excited levels.
The first term in (\ref{pizero}) therefore compensates
for this difference between the zero-mode states
and the excited states.
Moreover, as we discussed above, at energy scales above $\mu_0$ we may
treat the sector of the MSSM which has Kaluza-Klein excitations
as being effectively in $D$ flat spacetime dimensions.
The powers of $\Lambda$ and $\mu_0$ that appear in the
second term of (\ref{pizero}) thus
result from the higher-dimensional loop-momentum integration.
Equivalently, these powers may be viewed as the ``classical scaling'' behavior
that we expect the gauge couplings to experience due to their
enhanced classical mass dimensions in (\ref{classdims}).
The factor $\delta^{-1}$ in (\ref{pizero}) ensures that the
formal $\delta\to 0$ limit of the second term reproduces
the expected logarithmic behavior, and the overall factor
of $\mu_0^{-\delta}$ is required on dimensional grounds.

In Appendix~A,
we shall provide a rigorous expression for $\Pi(0)$ which does
not make the approximation of using $D$ flat dimensions for the
appropriate subsector of the MSSM,
but which instead incorporates the effects
of infinite towers of Kaluza-Klein states at all energy scales.
This will enable us to explicitly demonstrate
that our approximation (\ref{pizero}) for $\Pi(0)$ is an excellent one.

Perhaps the most subtle feature of (\ref{pizero}) is the coefficient
$X_\delta$ that appears in the second term.
Note that this coefficient
can be interpreted as providing
a {\it normalization}\/ for the cutoffs $\Lambda$ and $\mu_0$ that
appear in the $D$-dimensional integrations.
Thus, it becomes immediately apparent that within the context of our
non-renormalizable field theory, the coefficient $X_\delta$
is essentially cutoff- and regulator-dependent.
Even if we assume that our cutoff $\Lambda$ is to be associated with
an underlying physical mass scale $M$, as discussed in Sect.~2,
the relation between $\Lambda$
and $M$ would still involve an overall unknown proportionality constant
which would pollute any value of $X_\delta$
that we might otherwise calculate via phase-space arguments.
Thus, in some sense,
all of the uncertainties
inherent in working with a non-renormalizable field theory can ultimately
be embodied in our inability to determine a precise value for $X_\delta$.

Fortunately, as we discussed in Sect.~2, we can circumvent this
problem completely by making recourse
to our ``equivalent'' truncated Kaluza-Klein theory.
In this way we are then able
to make a direct comparison between the two theories in order
to precisely calculate $X_\delta$
in (\ref{pizero}).  As we shall show in Appendix~B,
this leads to the result
\beq
     X_\delta~=~ {\pi^{\delta/2}\over \Gamma(1+\delta/2)} ~=~
       {2 \pi^{\delta/2} \over \delta \Gamma(\delta/2)}~
\label{Xdef}
\eeq
where $\Gamma$ is the Euler gamma function satisfying
$\Gamma(n)=(n-1)!$, $\Gamma(1)=1$, and $\Gamma(1/2)=\sqrt{\pi}$.
Thus, $X_0=1$ (as expected), while $X_1=2$, $X_2= \pi$, $X_3=4\pi/3$,
and so forth.
In fact, $X_\delta$ is nothing but the volume $V_\delta$ of a
$\delta$-dimensional unit sphere!
Thus, in the remainder of this paper, we shall use
this precise value for $X_\delta$.  This will enable us to
interpret $\Lambda$ literally as the physical mass scale $M$ at which
new physics appears beyond our effective $D$-dimensional field theory.

Our next step is to resum these vacuum polarization diagrams
in order to obtain the full one-loop-corrected gauge couplings $g_i(\Lambda)$.
This resummation proceeds as in the usual case, yielding
\beq
      g_i(\Lambda) ~=~ \left({1\over 1-\Pi(0)}\right)^{1/2}
          \, g_i~,
\eeq
or equivalently
\beq
     \alpha_i^{-1}(\Lambda) ~=~
      \alpha_i^{-1}
      ~-~ {b_i-\tilde b_i \over 2\pi} \,\ln{\Lambda\over\mu_0}
           ~-~ {\tilde b_i X_\delta\over 2\pi \delta} \,\left\lbrack
                 \left({\Lambda\over \mu_0}\right)^\delta -1\right\rbrack~.
\label{geffd}
\eeq
The second and third terms on the
right side are the finite one-loop corrections
which depend explicitly on the cutoff $\Lambda$.

Finally, we impose our matching condition
that $\alpha_i$, the uncorrected value of the effective four-dimensional
coupling, must agree with the value of the true four-dimensional coupling
$\alpha_i(\mu_0)$ at the scale $\mu_0$.
Substituting for $\alpha_i^{-1}(\mu_0)$ from (\ref{oneloopRGEfour})
then yields our final result, valid for all $\Lambda\geq \mu_0$:
\beq
     \alpha_i^{-1}(\Lambda) ~=~
      \alpha_i^{-1}(M_Z)
      ~-~ {b_i \over 2\pi} \,\ln{\Lambda\over M_Z}
      ~+~ {\tilde b_i \over 2\pi} \,\ln{\Lambda\over \mu_0}
           ~-~ {\tilde b_i X_\delta\over 2\pi \delta} \,\left\lbrack
                 \left({\Lambda\over \mu_0}\right)^\delta -1\right\rbrack~.
\label{newsoln}
\eeq

Although this result bears a resemblance to
a renormalization group equation,
we stress that its physical interpretation is entirely different.
Rather than describe the higher-dimensional ``running''
of a higher-dimensional gauge coupling, this equation instead
expresses the dependence that such a coupling exhibits
on the value of the cutoff  $\Lambda$.
Thus, given any values for $\mu_0$,  $\delta$, and $\eta$ (which are
the parameters that describe the underlying non-renormalizable theory),
and given any value for $\Lambda$ (a parameter which is
external to the theory and which is interpreted as the energy scale at which
a new theory is presumed to appear),
$\alpha_i^{-1}(\Lambda)$
is the value of the one-loop corrected effective four-dimensional coupling.

It is clear from (\ref{newsoln}) that the
presence of the extra dimensions clearly has a profound effect on
the values of these gauge couplings.
Remarkably, however,
it turns out that there always exists a value
of $\Lambda$ for which the gauge couplings unify!
Indeed, this property is robust, occurring independently
of the number $\delta$
of extra dimensions,  independently of the scale $\mu_0$ at which they
appear, and independently of the number $\eta$ of chiral MSSM generations that
feel these extra dimensions!

\begin{figure}%[ht]
\centerline{ \epsfxsize 3.25 truein \epsfbox {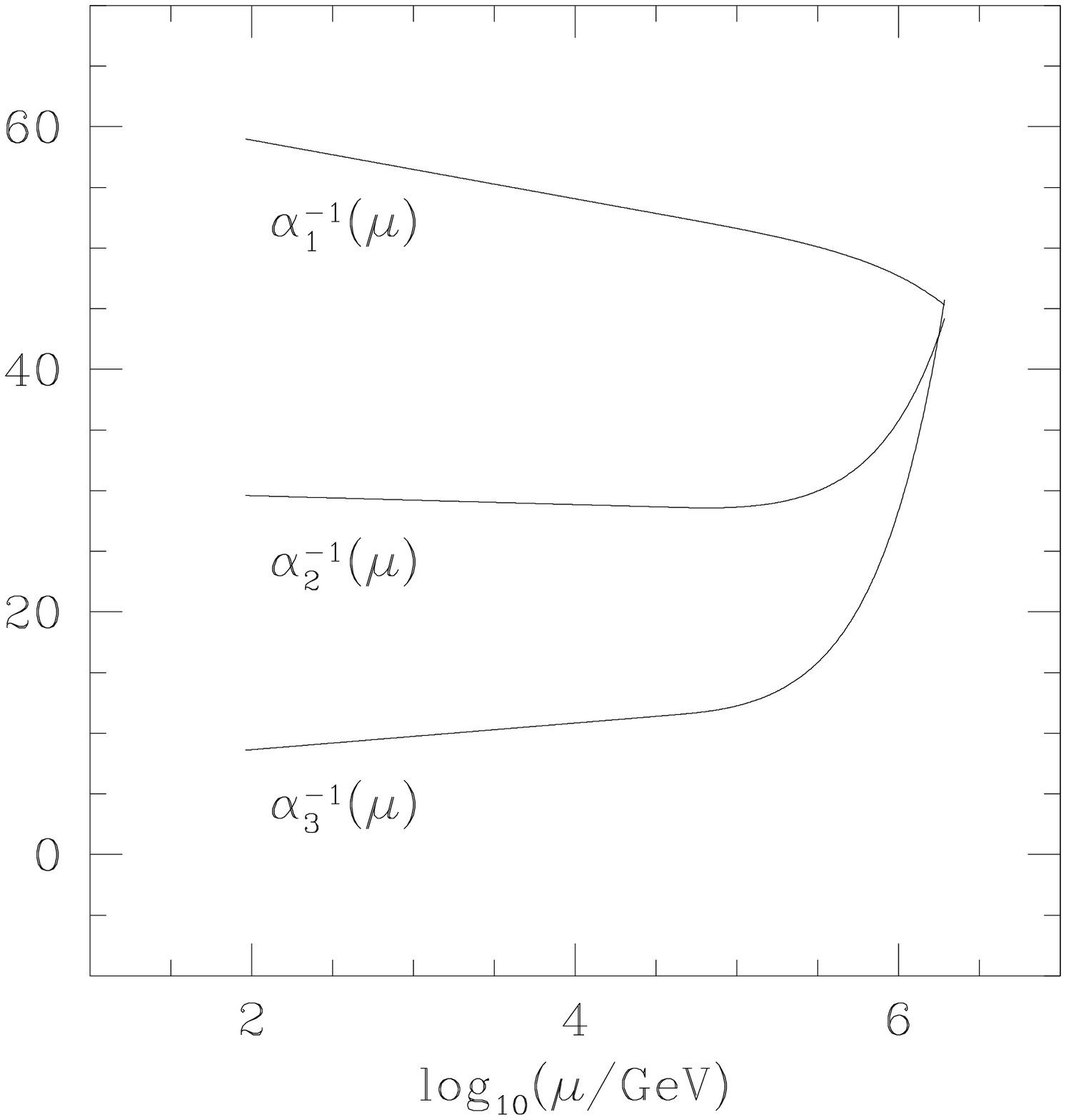}
             \epsfxsize 3.25 truein \epsfbox {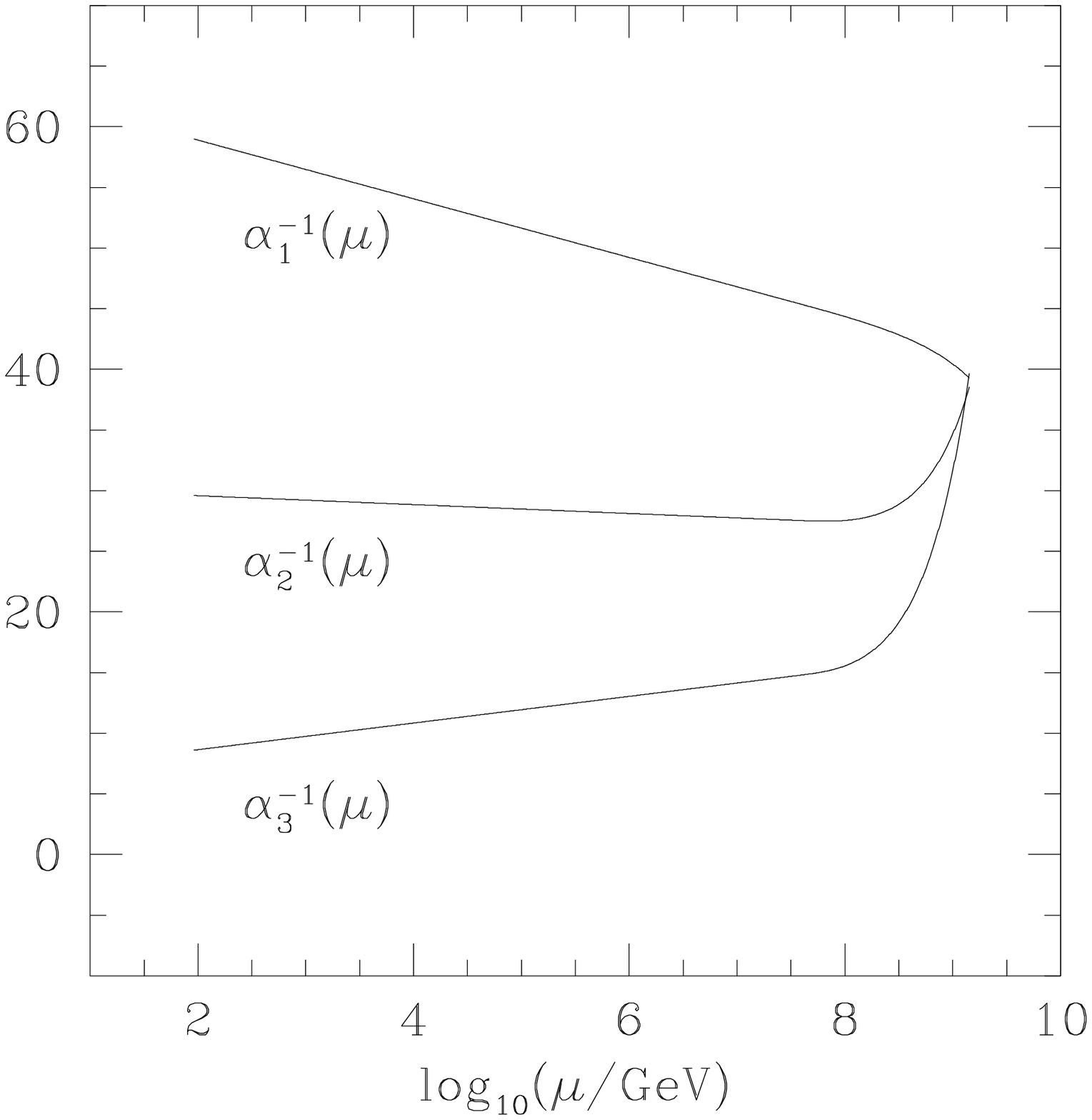}}
\centerline{ \epsfxsize 3.25 truein \epsfbox {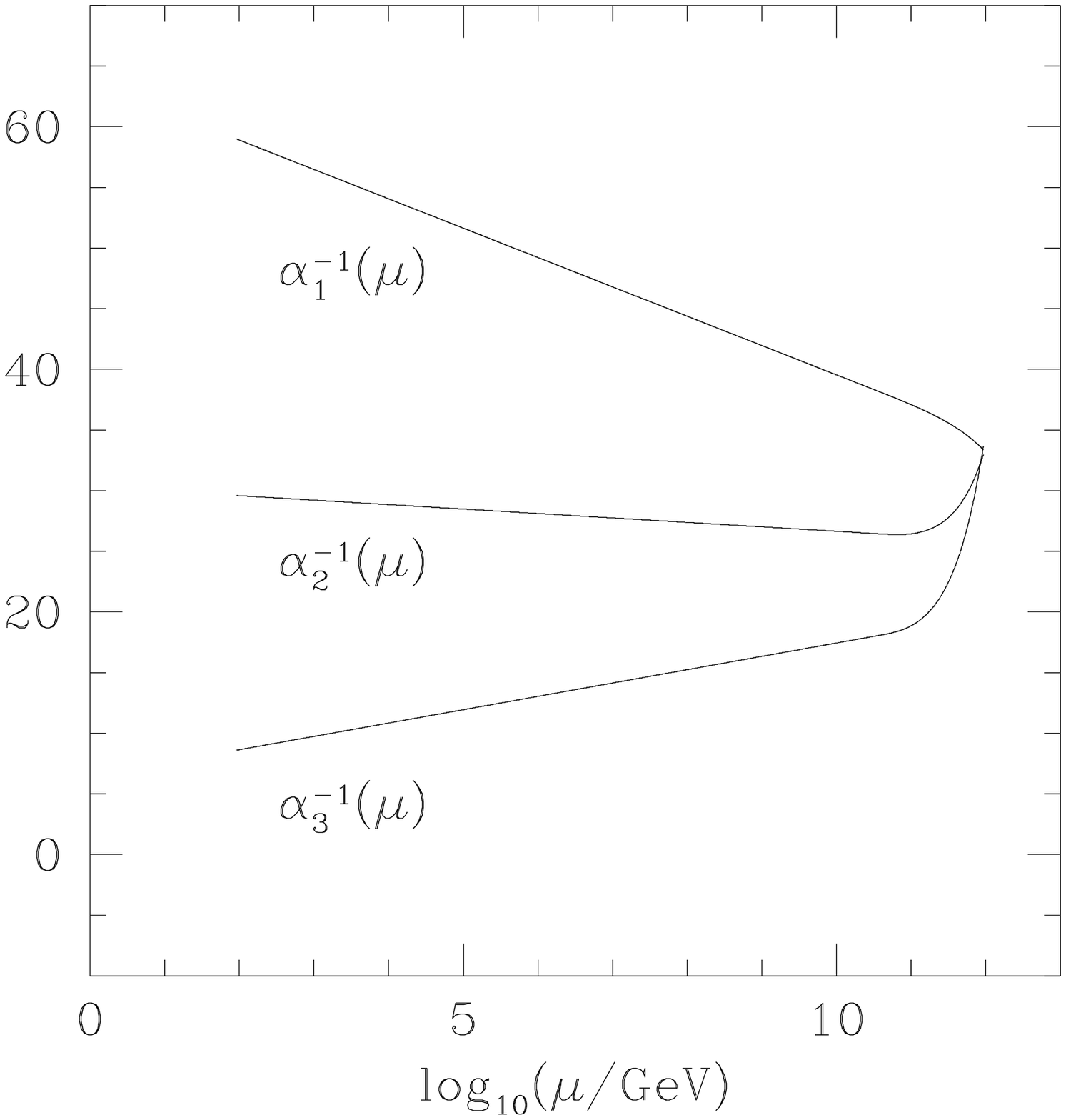}
             \epsfxsize 3.25 truein \epsfbox {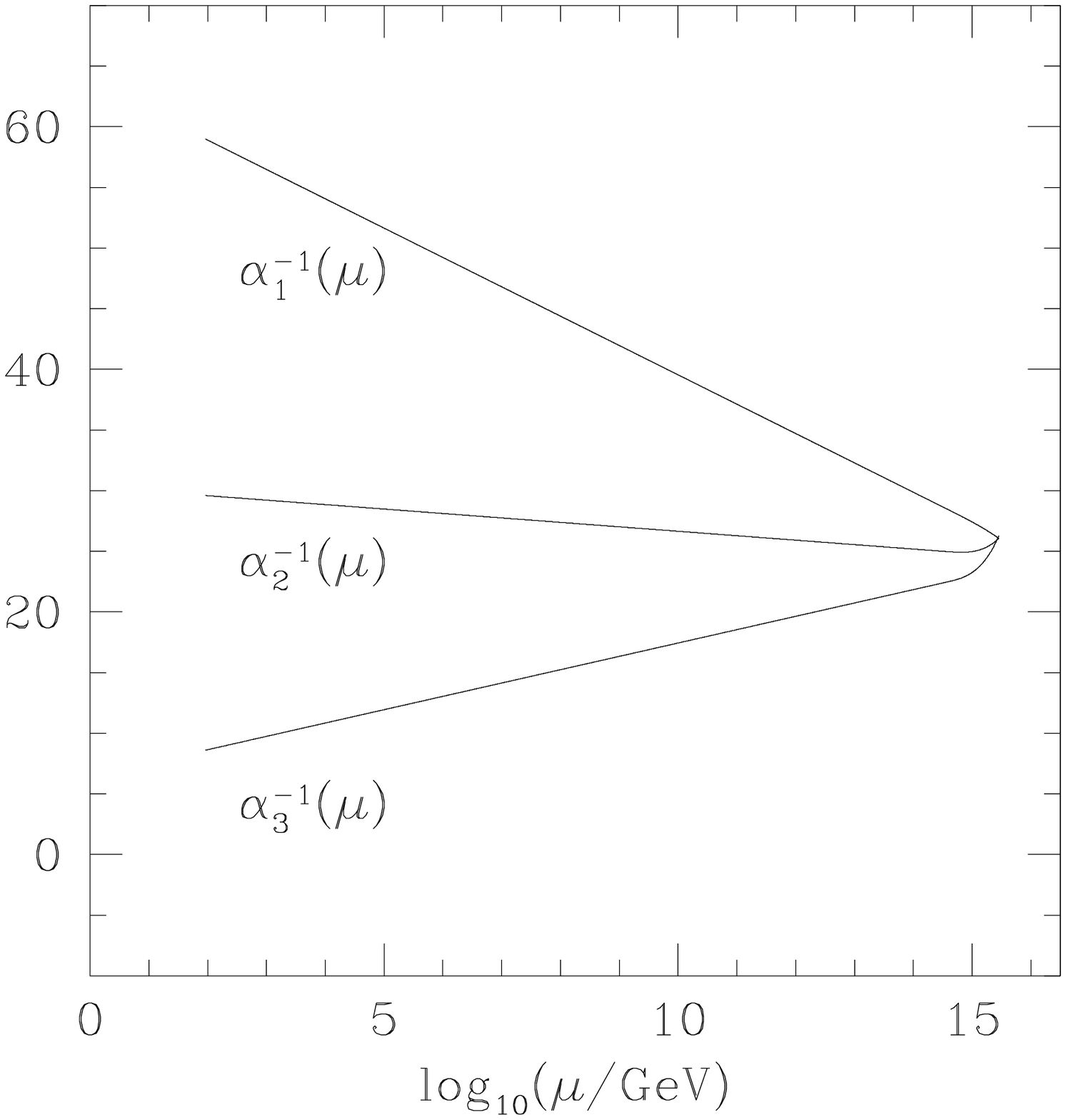}}
\caption{Unification of gauge couplings in the presence of
     extra spacetime dimensions.
     We consider four representative cases:
          $\mu_0 =  10^{5}$ GeV (top left),
          $\mu_0 =  10^{8}$ GeV (top right),
          $\mu_0 =  10^{11}$ GeV (bottom left),  and
          $\mu_0 =  10^{15}$ GeV (bottom right).
      In each case we have taken $\delta=1$ and $\eta=0$.  }
\label{unifII}
\end{figure}

\begin{figure}[ht]
\centerline{ \epsfxsize 4.0 truein \epsfbox {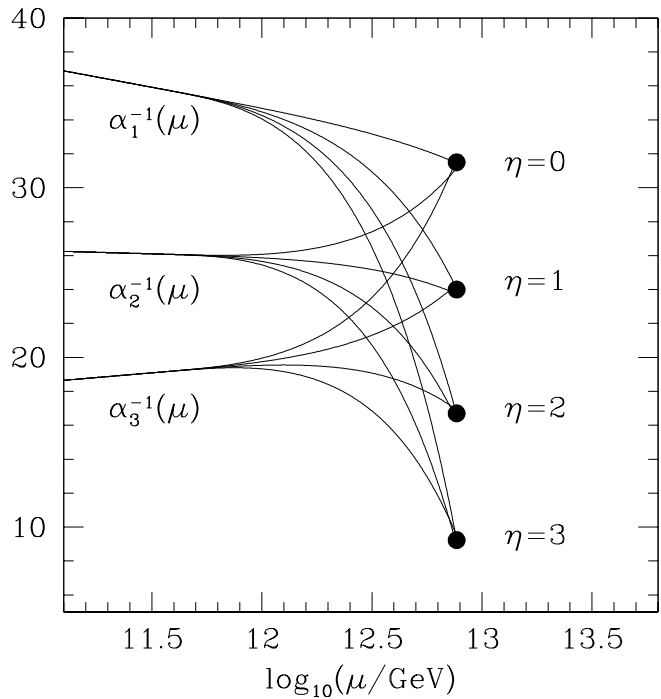}}
\caption{Unification of gauge couplings in the presence of
     extra spacetime dimensions.
     Here we fix $\mu_0= 10^{12}$ GeV, $\delta=1$,
      and we vary $\eta$.
      For this value of $\mu_0$, we see that the unification
      remains perturbative for all $\eta$.}
\label{unifnew}
\end{figure}

\begin{figure}[ht]
\centerline{ \epsfxsize 4.0 truein \epsfbox {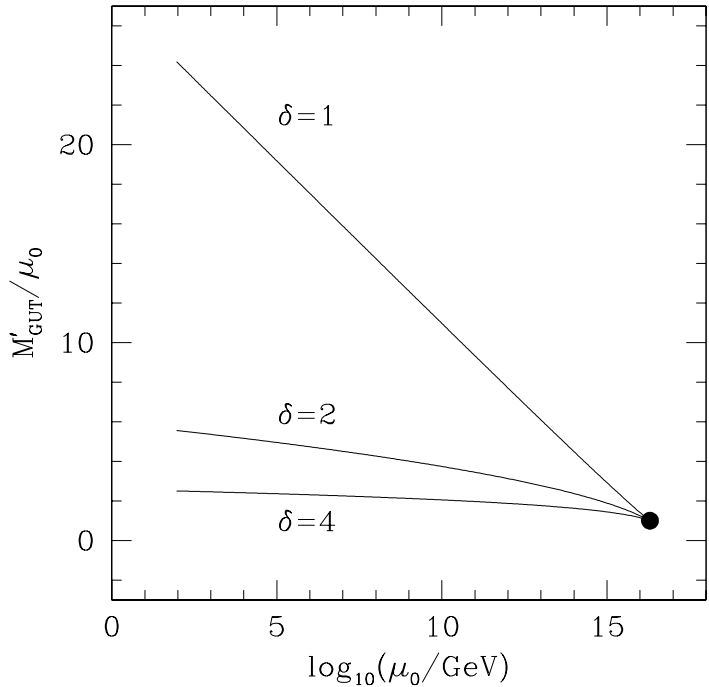}}
\caption{The ratio of the unification scale $M'_{\rm GUT}$
       to the scale $\mu_0$ at which $\delta$ extra spacetime
       dimensions appear, as a function of $\mu_0$.  This ratio
       describes the size of the energy range over which our
       effective higher-dimensional field theory is meant to apply.
       This curve is independent of the value of $\eta$.
       The limit of the usual four-dimensional MSSM is indicated
       with a dot.  }
\label{gaugetwo}
\end{figure}

\begin{figure}[ht]
\centerline{ \epsfxsize 4.0 truein \epsfbox {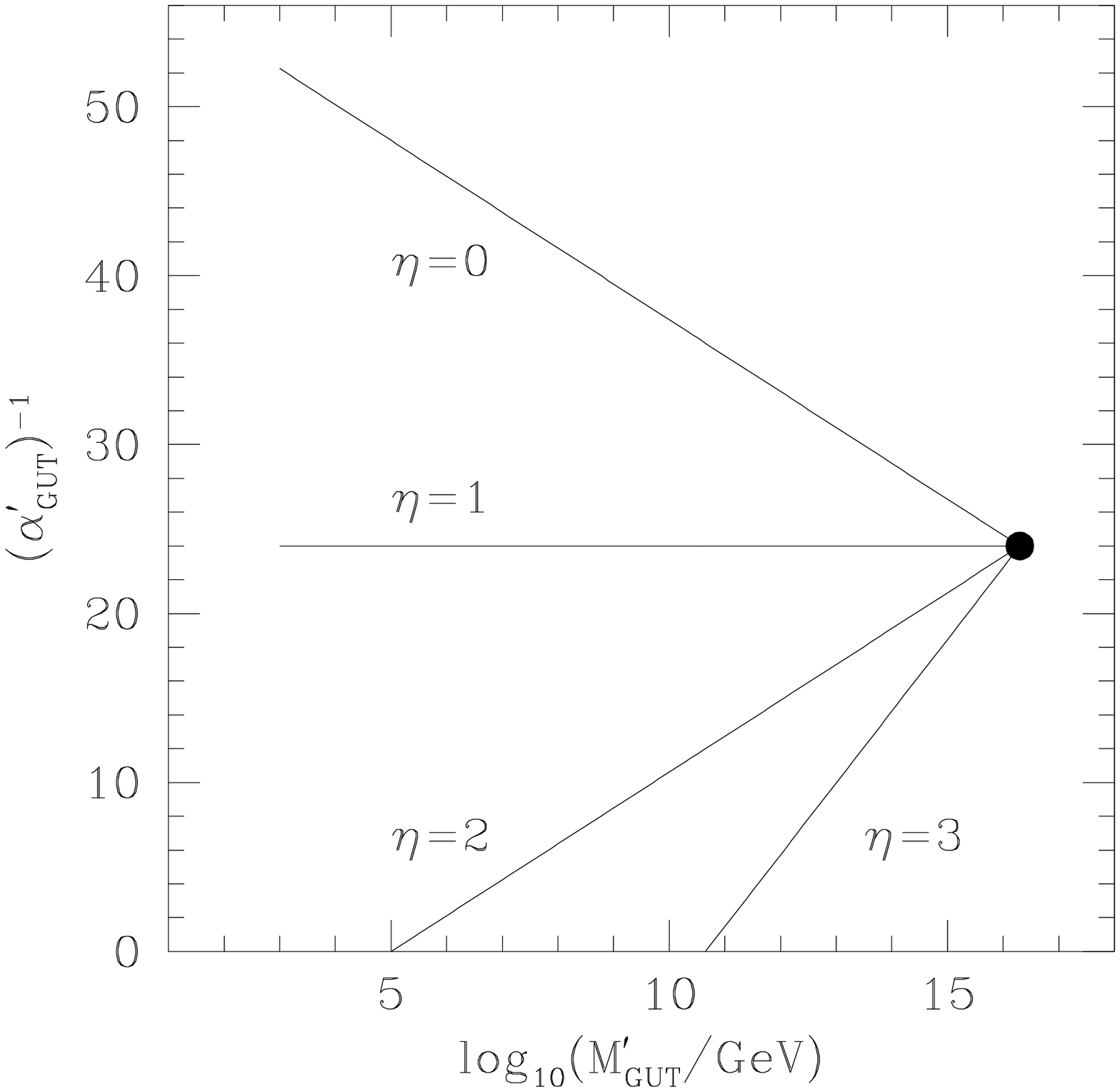}}
\caption{The unified coupling $(\alpha'_{\rm GUT})^{-1}$ as a function
       of the unification scale $M'_{\rm GUT}$, for $\eta=0,1,2,3$.
       This curve is independent of the number of extra spacetime dimensions,
       and the limit of the usual four-dimensional MSSM is indicated with
       a dot.  Note that the unified coupling is independent
       of the unification scale for $\eta=1$,
       while it is weaker than the MSSM value for $\eta=0$ and
       stronger if $\eta>1$.  It is also clear from this figure that there are
       natural lower bounds on the possible radii of extra dimensions in
       the $\eta=2,3$ cases if we wish the couplings to unify before
       diverging.  These bounds correspond
        to $M'_{\rm GUT}\gsim 100$ TeV
        and $M'_{\rm GUT} \gsim 3\times 10^{10}$ GeV
        for $\eta=2,3$ respectively.}
\label{gaugethree}
\end{figure}

This unification is illustrated in Figs.~\ref{unifII},
\ref{unifnew}, \ref{gaugetwo}, and \ref{gaugethree}.
For $\mu<\mu_0$,
we are plotting the usual running four-dimensional gauge couplings.
For $\mu>\mu_0$, however, we are treating $\mu$ as the cutoff $\Lambda$
and plotting the values of the
gauge couplings as functions of this cutoff.
It is our matching condition at $\mu_0$ which guarantees that
this procedure results in continuous curves of dimensionless numbers.
We see that below $\mu_0$, the gauge couplings run in the usual logarithmic
fashion.  Above $\mu_0$, by contrast, the appearance of the extra spacetime
dimensions causes this ``running'' to accelerate, quickly leading to
a unification.

%======================================================================
%======================================================================

%======================================================================
%======================================================================

%======================================================================
%======================================================================

%======================================================================
%======================================================================

It is natural to wonder if this unification might simply
be an artifact of our approximation
of treating the physics as being in four flat dimensions below
the scale $\mu_0$ and in $D$ flat dimensions above this scale,
all while ignoring Kaluza-Klein modes.
However, it is possible to do a rigorous calculation
which assumes only four flat dimensions at all energy scales
and which explicitly allows the complete infinite towers of Kaluza-Klein states
to circulate in the one-loop wavefunction renormalization diagram.
The details of this calculation are given in Appendix~A, and
the results are virtually identical to those in Figs.~\ref{unifII} and
\ref{unifnew}.
Thus, we see that our conclusion is unaltered:
gauge coupling unification continues to occur.

It is possible to understand physically why this gauge coupling
unification occurs.
Let us first imagine that it had been the case
that $\tilde b_i=b_i$ for all $i$.
This would have occurred, for example, if {\it all}\/ of the
MSSM states had had Kaluza-Klein excitations that exactly
matched the zero-modes,
with masses $m_n$ given in (\ref{KKmasses}).
If this had been the case, then
each energy level $\lbrace n_i\rbrace$
would have effectively provided a heavier duplicate copy of
the entire chiral MSSM particle content.  However, within the MSSM,
gauge coupling unification is independent of the number of
generations because each generation provides
extra matter in complete $SU(5)$ multiplets.
Therefore, we would have found that
gauge coupling unification is preserved regardless
of the number of extra dimensions or the scale at which they appear.

Of course, in the present case we do not have Kaluza-Klein excitations
for all of the MSSM states, and consequently $\tilde b_i\not= b_i$.
However, in order to preserve unification, we need not demand that
$\tilde b_i=b_i$:  we simply need to demand that the ratios
\beq
          B_{ij} ~\equiv~ {\tilde b_i-\tilde b_j\over b_i-b_j}
\label{Bdef}
\eeq
be independent of $i$ and $j$.
In other words, we want
\beq
          {B_{12}\over B_{13}} ~=~
          {B_{13}\over B_{23}} ~=~ 1~.
\eeq
It is easy to check that although these relations are not satisfied
 {\it exactly}\/ in our case, they are nevertheless
 {\it approximately}\/ satisfied:
\beq
          {B_{12}\over B_{13}} ~=~
            {72\over 77} ~\approx~ 0.94 ~,~~~~~~~~
          {B_{13}\over B_{23}} ~=~ {11\over 12} ~\approx~ 0.92~.
\eeq
This remains true independently of the value of $\eta$, which
shifts all $\tilde b_i$ by a fixed amount.\footnote{
   We thank C.~Wagner for questions and comments
   that prompted us to investigate the general $\eta$ scenario.}
Thus, we expect that gauge coupling unification will
continue to hold to a good degree of accuracy.  In fact, it is apparent
from Fig.~\ref{unifII} that the unification is quite precise.
Thus, given the large experimental uncertainties in the measured
value of $\alpha_3(M_Z)$  quoted in (\ref{lowenergycouplingsa}),
we see that the unification is essentially
preserved\footnote{
    In fact, given the well-known (slight) discrepancy
    between the low-energy values of $\alpha_3(M_Z)$ and the
    value required for unification within the MSSM,
    we might even be a bit bolder and hypothesize that
    it is the MSSM beta-function coefficients which lead to a
    failure of unification,
    and that our scenario actually {\it fixes}\/ unification!
    Similarly, light SUSY threshold effects might also be
    accommodated more naturally in our scenario than in the usual
     MSSM.}
for all values of $\mu_0$, $\delta$, and $\eta$.
It is also easy to see why the unification scale is independent
of $\eta$:  increasing the value of $\eta$ simply amounts to
adding extra complete $SU(5)$ multiplets to the
spectrum at each excited Kaluza-Klein mass level.
However, adding extra complete $SU(5)$ multiplets always preserves
the unification scale to one-loop order, and only shifts the unified
coupling towards stronger values.  This then explains the behavior
shown in Fig.~\ref{unifnew}.

We shall refer to the value of $\mu$ or $\Lambda$
for which this gauge coupling unification occurs
as the ``new unification scale'' $M'_{\rm GUT}$.
Indeed, it is natural to interpret such a cutoff $\Lambda$ for
which the gauge couplings unify
as the energy scale at which we would expect a more fundamental
grand-unified theory to appear.
In any case, as we have mentioned, we are always free to pass
to a description in terms of our
equivalent renormalizable truncated Kaluza-Klein theory.
In such a case, these figures can indeed be interpreted as describing
the running of gauge couplings in the usual sense, and $M'_{\rm GUT}$
can indeed be interpreted as the scale of unification.
Thus, we see that our scenario
naturally predicts the emergence of
a $D$-dimensional GUT at the scale $M'_{\rm GUT}$.

Despite the interpretation of $M'_{\rm GUT}$ as a
potential grand-unification scale,
it is evident from Fig.~\ref{unifII} that
this new unification scale $M'_{\rm GUT}$ is generally {\it not}\/ the
usual unification scale $M_{\rm GUT}\equiv 2\times 10^{16}$ GeV
that appears in the ordinary MSSM.
Instead, it is a good deal lower.
In order to solve the equations (\ref{newsoln}) for the
unification parameters, let us first define
\beq
       B~\equiv~ {1\over 3}(B_{12} + B_{23} + B_{13}) ~=~
             {233\over 336} ~\approx~ 0.69~
\eeq
where the $B_{ij}$ are defined in (\ref{Bdef}).
We will then make the approximation that
$B_{ij}=B$ for all $(i,j)$, as required by our
assumption of unification.
With this assumption, it is then possible to solve (\ref{newsoln})
at the unification point.
For any value of $\mu_0$ and $\delta$, we find that $M'_{\rm GUT}$
is approximately given by
\beq
       M'_{\rm GUT} ~\approx~ \mu_0\, f^{1/\delta}
\label{newgutscale}
\eeq
where $f$ is an exponential enhancement factor, defined as
\beq
       f~\equiv~ 1 +  {\delta \over X_\delta B}\,
                \ln {M_{\rm GUT} \over \mu_0} ~  \geq ~ 1~.
\label{fdef}
\eeq
As $\delta\to 0$ or as $\mu_0\to M_{\rm GUT}$,
we see that $f\to 1$ and $M'_{\rm GUT}\to M_{\rm GUT}$ as well.
This is the MSSM limit.
In all other cases, however,
we see that our new unification scale $M'_{\rm GUT}$
can be reduced relative to the usual $M_{\rm GUT}$, and
ultimately depends on the chosen values of $\delta$ and $\mu_0$.
This behavior is shown in Fig.~\ref{gaugetwo}.

The new unified coupling $\alpha'_{\rm GUT}$
also generally differs from its MSSM value $\alpha_{\rm GUT}\approx 1/24$.
We find the approximate analytical result
\beq
      (\alpha'_{\rm GUT})^{-1} ~\approx~
       \alpha_{\rm GUT}^{-1}  ~+~
       {2\over \pi B}\, (1-\eta)\,\ln {M_{\rm GUT}\over M'_{\rm GUT}}~.
\label{alphagutprime}
\eeq
This behavior is shown in Fig.~\ref{gaugethree}.
Thus, we see that in the $\eta=0$ ``minimal'' scenario,
$\alpha'_{\rm GUT}$
is always less than $\alpha_{\rm GUT}$ --- \ie, our theory is always
 {\it more perturbative}\/ than the usual MSSM.
Likewise, for $\eta=1$, the theory is always {\it exactly as perturbative}\/
as the usual MSSM, and the unified coupling exhibits an intriguing
invariance under changes in the unification scale 
(a sort of ``conformal'' symmetry).
For $\eta=2$, the theory is less perturbative than the usual MSSM,
and in fact the unified coupling becomes infinite near
$M'_{\rm GUT}\approx 100$ TeV.  Thus, for $\eta=2$, we see
that there is a {\it natural lower bound}\/ on the radii of the extra
dimensions
if we wish to have the gauge couplings unify before they diverge.
Similarly, for $\eta=3$, the theory hits an infinite unified coupling at
$M'_{\rm GUT}\approx 3\times 10^{10}$ GeV, which again
determines a natural lower bound
on the radii of extra dimensions in this scenario.
Note that this behavior for $\alpha'_{\rm GUT}$
as a function of $M'_{\rm GUT}$ is
independent of $\delta$.

One might worry that our calculation has
only been performed to one-loop order, and that higher-loop
corrections might destabilize the unification of gauge couplings
that we have achieved.  Indeed, this worry is particularly significant
in the case of our higher-dimensional theory
because the gauge couplings have power-law rather than logarithmic
behavior.  Thus, two- and higher-loop effects might be expected
to be particularly large.  
Moreover, there are also various subtleties involved in assessing
the perturbativity of a higher-dimensional theory:  
although the unified {\it gauge coupling}\/ is clearly perturbative
in our minimal $\eta=0,1$ scenarios,
the actual expansion parameter of the higher-dimensional
perturbation series is typically the gauge coupling 
multiplied by the effective number of Kaluza-Klein states
in the theory.
Thus, once again, one might suspect that higher-loop corrections
to the unification might be sizable.
However, it is easy to see that all
higher-order corrections to the gauge couplings are at most logarithmic.
This is because
the excited Kaluza-Klein states fall into $N=2$ supermultiplets,
thereby guaranteeing that the corresponding power-law corrections
to the gauge couplings must vanish identically beyond one-loop.\footnote{
        For this purpose, it is necessary to verify that the orbifolding
        procedure discussed in Sect.~2 actually preserves the $N=2$
        supersymmetry
        of the excited Kaluza-Klein states.  This in turn
        depends on certain model-dependent details concerning the manner
        in which the MSSM is embedded into higher dimensions.
        This issue will be discussed further in Sect.~7.
        However, as we shall see, it is not difficult to ensure
        $N=2$ supersymmetry at the massive Kaluza-Klein levels.
        We shall therefore assume unbroken $N=2$ supersymmetry
        at the massive levels, both here and in subsequent sections.}
Indeed, the only corrections that might exist beyond one-loop order
are the logarithmic corrections that come from the zero-modes.
Thus, all higher-loop corrections to the gauge
couplings will be small (as they are in the usual MSSM),
and consequently our gauge coupling unification
can be expected to survive beyond one-loop order.

Thus, to summarize this section, we see that
the appearance of extra spacetime dimensions
offers the interesting possibility of having
gauge coupling unification at scales that
are reduced, in some cases substantially, relative to the usual GUT scale.
In particular, we see that it is no longer necessary for gauge coupling
unification to occur at the usual, comfortably remote energy
scale $M_{\rm GUT}$.
Moreover, we see that the unified gauge coupling in our ``minimal'' $\eta=0$
scenario is always less than its value within the usual MSSM.
Thus, for all values of $\mu_0$ and $\delta$,
the ``minimal'' theory is {\it even more perturbative}\/ than the MSSM.
Similarly, the $\eta=1$
scenario is  always {\it exactly as perturbative}\/ as the MSSM,
and  even the $\eta=2$ and $\eta=3$ scenarios can lead to perturbative
unification for suitable values of $\mu_0$.

Such scenarios can be expected to have a variety of
striking implications, ranging from
new mechanisms for suppressing proton decay
to possible explanations for the fermion mass hierarchy.
We shall therefore now turn our attention to these important issues.

%========================================================================
%  \vfill\eject
\section{Extra dimensions and proton decay}
\setcounter{footnote}{0}

Perhaps the most immediate question that arises
in our scenario is the question of proton decay.
In this section we shall show that for the ``minimal''
scenario with $\eta=0$,
there is a {\it higher-dimensional}\/ mechanism involving
Kaluza-Klein selection rules which enables us to
cancel the usual proton-decay diagrams to all orders
in perturbation theory.
We shall also show that the non-minimal scenarios with
higher values of $\eta$ also lead to higher-dimensional
suppression mechanisms for proton decay.

In our scenario, $M_{\rm GUT}$ is lowered by an amount 
which depends on $\mu_0$ (the
scale of new dimensions) and $\delta$ (the number of extra dimensions).
However, in our minimal $\eta=0$ scenario,
the unified gauge coupling is weaker.
Thus, {\it a priori}\/,
relative to the usual supersymmetric GUT amplitude,
the leading amplitude for proton decay in our scenario
is larger by a factor of
\beq
     \left( \alpha'_{\rm GUT} \over \alpha_{\rm GUT}\right) \,
     \left( M_{\rm GUT} \over M'_{\rm GUT}\right)^2  ~\approx~
       f^{-2/\delta}\, \left\lbrack 1+ {X_\delta\over 12
      \pi\delta}(f-1)\right\rbrack^{-1}\,
              \exp\left\lbrack {\delta\over X_\delta B} (f-1)\right\rbrack
\label{badfactor}
\eeq
where $M'_{\rm GUT}$, $f$, $X_\delta$, and $B$ are defined in 
 (\ref{newgutscale}), (\ref{fdef}),
(\ref{Xdef}), and (\ref{Bdef}) respectively.
Using the experimental bounds on the
proton lifetime, we can then derive bounds on $\mu_0$ and $\delta$.
We obtain
$\mu_0\gsim 1\times 10^{14}$ GeV for $\delta=1$,
$\mu_0\gsim 3\times 10^{14}$ GeV for $\delta=2$,
and $\mu_0\gsim 8\times 10^{14}$ GeV for $\delta=3$.
Thus, as long as $\mu_0$ is sufficiently large,
the usual proton decay bounds
can be satisfied in each case.

The above calculation exactly mimics the usual calculation of
proton decay, and solves
the proton decay problem by pushing the scale of grand unification
back up to the usual neighborhood near $10^{16}$ GeV.
However, this would then ruin much of the attractiveness of
our scenario.
Is there a better way?

One observation is that above the scale $\mu_0$, the
physics (and in particular our presumed grand-unified theory) is
higher-dimensional.  Thus, it may be possible
to take a course that cannot be followed within the usual MSSM,
namely to find {\it an intrinsically higher-dimensional
solution to the proton-decay problem}\/.

Let us now see how such a higher-dimensional mechanism
might work.  As we have already discussed in
Sect.~2, 
we can 
embed the MSSM into higher dimensions
only by compactifying our higher-dimensional
theory on an {\it orbifold}\/.
In the case of a $\IZ_2$ orbifold,
this requires that we decompose all of our higher-dimensional
quantum fields $\Phi(x)$ into even and odd functions $\Phi_\pm(x)$
of these extra coordinates, as in (\ref{KKcossin}).
Given that such an orbifold is already necessary on the grounds
of the MSSM alone, let us now consider how this same orbifold
might be exploited in the case of proton decay.
Let us begin by
separating all of the quantum fields of our grand-unified theory
into two groups depending on
whether they are present in the MSSM alone, or whether they appear
only in the full GUT theory.
In a purely schematic notation,
we shall denote by $\Phi(x)$
any field which is present in the MSSM alone,
and use $\Psi(x)$ to denote any field which only appears
at the level of the GUT.
Thus, for example, the quarks, leptons,
gluons, $W^\pm$, $Z$, and Higgs doublets are all fields of the
$\Phi(x)$ variety, while the $\Psi(x)$ fields
include $X$-bosons and Higgs triplets.
It is the appearance of the $\Psi(x)$ fields that leads to
proton decay.

We have already discussed in Sect.~2 what the symmetry properties
of the $\Phi$ fields must be with respect to the compactified
coordinates $y_i$, $i=1,...,\delta$.
The fact that we do not expect to see the $\Psi$ fields
at low energies (for which we demand only the MSSM gauge
group and spectrum) suggests that we take the $\Psi$ fields
to be odd functions of the $y_i$, so that we retain only
the fields $\Psi_-$ as in (\ref{KKcossin}).
This choice preserves the
gauge symmetries of the MSSM while simultaneously reflecting the
breaking of the GUT symmetry below $M'_{\rm GUT}$,
and guarantees that the $\Psi$ fields have no zero-modes
which could be observable at low energies.
Thus, with this choice, all of the fields that mediate proton decay
will be odd functions
of these extra spacetime coordinates.

%======================================================================
\begin{figure}[th]
\centerline{ \epsfxsize 5.0 truein \epsfbox {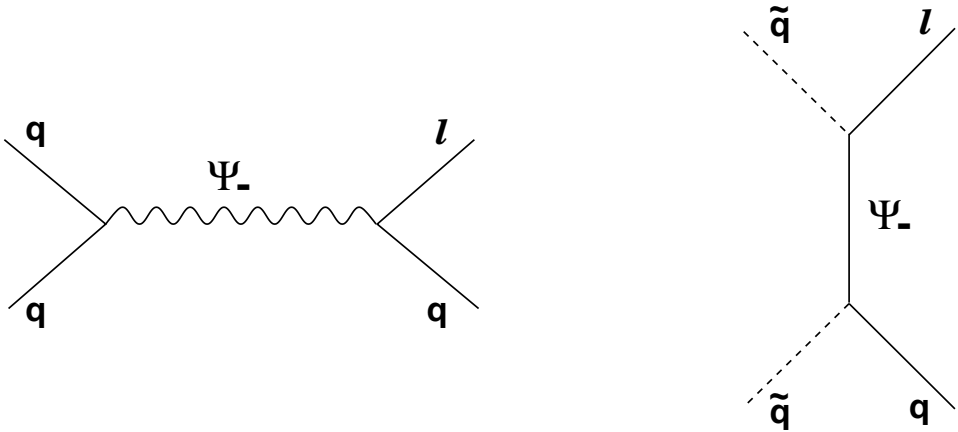}}
\caption{Typical diagrams that can mediate proton decay.
   Here the external lines correspond to the MSSM (s)quarks and leptons,
   while the internal $\Psi$ fields correspond to $X$-bosons
   (as in the diagram on the left) and
   Higgsino fields (as in the diagram on the right).
   We choose the wavefunctions of the $\Psi$ fields to be
   odd $(-)$ functions of the extra spacetime coordinates.
   With this choice, all vertices between the $\Psi$ fields
   and the chiral MSSM fermions vanish identically in the ``minimal''
    $\eta=0$ scenario, and
   all possible proton-decay diagrams vanish to all orders in
   perturbation theory.}
\label{protondecayfig}
\end{figure}
%======================================================================

This fact has dramatic consequences for proton decay.
Let us consider a typical diagram that can mediate proton decay,
as illustrated in Fig.~\ref{protondecayfig}.
We have already seen in Sect.~2 that
in the ``minimal'' $\eta=0$ scenario,
the MSSM fermions are restricted
to the fixed points of the orbifold.
However,
if the $\Psi_-$ fields are odd functions of the compactified
coordinates,
then their {\it wavefunctions
vanish at the orbifold fixed points}.
Indeed, this property holds for {\it all}\/ of the Kaluza-Klein
modes of the $\Psi_-$ fields.
Thus, to all orders in perturbation theory,
there is simply no coupling of the $\Psi_-$ fields
to the low-energy quarks and leptons of the MSSM.
In other words, {\it all}\/ such perturbative
proton-decay diagrams, such as the diagram in Fig.~\ref{protondecayfig},
vanish identically.
Note that this result
holds not only to all orders of perturbation theory,
but also independently of the number of extra dimensions
or the energy scale at which they appear.

Implicit in this proposal is also a solution to the famous
doublet-triplet splitting problem.
Rather than make the Higgs triplets much heavier than the Higgs
doublets, which is the situation required in the usual GUT scenarios,
we instead can allow the Higgs triplets to remain relatively light
because they simply do not couple to the chiral MSSM fermions.
As with the $X$-bosons, the Higgs triplets do not couple because
their wavefunctions vanish at those locations in the fifth dimension
at which the chiral MSSM fermions are located.
Thus, no large mass ``splitting'' is required at all.

It is tempting to think of this suppression mechanism
as simply Kaluza-Klein momentum conservation.
After all, such an argument would state that $\Psi_-$
fields have no zero-modes,
whereas the MSSM fermions are essentially zero-mode states.
Thus, conservation of the Kaluza-Klein momentum
at the vertex would seem to imply that any such tree-level vertex
must vanish.
However, such an argument is ultimately incorrect because
we cannot impose Kaluza-Klein momentum conservation in our theory
because we have explicitly broken translational invariance
with respect to the compactified coordinates $y_i$ when we introduced
our orbifold relations $y_i\to -y_i$.  Indeed, as we have
seen, this action results in fixed points $y_i=0$
and $y_i=\pi R$ which are not translationally invariant.
Moreover, even if we could impose Kaluza-Klein momentum conservation,
such a simple mechanism would at best only suppress tree-level
proton-decay diagrams;  the simplest one-loop box diagram would
easily evade such a constraint.
Our wavefunction mechanism, by contrast, is more robust, for it
completely decouples the low-energy quarks from all of the
proton-decay mediating effects of the full GUT theory to all orders
in perturbation theory.
Indeed, as we shall see in Sect.~7, this mechanism continues to work
even in the more general cases of open-string theories. 

Thus, we conclude that by making a judicious choice
for the modings of the GUT fields with respect to the
$\IZ_2$ orbifold that produces our $N=1$ MSSM states
in the $\eta=0$ minimal scenario,
it is possible to exploit the higher-dimensional nature of the
grand-unified theory in such
a way as to completely eliminate proton decay to all orders
in perturbation theory.
This is clearly a symmetry argument which relies on the presence
of the extra compactified dimension, and thus has no
analogue in ordinary four-dimensional grand unification.

Finally, let us briefly consider the cases with $\eta\geq 1$.
In such cases, not all of the chiral MSSM generations are restricted
to orbifold fixed points (or three-branes of an open string theory),
and consequently there can be couplings between these fermions
and the $\Psi_-$ fields that mediate proton decay.  However, for $\eta<3$,
there will typically be new, large, higher-dimensional suppression
factors associated with proton-decay amplitudes.  For example,
let us consider the case $\eta=1$, and assume that only the third
generation has Kaluza-Klein excitations.
In this case, the $\Psi_-$ fields can couple only to the fermions
of the third generation, and therefore proton decay can proceed
only through a higher-order diagram which will be suppressed not only
by extra loop factors but also by products of small CKM matrix
elements.
Thus, the $\eta=1$ scenario probably does not have a problem with
proton decay for $\mu_0\gsim 10^{12}$ GeV.
Similar arguments (leading to a more stringent bound) would
also apply to the $\eta=2$ case.
In any case, we point out that within the context of string theory,
it is also possible to circumvent problems of proton decay through other
model-dependent mechanisms, such as the selection rules corresponding to
hidden-sector discrete symmetries.

%  \vfill\eject

%========================================================================
\section{Extra dimensions and the fermion mass hierarchy}
\setcounter{footnote}{0}

We now address the question of understanding the
fermion mass hierarchy
within the context of our extra-dimensional
theory.  In fact, it is in attempting to explain
the fermion mass hierarchy that our scenario
is particularly powerful, for (unlike the situation with the
gauge couplings) we now are faced with attempting to unify
a set of Yukawa couplings whose values at low energies
differ by many orders of magnitude.
Specifically, if we let $y_F$
$(F=u,d,s,c,b,t,e,\mu,\tau)$
denote the different Yukawa couplings for the quarks and leptons
and define $\alpha_F\equiv y_F^2/4\pi$ in analogy with the gauge
couplings, we are faced with the task of explaining (\ie, approximately
unifying) low-energy values ranging
in order of magnitude from $\alpha^{-1}_t\approx 1$ to
$\alpha_e^{-1}\approx 10^{12}$.
This is extremely difficult to reconcile within the context
of the usual grand-unification scenario in which the Yukawa couplings
run only logarithmically.

\subsection{The effects of extra dimensions}

Once again, our fundamental idea is that the presence
of extra dimensions causes the Yukawa couplings to evolve
exponentially rather than linearly as a function of $\log \mu$.
Under the proper conditions,
this exponential evolution
might therefore be capable of producing a
relatively  large fermion mass hierarchy over
a relatively small energy scale interval.

We begin by recalling how the nine
Yukawa couplings $y_F$
(with $F=e,\mu,\tau,u,d,s,c,b,t$)
run within the usual four-dimensional MSSM.
These Yukawa couplings are defined in relation to the corresponding
fermion masses $m_F$ via
\beq
           m_F ~=~ y_F\, \times \, v\, \times\, \cases{
               \cos \beta & for down-type quarks and leptons\cr
               \sin \beta & for up-type quarks\cr}
\eeq
where $v\approx 174$ GeV and where $\tan\beta$
is the usual ratio of up- and down-type Higgs VEV's in the MSSM.
These couplings
then appear in the superpotential, which
takes the generic form
\beq
          W ~=~ \sum_F \, y_F\, F \overline{F}\,H_{u,d}~.
\label{MSSMsuperpot}
\eeq
If we define $\alpha_F\equiv y_F^2/4\pi$, then just like the gauge
couplings $\alpha_i$ in (\ref{diffeqfour}),
these Yukawa couplings run in the usual MSSM according
to a one-loop RGE of the form
\beq
        {d\over d \ln \mu} \, \alpha^{-1}_F(\mu) ~=~ -{ b_F(\mu) \over 2\pi}~.
\label{yukrunning}
\eeq
Indeed, the only difference relative to the gauge couplings is that
the one-loop beta-function ``coefficients'' $b_F(\mu)$ are not constants,
but instead depend on the scale $\mu$ through extra terms that depend
on the couplings themselves.  For example, within the usual MSSM,
$b_t(\mu)$ is given by:
\beq
   b_t ~\equiv~  6 ~+~ {1\over \alpha_t}\, \left( \alpha_b + 3\alpha_u +
3\alpha_c
             -{ 16\over 3}\, \alpha_3 - 3 \alpha_2 - { 13 \over 15}\,
\alpha_1\right)~,
\eeq
and each of the other $b_F(\mu)$ has a similar form.

%======================================================================
\begin{figure}[ht]
\centerline{ \epsfxsize 4.5 truein \epsfbox {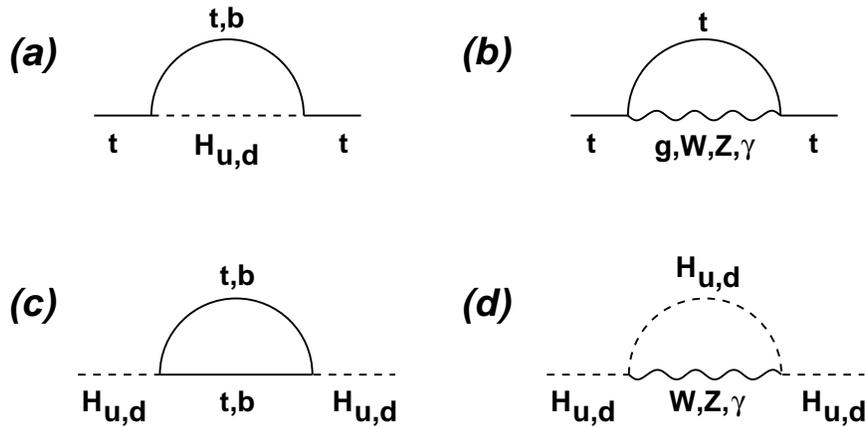}}
\caption{Typical classes of superfield wavefunction renormalization
       diagrams in the MSSM.
   Diagrams (a,b) contribute to $Z_{t}$ and $Z_{\overline t}$,
   while diagrams (c,d) contribute to $Z_{H_{u,d}}$.
    }
\label{yukfigs}
\end{figure}
%======================================================================

It will be important for us to recall how (\ref{yukrunning}) is derived.
Since the Yukawa coupling $y_F$ describes the Higgs/fermion/antifermion
interaction  term (\ref{MSSMsuperpot}),
the renormalization of $\alpha_F$ depends on the wavefunction renormalization
factors $Z_i$ of each of these three fields:
\beq
             \alpha_F^{-1}(\mu) ~=~ Z_H Z_F Z_\Fbar ~\alpha_F^{-1}(\mu_0)~.
\label{ZZZ}
\eeq
These three wavefunction renormalization factors $Z_i$ are calculated
by evaluating diagrams such as those shown in Fig.~\ref{yukfigs}.
In general, we obtain a result of the form
\beq
      Z_i ~=~ 1 ~- ~ {\gamma_i\over 2\pi}\, \ln {\mu\over \mu_0}~
\eeq
where $\gamma_i$ is the anomalous dimension of the field $i$.
For example, in the case of the Higgs and top quarks, we find
\beqn
          \gamma_t &=& \alpha_t + \alpha_b - {1\over 30}\alpha_1 -{3\over
2}\alpha_2
                          -{8\over 3}\alpha_3\nonumber\\
          \gamma_{\overline t} &=& 2\alpha_t - {8\over 15}\alpha_1
                          -{8\over 3}\alpha_3\nonumber\\
          \gamma_{H_u} &=& 3\alpha_t + 3\alpha_c +3\alpha_u
            - {3\over 10}\alpha_1 -{3\over 2}\alpha_2~.
\eeqn
Combining these factors within (\ref{ZZZ}) and keeping terms at most linear
in the logarithms then yields (\ref{yukrunning}), with
\beq
      \alpha_F\,b_F ~\equiv~ \gamma_F +\gamma_{\Fbar} +
                         \gamma_{H_i}~.
\eeq

Let us now consider how this calculation is modified
in the presence of extra spacetime dimensions.
As before, we shall assume that a certain number $\delta$ of
extra spacetime dimensions
appear at an energy scale $\mu_0\equiv R^{-1}$, and for simplicity
we shall concentrate on the ``minimal'' scenario with $\eta=0$.
Below the scale $\mu_0$, the Yukawa couplings run according
to (\ref{yukrunning}), as in the usual four-dimensional MSSM.
Above the scale $\mu_0$, however, the Yukawa couplings
instead receive finite one-loop corrections whose
size is a function of the cutoff $\Lambda$.
By comparison with our prior results for the gauge couplings,
it is straightforward to
write down the expected general form for these corrections:
\beq            \alpha_F^{-1}(\Lambda) ~=~
              Z_H Z_F  Z_\Fbar ~\alpha_F^{-1}(\mu_0)~
\label{ZZZnew}
\eeq
where now we expect
our wavefunction renormalization factors to have
the general form
\beq
      Z_i ~=~ 1 ~- ~
          {\gamma_i(\mu_0)-\tilde\gamma_i(\mu_0)\over 2\pi}\, \ln {\Lambda
\over \mu_0}~
          ~- ~{\tilde \gamma_i(\mu_0)\over 2\pi}\,  {X_\delta\over \delta}\,
                        \left\lbrack
                \left({\Lambda\over \mu_0}\right)^\delta -1\right\rbrack~.
\label{Zexpectedform}
\eeq
In this expression for $Z_i$, the power-law term is expected to arise
from the summation over Kaluza-Klein states in the loops of the diagrams
in Fig.~\ref{yukfigs}, reflecting the higher-dimensional nature of such
contributions.  In general, the anomalous dimensions $\tilde \gamma_i$
corresponding
to the excited Kaluza-Klein modes can differ from the
anomalous dimensions $\gamma_i$
corresponding to the zero-mode ground states.  This difference would then
give rise to the logarithm term in (\ref{Zexpectedform}).

One important difference relative to the gauge coupling case
concerns the scale-dependence of these anomalous dimensions $\gamma_i$
(or equivalently the ``beta-functions'' $b_F$).
Note that in (\ref{ZZZnew}) and (\ref{Zexpectedform}),
these coefficients
are evaluated at the fixed scale $\mu_0$.
This is because we are working within the context of a non-renormalizable
theory and calculating corrections to quantities evaluated
at the fixed scale $\mu_0$.  

Given these general expressions,
the task then remains to calculate the functions $\tilde \gamma_i$
which parametrize the one-loop contributions from each massive level in the
Kaluza-Klein tower.
Let us begin by considering a diagram of the form shown in
Fig.~\ref{yukfigs}(a).
Within the loop, we have both a fermion and a Higgs field.
However, as we discussed in Sect.~2, the MSSM fermions do not have Kaluza-Klein
towers in the minimal $\eta=0$ scenario, so the only fermion in
the loop is the zero-mode MSSM fermion itself.
Furthermore, at each excited level, the Kaluza-Klein tower
corresponding to the Higgs fields exactly mirrors the
MSSM Higgs (zero-mode) ground state.
Consequently, the contribution to $\tilde \gamma_F$
from this diagram is exactly the same as it is in the usual four-dimensional
MSSM.
Turning our attention to diagram in Fig.~\ref{yukfigs}(b), we see that
a similar situation exists here too.
Only the zero-mode MSSM fermion itself can propagate in the loop.
Likewise, although the Kaluza-Klein towers for the gauge bosons are
$N=2$ supersymmetric, {\it only their $N=1$ supersymmetric components
can couple to the MSSM fermions}\/.
This restriction arises because, as discussed in Sect.~2, the
wavefunctions of the additional
$\phi$ and $\psi$ fields
which fill out the $N=2$ gauge-boson Kaluza-Klein tower
are {\it odd}\/ functions of the compactified coordinates;
these wavefunctions thus vanish at the orbifold fixed points at which
the chiral MSSM fermions are found.  Thus, once again, the
contribution to the fermion anomalous dimension from
this diagram is exactly the same as it is in the usual four-dimensional MSSM.
We therefore find
\beq
              \tilde\gamma_F ~=~ \gamma_F~,~~~~~~~~
              \tilde\gamma_{\overline F} ~=~ \gamma_{\overline F}~.
\label{gammaFFbar}
\eeq
for all fermions $F$.

Finally, let us consider the
anomalous dimensions of the Higgs fields.
These diagrams are shown in Figs.~\ref{yukfigs}(c,d).
In Fig.~\ref{yukfigs}(c), only the chiral MSSM fermions can
propagate in the loop;  there are no excited Kaluza-Klein states
which can propagate.
Thus, this diagram is immune to the effects of the extra dimensions,
and makes no contribution to $\tilde \gamma_H$.
In Fig.~\ref{yukfigs}(d), by contrast, the full $N=2$ set of
Kaluza-Klein states for both the Higgs fields and the gauge bosons
can propagate in the loop.
Note that we must impose Kaluza-Klein momentum conservation at each of 
these vertices.
Because the appropriate
Kaluza-Klein states for this diagram fall into $N=2$ supermultiplets,
there is again no net contribution to $\tilde \gamma_H$.
(This is a consequence of the general fact that in unbroken $N=2$
supersymmetric
theories, hypermultiplets do not receive any wavefunction renormalizations.)
Consequently, combining our contributions, we find
\beq
                   \tilde \gamma_{H_i}~=~ 0~.
\label{gammaH}
\eeq
The Higgs wavefunction renormalization is
therefore completely immune to the effects of the
extra spacetime dimensions.

Given the results (\ref{gammaFFbar}) and (\ref{gammaH}), we can now examine
the evolution of the Yukawa couplings in the presence of extra spacetime
dimensions.
Below the scale $\mu_0\equiv R^{-1}$, the Yukawa couplings run logarithmically,
as in the usual four-dimensional MSSM.
Above this scale, however, the form of the
fermion and antifermion $Z$-factors in (\ref{Zexpectedform})
implies that
the Yukawa couplings start evolving 
with a power-law dependence on the energy scale (cutoff) $\Lambda$.
Because the anomalous dimensions $\tilde \gamma_i$ tend to be dominated
by their gauge contributions, we typically find $\tilde \gamma_i(\mu_0)<0$.
Thus the $Z$-factors in (\ref{ZZZnew}) each tend to be positive and grow
quickly with $\Lambda$.  This in turn drives the Yukawa couplings
dramatically towards extremely weak values.
Specifically,
for $\Lambda\gg\mu_0$ and
neglecting the logarithmic contributions to $Z_{H_i}$ in (\ref{ZZZnew}),
we find
\beq
           {\alpha_F(\Lambda)\over \alpha_F(\mu_0)} ~\approx~
                  {4\pi^2\over \gamma_F\gamma_\Fbar}
                  \left(\delta\over X_\delta\right)^2 \,
                   \left(\Lambda\over \mu_0\right)^{-2\delta}~.
\label{firstbehavior}
\eeq
Thus, we see that the effect of the extra dimensions is to drive
all of the Yukawa couplings (including the Yukawa coupling of the
top quark) towards weak values.

This effect might be extremely useful for various phenomenological
purposes (\eg, avoiding the Landau poles that often arise in the usual
MSSM).   Unfortunately, however, it is easy to see that this effect cannot
be used to explain the fermion mass {\it hierarchy}\/.  According to
(\ref{firstbehavior}), the {\it ratio}\/
between any two Yukawa couplings for different fermions evolves as
\beq
       {\alpha_{F_1}(\Lambda)\over
        \alpha_{F_2}(\Lambda)} ~\approx~
          \left({ \gamma_{F_2} \gamma_{\overline{F_2}} \over
           \gamma_{F_1} \gamma_{\overline{F_1}}}\right)
                 ~{\alpha_{F_1}(\mu_0)\over
                  \alpha_{F_2}(\mu_0)}~
\label{firsthierarchy}
\eeq
for $\Lambda\gg \mu_0$.
Thus, the hierarchy is affected only by scale-independent factors of
order one.

In order to do better,
let us now consider a slight generalization
of the previous scenario.
Let us imagine that in addition to the usual MSSM Yukawa terms,
there also exists a new $N=1$ chiral gauge-singlet scalar
superfield $S$ which couples to the Higgs fields via a
superpotential term of the form
\beq
                  W ~=~ \lambda\, H_u H_d S~.
\eeq
Here $\lambda$ is a coupling which we can make arbitrarily small.
For the purposes of this discussion, we shall consider $S$ to
be a light field which lacks Kaluza-Klein excitations (\eg, this
field might arise from a string twisted sector and therefore exist
only at orbifold fixed points).
We shall also ignore the running of $\lambda$,
and consider this to be a fixed parameter;  this assumption
will not affect the qualitative features of our results.

%======================================================================
\begin{figure}[th]
\centerline{ \epsfxsize 2.5 truein \epsfbox {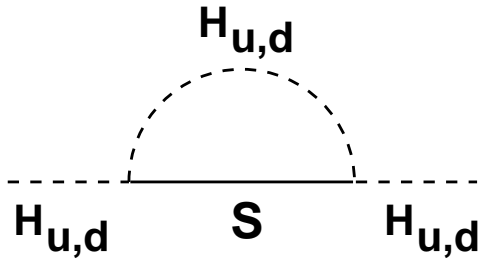}}
\caption{New superfield diagram, where $S$ is a light $N=1$ scalar
    singlet superfield.  This diagram allows the Higgs wavefunction
    renormalization to feel the extra spacetime dimensions,
    which in turn causes the fermion Yukawa couplings to become strong. }
\label{yukfigstwo}
\end{figure}
%======================================================================

Because of the extra dimensions, such a coupling (no matter how weak)
has a profound effect.
There is a new diagram,
as shown in Fig.~\ref{yukfigstwo},
which must also be considered.
Because the scalar field lacks a Kaluza-Klein tower, this component
of the diagram is only $N=1$ supersymmetric.
Thus, the Higgs hypermultiplet can now experience wavefunction
renormalization.
Moreover, thanks to the Kaluza-Klein tower of Higgs fields,
this diagram leads to non-zero power-law corrections to $Z_{H_i}$.
Specifically, we now find
\beq
              \tilde \gamma_{H_i} ~=~ \lambda^2/4\pi~.
\eeq
Note that $\tilde \gamma_{H_i}$ is strictly positive.
This means that $Z_{H_i}$ {\it decreases}\/ from unity, and rapidly
vanishes.

%======================================================================
\begin{figure}[th]
\centerline{ \epsfxsize 4.0 truein \epsfbox {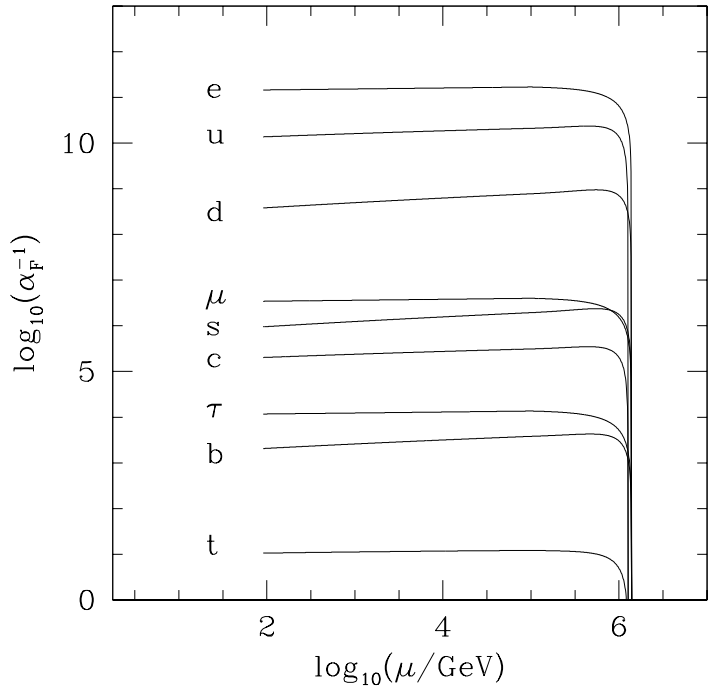}}
\caption{ The evolution of the Yukawa couplings
     $\alpha_F^{-1}\equiv 4\pi / y_F^2$
     within the singlet-enhanced MSSM,
     assuming the presence of a single extra
     dimension at $\mu_0=R^{-1}=100$ TeV.
     We have taken $m_t=180$ GeV and $\tan\beta=3$ as a
     representative case.  Note that we are plotting the Yukawa
     couplings on a {\it logarithmic}\/ scale
     in order to display them all simultaneously. It is evident that all of
     the Yukawa couplings simultaneously and independently
     approach a common Landau pole which precisely
     agrees with the scale at which the gauge couplings unify.}
\label{yukplot1}
\end{figure}
%======================================================================

This in turn implies that all of the fermion Yukawa couplings
quickly become strong rather than weak!
We stress that this change in behavior occurs regardless of how
small we take $\lambda$, since power-law behavior always eventually
dominates over logarithmic behavior.
Indeed, if we neglect the logarithmic contributions to $Z_{H_i}$ entirely,
we see that the power-law contributions to $Z_{H_i}$ have the right
signs and magnitudes to bring all of the Yukawa couplings
simultaneously to a common Landau pole scale.
This can be explicitly seen in
Fig.~\ref{yukplot1}, for which
we have taken $\mu_0=R^{-1}=100$ TeV and
$\lambda^2/4\pi\approx 1/4$.
It is clear that above the scale $\mu_0$, the power-law term coming from the
Kaluza-Klein states dominates the evolution, and the Yukawa couplings
tend towards a common large Yukawa coupling (\eg, towards a common
Landau pole defined by the equation $Z_{H_i}=0$).
Note that because the power-law corrections for each fermion
are not coupled to those of the other fermions, as would have
been the case for the usual renormalization group equations,
each fermion {\it independently}\/ tends towards the common Landau pole.
Moreover, for appropriate values of the coupling $\lambda$,
this Yukawa ``unification'' scale agrees precisely
with the scale $M'_{\rm GUT}$ at which the gauge couplings unify.

\begin{figure}[ht]
\centerline{
     \epsfxsize 3.0 truein \epsfbox {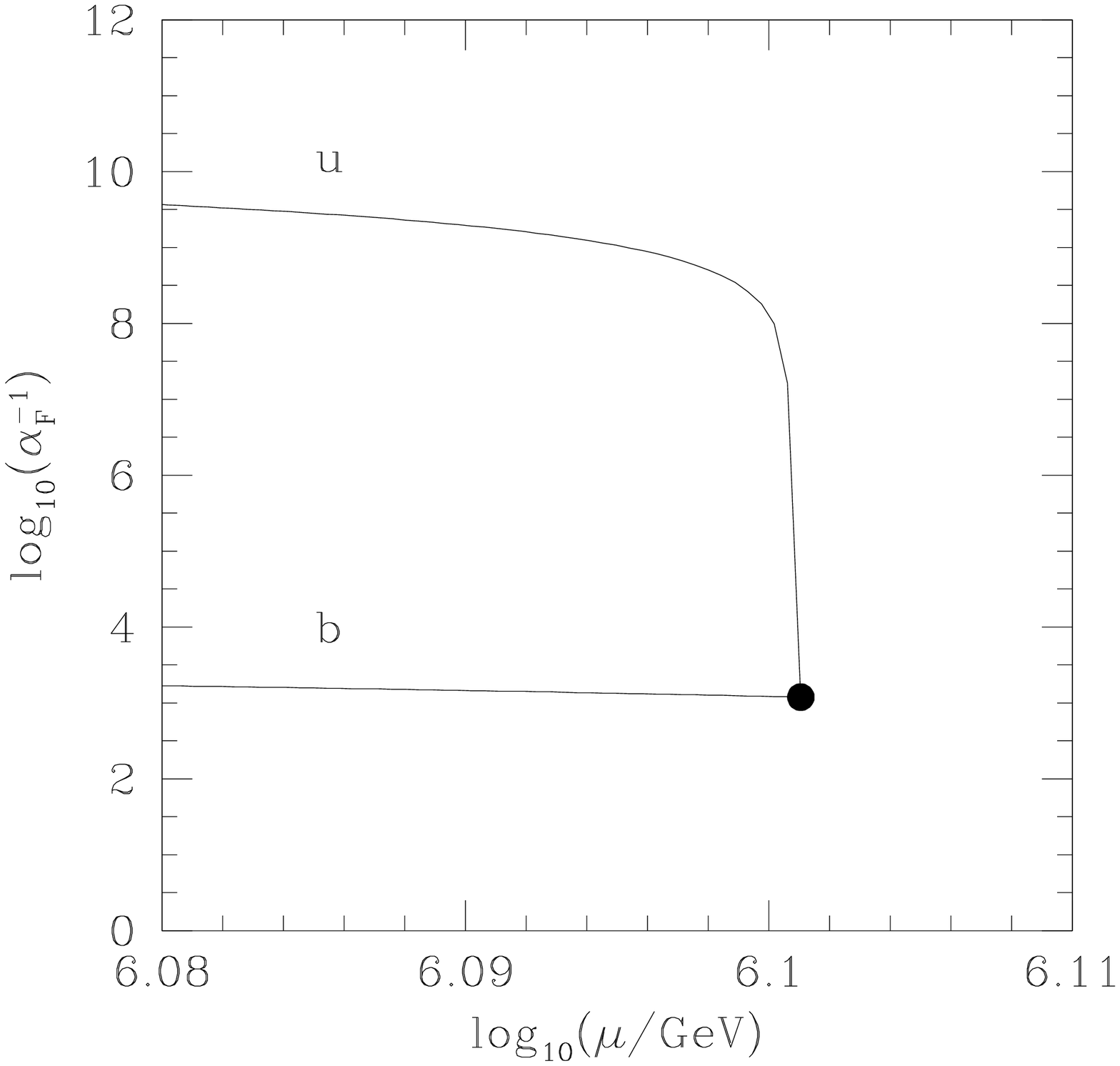}
     \epsfxsize 3.0 truein \epsfbox {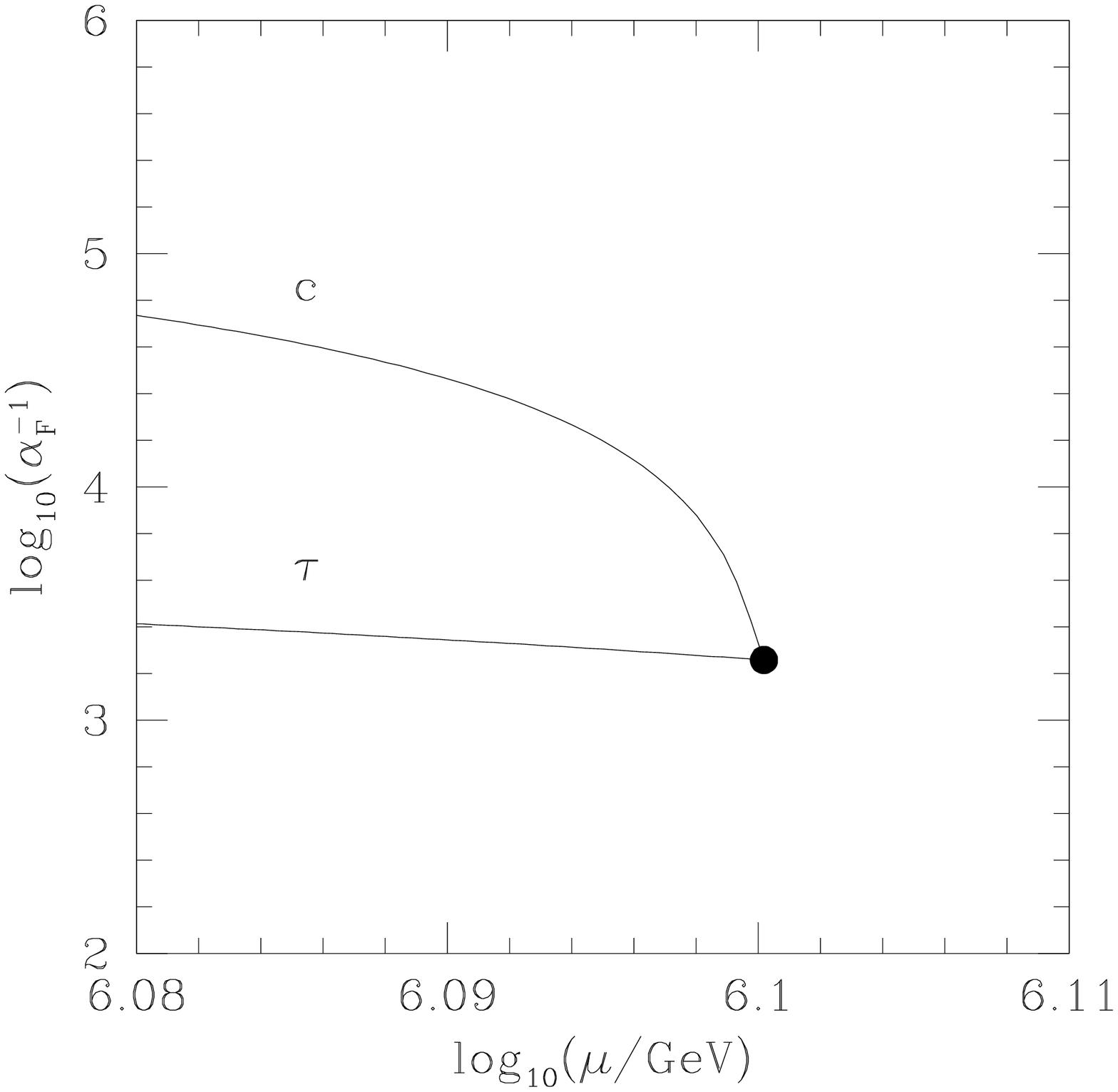}
         }
\caption{Two representative pairwise
      unifications of Yukawa couplings within the singlet-enhanced MSSM.
     In both cases we have taken a single extra dimension
      at $\mu_0=R^{-1}=100$ TeV, with
     $m_t=180$ GeV and $\tan\beta=3$.
     Once again, we plot the Yukawa
     couplings on a {\it logarithmic}\/ scale.
     It is clear from these plots that extra
     spacetime dimensions can
     explain hierarchies of many orders of magnitude,
     and produce Yukawa coupling unifications
     while still in the perturbative regime.  }
\label{yukplot2}
\end{figure}

This behavior might also be interesting for a
number of phenomenological purposes.
However, to what extent does this ``unification''
actually solve the fermion mass hierarchy problem?
Once again, because the behavior of the extra dimensions is universal
for all fermions, it might seem that no relative fermion
hierarchy can be explained.
This is not true, however:  certain {\it partial}\/
hierarchies can indeed be explained.
To see this, let us consider the precise positions of the Landau
poles.  For each fermion, the position of the Landau pole is determined
as the solution to the equation
\beq
           Z_{H_i} ~=~0
\eeq
for the appropriate Higgs field.  However, thanks to their different
logarithmic contributions to $Z_{H_i}$, the position of the Landau pole
for the up-type quarks
is slightly shifted relative to that for the bottom-type quarks
and leptons.  Ordinarily, this might
seem to be an insignificant observation.
However, {\it in the presence of extra dimensions}\/,
this difference amounts to a huge relative shift in the
Yukawa couplings between (up/down)-type pairs of fermions.
This allows pairs of up-type and down-type Yukawa couplings
to intersect, thereby providing a true unification of pairs of
Yukawa couplings and completely eliminating the mass hierarchy between
the corresponding fermions.
Some of these dramatic unifications
are shown in Fig.~\ref{yukplot2}.
Note that all of these unifications occur while
the corresponding Yukawa couplings are still weak, so perturbation
theory remains valid.

%======================================================================
%======================================================================

Even though we have managed to achieve pairwise Yukawa coupling
unifications in this way, we still seek a more complete solution
to the fermion mass hierarchy problem.
It is clear, of course, that the only true way to
explain a fermion mass hierarchy is through
a {\it flavor}\/-dependent coupling  (which we have so far not
introduced).
Extra dimensions, by themselves, cannot be expected to achieve
this since they are universal, and affect all fermions equally.
However, even if we must introduce a flavor-dependent coupling,
this flavor-dependence need not be very strong,
for the power-law effects of the extra dimensions
can easily {\it enhance}\/ or {\it amplify}\/ the effects
of even a relatively mild flavor-dependence.
Thus, by introducing a relatively mild flavor-dependent
coupling, we can exploit the effects of extra dimensions
in order to achieve a complete explanation of the fermion
masses.

In order to illustrate this point, we shall consider one final scenario.
Let us go back to the usual MSSM, and assume the existence of
an additional heavy MSSM-singlet scalar field $\Phi$
which modifies the form of the Yukawa interactions from
(\ref{MSSMsuperpot}) to
\beq
         W ~=~ \sum_F \, \hat y_F\, \Phi^{n_F}\, F\overline{F} H_{u,d}
\label{newsuperpot}
\eeq
where $n_F\in \IZ$.
Here $\hat y_F$ are a set of {\it dimensionful}\/ Yukawa couplings,
and we shall assume that $\Phi$ is endowed with a
corresponding $N=2$ supersymmetric tower of Kaluza-Klein states.
For the purposes of our discussion, the precise mass of $\Phi$ is
not important as long as it exceeds observable bounds;
for convenience we may assume $m_\Phi \approx \mu_0$.
The important point is that the exponents $n_F$ will be
assumed to be flavor-dependent,
taking different values for different fermions.
This will serve as the ``input'' flavor-dependence that
is ultimately required for addressing
the fermion mass hierarchy.  Although this setup 
is reminiscent of the Froggatt-Nielsen 
scenario~\cite{FN}, we will see that the role of $\Phi$
is different:  for example, we shall make absolutely no assumptions
about its vacuum expectation value, and in particular we shall treat
$\Phi$ as a fully dynamical field above the scale $\mu_0$.
Rather, our goal will be to see how extra dimensions
themselves can amplify this flavor-dependence into a
fermion mass hierarchy.

%======================================================================
\begin{figure}[ht]
\centerline{ \epsfxsize 3.5 truein \epsfbox {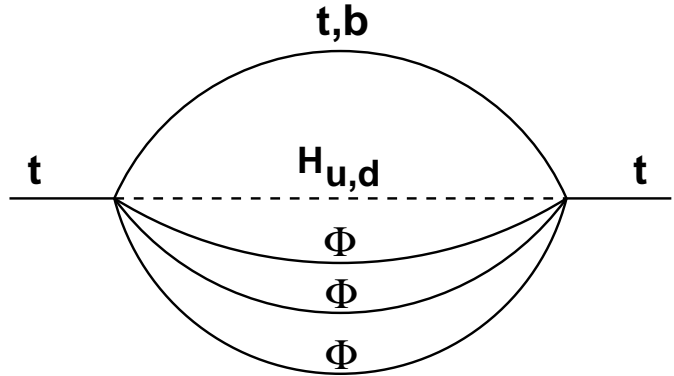}}
\caption{The generalization of Fig.~\protect\ref{yukfigs}(a)
           in the presence of a superpotential of the form
           (\protect\ref{newsuperpot}).
       }
\label{newyukfig}
\end{figure}
%======================================================================

How does the presence of the field $\Phi$
affect the analysis?  Above the scale $\mu_0$, the usual diagrams
in Fig.~\ref{yukfigs} must be replaced
by new diagrams in which the dynamical $\Phi$ field also propagates in loops.
For example, Fig.~\ref{yukfigs}(a) is replaced with Fig.~\ref{newyukfig},
where there are $n_F$ independent $\Phi$ propagators.  Because the $\Phi$
field is endowed with infinite towers of Kaluza-Klein states,
we see that we now have a total of $(n_F+1)\delta$ {\it independent towers
of Kaluza-Klein states that can propagate in the loop}\/.  These consist
of the $\delta$ towers for each compactified direction
for the Higgs field, and $n_F\delta$ towers for each compactified direction of
each of the $n_F$ different $\Phi$ propagators.
Thus, the {\it effective number of extra spacetime dimensions}\/
felt by such a diagram is shifted in a flavor-dependent way:
\beq
        \delta ~\to~ \Delta_F \,\equiv\, (n_F+1)\,\delta~.
\eeq
Of all the diagrams in Fig.~\ref{yukfigs}, this is the dominant
shift that arises.
(For example, there is also a contribution from
the generalization of Fig.~\ref{yukfigs}(c),
but this contribution has fewer internal Kaluza-Klein propagators.)
This in turn implies that
the wavefunction renormalization factors
$Z_F$ will have the dominant form
\beq
         Z_F ~=~ 1  ~-~ { c_F(\mu_0)\over 2\pi} {X_{\Delta_F}\over \Delta_F}
               \left\lbrack
            \left({\Lambda\over \mu_0}\right)^{\Delta_F} - 1 \right\rbrack~,
\label{ZFphi}
\eeq
where we have neglected possible logarithmic contributions.
Here
\beq
         c_F(\mu_0)~\sim~ \Lambda^{2 n_F}\,
         \left\lbrack\hat \alpha_F(\mu_0) \,+\,...\right\rbrack ~>~0~
\eeq
is a (dimensionless) anomalous dimension which
must be computed in the presence of the $\Phi$ field, and whose
precise value is irrelevant for our purposes.
Note, however, that $c_F(\mu_0)$ scales as the cutoff
$\Lambda^{2 n_F}$.
This is evident from simple power-counting in the $(n_F+1)$-loop diagram
of Fig.~\ref{newyukfig},
since there are $n_F+1$ internal loop four-momenta $p_i$ and
$n_F+2$ internal superfield propagators (each of which contributes
$p_i^{-2}$).

The result (\ref{ZFphi}) implies that the
Yukawa couplings $\hat y_F$ will evolve
according to the general power-law form
\beq
        \hat\alpha_F^{-1}(\Lambda) ~=~
         \hat\alpha_F^{-1}(\mu_0)  -
          {\hat c_F\over 2\pi} {X_{\Delta_F}\over \Delta_F}
               \left\lbrack
            \left({\Lambda\over \mu_0}\right)^{\Delta_F} - 1 \right\rbrack
\label{intermediatestep}
\eeq
where $\hat c_F(\mu_0)\equiv c_F(\mu_0)/\hat\alpha_F(\mu_0)
\sim \Lambda^{2 n_F} (1+...)$
is a dimensionful beta-function coefficient and
where we are again neglecting possible logarithmic contributions.
Thus, we see that the presence of the extra dimensions
(\ie, $\delta\not= 0$) is capable of
producing a flavor-dependent power-law hierarchy!
Specifically, since $c_F(\mu_0)$ is positive, our Yukawa couplings
are all once again driven towards a simultaneous Landau pole
at which $\hat \alpha_F^{-1}(\Lambda)\to 0$.
However, unlike the previous case, these Yukawa couplings approach
the Landau pole in a flavor-dependent manner,
and have the potential to actually unify on the way.
Equivalently, assuming a unification near the Landau pole
and solving for $\hat\alpha_F^{-1}(\mu_0)$ from (\ref{intermediatestep}),
we find
\beq
       \hat\alpha_F^{-1}(\mu_0) ~\approx~
          {\hat c_F(\mu_0)\over 2\pi} {X_{\Delta_F}\over \Delta_F}
               \left\lbrack
            \left({\Lambda\over \mu_0}\right)^{\Delta_F} - 1 \right\rbrack~.
\eeq
Identifying the physical Yukawa coupling $y_F(\mu_0)$
at the scale $\mu_0$ via $y_F(\mu_0)\equiv \mu_0^{n_F} \hat y_F(\mu_0)$
then yields the result
\beq
       \alpha_F^{-1}(\mu_0) ~\approx~
          {X_{\Delta_F}\over 2\pi \Delta_F}
            \left( {\Lambda\over \mu_0}\right)^{2 n_F}
               \left\lbrack
            \left({\Lambda\over \mu_0}\right)^{\Delta_F} - 1 \right\rbrack~
     ~\approx~
          {X_{\Delta_F}\over 2\pi \Delta_F}
            \left( {\Lambda\over \mu_0}\right)^{\Delta_F+ 2 n_F}~.
\eeq
Thus, we see that
as a simple result of having extra spacetime dimensions ($\delta >0$),
the flavor-dependent coupling
(\ref{newsuperpot}) has now been amplified, ultimately producing
a large power-law Yukawa mass hierarchy with total exponent
\beq
            \Delta_F+2 n_F ~=~ n_F(2+\delta) + \delta ~.
\label{totalexponent}
\eeq
This exponent can also be interpreted physically as the mass dimension
of the Yukawa coupling $\hat \alpha_F$ in $4+\delta$ flat spacetime
dimensions.

Note that this scenario differs from the usual Froggatt-Nielsen
scenario~\cite{FN} in that
we were not forced to introduce
an arbitrary small number to parametrize the VEV of the $\Phi$ field.
In our scenario, by contrast, the
hierarchy is generated purely as a result of a smooth power-law
evolution of Yukawa couplings
between the scale of the extra dimensions
and the unification scale.
Indeed, because the Yukawa couplings are driven to strong coupling
in this scenario, it is even possible to imagine that more refined results
could be obtained using a fixed-point analysis.
Finally, we note that because of the naturally
large exponent (\ref{totalexponent}), the Yukawa coupling hierarchy
can be explained without the {\it a priori}\/ large values
of $n_F$ that would have been needed in the usual Froggatt-Nielsen
scenario.  For example, taking $\delta=1$ and $\mu_0$ in the TeV range implies
(see Fig.~\ref{gaugetwo}) that $\Lambda/\mu_0 \approx 20$.
We can therefore explain the {\it entire hierarchy}\/ factor of $10^{12}$
between the electron and
the top quark simply by taking $n_t=0$ and $n_e=3$.

Finally, we remark that the possibility of extra spacetime
dimensions also opens up new scenarios for generating a
fermion mass hierarchy which do not require
the arbitrary introduction
of low-energy flavor-dependent couplings, or the introduction
of new low-energy matter fields.
Such new mechanisms rely on the non-perturbative behavior
of open-string theories in the presence of extra large dimensions,
and will be discussed in detail in Sect.~8.

%========================================================================
%  \vfill\eject
\section{Unification without supersymmetry}
\setcounter{footnote}{0}

Until this point, our discussion has focused on the effects
of extra large spacetime dimensions in theories with supersymmetry.
However, given that the observed low-energy world is non-supersymmetric,
and given that our extra dimensions can appear at a scale which
is comparable (and perhaps even lower) than the superpartner scale,
it also makes sense to consider the corresponding {\it non-supersymmetric}\/
situation.
Indeed, two of the primary motivations for introducing supersymmetry
at all are the gauge hierarchy problem and the unification of gauge couplings
within the MSSM.  However, if new dimensions populate the desert
between the electroweak scale and the usual GUT scale, then the gauge
hierarchy problem is greatly ameliorated, and (in the case
of TeV-scale extra dimensions)  no longer exists.
Thus, it remains to consider whether extra dimensions can also
give rise to gauge coupling unification when supersymmetry is absent.

 {\it A priori}\/,
embedding the Standard Model into higher dimensions is
more straightforward than embedding the MSSM.
This is because our Kaluza-Klein states no longer need
to form $N=2$ multiplets.  As a ``minimal'' scenario,
we shall assume (as we did for the MSSM) that
the gauge bosons and Higgs field have Kaluza-Klein excitations,
while the chiral Standard Model fermions do not.
Note that at each excited Kaluza-Klein level,
this requires the introduction of an additional real scalar
transforming in the adjoint of each gauge group
(in order to make the corresponding gauge bosons massive).
This then implies the beta-function coefficients
\beqn
      (b_1,b_2,b_3) &=& (41/10, -19/6, -7)~\nonumber\\
      (\tilde b_1,\tilde b_2,\tilde b_3) &=& (1/10, -41/6, -21/2)~
\eeqn
where as before we have defined $b_1\equiv (3/5) b_Y$.

\begin{figure}[ht]
\centerline{ \epsfxsize 4.0 truein \epsfbox {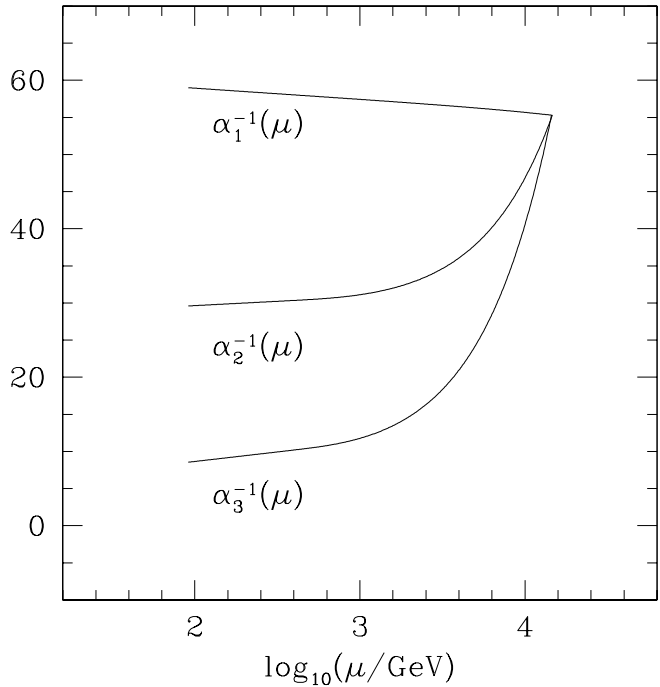}}
\caption{Gauge coupling unification without supersymmetry can
       also be achieved at very low energy scales.
       Here we have considered a ``minimal'' embedding of the
       non-supersymmetric Standard Model into five dimensions,
       supplemented with three real scalars transforming in
       the adjoint of $SU(2)$.
       We have taken our single extra dimension
       to have radius $R^{-1}\equiv \mu_0=1$ TeV.
       Thus this scenario also avoids the gauge hierarchy problem.}
\label{gauge_nonsusy}
\end{figure}

Unfortunately, this simple-minded scenario does not lead to gauge
coupling unification.  However, it is not hard to find extended
scenarios that do.  For example, let us imagine adding
three extra real scalar fields
transforming in the adjoint of $SU(2)$.  We shall assume that
the wavefunctions of these extra scalar fields are
odd functions of the coordinates $y_i$ of the compactified dimensions.
Indeed, such sorts of extra states are extremely natural from
the point of view of string theory, and the fact that they are
odd functions of $y_i$ guarantees that they have no light zero-modes.
We then have
\beq
      (\tilde b_1,\tilde b_2,\tilde b_3) ~=~ (1/10, -35/6, -21/2)~,
\eeq
which leads to the gauge coupling unification shown in
Fig.~\ref{gauge_nonsusy}.

%======================================================================
%======================================================================

Of course, this scenario is not unique.
In principle, there may exist other combinations of extra
matter that can also lead to gauge coupling unification without supersymmetry.
We may even choose to associate the scale $\mu_0$ with the supersymmetry-breaking
scale, \eg, by implementing a Scherk-Schwarz supersymmetry-breaking
mechanism~\cite{SS} using the same orbifold that prevents our chiral fermions
from having Kaluza-Klein excitations.
One finds that
the resulting string spectrum
always exhibits a hidden ``misaligned supersymmetry''~\cite{missusy}
which is responsible for maintaining the 
fundamental finiteness properties of the string,
even without supersymmetry.
Indeed, such supersymmetry-breaking scenarios using the $\IZ_2$ Scherk-Schwarz
mechanism have been investigated in a number of contexts~\cite{antoniadis,Dudas}.
Moreover, note that just as for the MSSM, we are free to 
consider non-minimal scenarios with $\eta >0$, since 
increasing $\eta$ does not disturb the unification at $\eta=0$.

However, the important point is that even without supersymmetry,
it is the power-law evolution
of the gauge couplings, induced by the extra spacetime dimensions,
that enables such extra matter to produce perturbative gauge coupling
unification
at such low energy scales.
Thus, through extra large spacetime dimensions, it may well be possible
to contemplate a scenario in which perturbative gauge coupling unification
is preserved and the gauge hierarchy problem is eliminated,
all without supersymmetry.
Indeed, one can even contemplate
explaining the Standard Model fermion masses at the same time,
again through the power-law behavior induced by the extra spacetime
dimensions.

%========================================================================
%  \vfill\eject
\section{Embedding our scenario into string theory}
\setcounter{footnote}{0}

In this paper, we have studied the effects of extra
large spacetime dimensions from a strictly
 {\it field-theoretic}\/ point of view.
Even though we have borrowed certain ideas from string
theory (\eg, the notion of orbifolds), we have
limited ourselves to questions that are
field-theoretic in nature, and for which a purely
field-theoretic analysis suffices.  For example,
we have concentrated on Kaluza-Klein momentum states but
neglected winding-mode states (which is a valid approximation
in the field-theory limit, since our radii are presumed
large);  we have not worried
about ensuring that the orbifold is a symmetry of the
full theory (since we have not constructed a full string
theory);
and we have not considered any extra states beyond the MSSM
(although string theory generically predicts such states).
We have not addressed these sorts of issues because
they tend to be extremely model-dependent.

There are also various questions that are {\it generic}\/.
For example, one of the attractive features of the
conventional GUT paradigm is that the unification scale
$M_{\rm GUT}\approx 2\times 10^{16}$ GeV
is close to (though not equal to~\cite{review}) the
perturbative heterotic string scale $M_{\rm string}\approx
5\times 10^{17}$ GeV.
This suggests that we might identify these two scales,
and imagine that our grand unified theory might be directly
embedded into string theory.
However, in this paper we have seen that extra large spacetime dimensions
can lower the unification scale quite substantially.
How then do we go about embedding our scenario into string theory?

In this section, we shall address a number of these questions.
Rather than attempt to construct a given realistic
string model which incorporates our scenario, we shall limit
the following discussion to {\it general}\/ comments concerning
various issues that
come into play when attempting to embed
our scenario into string theory.

\subsection{Implementing the orbifold projection}

We begin with a general comment concerning the orbifolding
projection discussed in Sect.~2.
As we have seen, it is necessary to
compactify our extra dimensions on 
orbifolds rather than circles
for two reasons:
we need the orbifold to break the $N=2$ supersymmetry
of the excited Kaluza-Klein states down to $N=1$
supersymmetry for the observable MSSM ground states;
and we need to ensure, in our ``minimal'' scenario,
that the chiral MSSM fermions do not have Kaluza-Klein
towers.  
Once this scenario is
embedded into a full string theory,
it becomes necessary to ensure that the $\IZ_2$ orbifold
action on the compactified coordinates $y_i\to -y_i$
is a symmetry of the full string theory.

It turns out that the ``minimal'' scenario that we examined
in Sects.~2 and 3 does not have this property, and must be
extended if it is to embedded into string theory.
Specifically, although we have joined the two $N=1$ supersymmetric
MSSM Higgs doublets together in (\ref{Thyp}) to form
a single $N=2$ supermultiplet, it is actually necessary for
one of these Higgs fields to be {\it odd}\/ under the
orbifold action $y_i\to -y_i$
in order for this action to be a symmetry of
the full theory.  This then prevents one of the Higgs fields from
having a light zero-mode.  Of course, this need not cause
any logical inconsistency, for it may well be that
the remaining MSSM Higgs field arises in a ``twisted'' sector.
Alternatively, one of the MSSM Higgs fields might arise as the first
excited Kaluza-Klein state of the odd tower, with mass $m\approx \mu_0$.
Gauge coupling unification
would then be slightly altered, though not significantly
damaged.  In any case, string theory also predicts new
so-called threshold corrections~\cite{thresholds,review}
which can easily accommodate such minor discrepancies.

A third alternative would be to introduce a new,
separate, odd-transforming Higgs field for each of
the MSSM Higgs fields.  This would then imply
the existence of {\it four}\/ Higgs fields at each
excited Kaluza-Klein level, while maintaining only two
Higgs doublets at the massless level.
Unlike the previous case, this choice would have a 
considerable effect on the unification of the gauge couplings.
However, since this is essentially a model-dependent
choice that would be dictated by the string model
in question, it is quite reasonable to expect that
string theory will provide even further
states that can once again restore gauge coupling unification.
We stress, however, that these are model-dependent questions
that can be addressed only in the context of a fully
constructed, realistic string model.

As a separate issue related to the orbifold, we remark
that in our ``minimal'' scenario, the MSSM chiral fermions
were taken not to have Kaluza-Klein towers.
The orbifold manages to accomplish this because, in
any closed string theory, modular invariance must be preserved.
Thus, modding out by the orbifold action
necessarily requires the introduction of so-called ``twisted
states'' whose wavefunctions are restricted to the orbifold
fixed points.
Hence our assertion that the MSSM chiral fermions have
no Kaluza-Klein towers is fully consistent within the context
of closed string theory.
For an {\it open}\/ string theory, by contrast,
the process of orbifolding generally does not lead
to such twisted sectors.  This difference arises 
because modular invariance is not required
to be a symmetry of open-string theories.
In such cases, we must assume
the chiral fermions to be restricted to certain three-branes
of the open-string theory.  This will be discussed further below.

\subsection{Including the extra GUT states}

The second issue that we shall discuss concerns the
appearance and subsequent breaking
of the GUT symmetry in string theory.
Throughout this paper, we have interpreted
the phenomenon of gauge coupling unification as signalling
the emergence of a grand unified symmetry at the scale
$M'_{\rm GUT}$.  Strictly speaking, this
terminology is field-theoretic, and implicitly assumes
a conventional Higgs mechanism for breaking the GUT symmetry.
Indeed, in the terminology appropriate for the Higgs mechanism,
the GUT symmetry can be said to exist  above the scale
of unification, and to be broken  below this scale.
However, as is clear from the discussion in Sect.~4,
in this paper we are actually imagining that the GUT symmetry is
broken in a {\it string-theoretic manner}\/, namely through
the orbifold that acts on the compactified dimensions.

This fact has a number of consequences.
The most important of these concerns the extra GUT states
that are part of the GUT symmetry but which are not
present in the MSSM itself.  These include, for example,
the $X$ and $Y$ gauge bosons and the colored Higgs
triplets;  these are the states which we collectively
denoted $\Psi$ in Sect.~4.
Of course, despite their odd symmetries under
$y_i\to -y_i$, these states still continue to
exist in the string spectrum,
with masses $m \sim n/R$, $n\in\IZ$, with $n\geq 1$.
As we stressed in Sect.~4, this does not cause a problem for
proton decay because these states do not couple to the chiral MSSM
fermions.  However, the presence of such states
with masses $n\mu_0$ means that these states  --- and
indeed their entire Kaluza-Klein towers --- should
actually be included in the evolution of the gauge couplings
between $\mu_0$ and $M'_{\rm GUT}$.

It may seem very counter-intuitive that we must include the effects
of the $X$ and $Y$ bosons and colored Higgs triplets
in the running of gauge couplings below the
unification scale.  However, this is precisely the
consequence of breaking the GUT symmetry via an orbifold projection.
Indeed, if we break a GUT symmetry in string theory via
an orbifold associated with a radius scale $\mu_0\equiv R^{-1}$,
it makes no sense at all to think of
the GUT symmetry as being ``restored'' above $M'_{\rm GUT}$
or broken below it.
Rather, our GUT symmetry is actually ``restored'' at
the scale $\mu_0$ in the sense that the states appearing
at masses $m\geq \mu_0$ fall into GUT multiplets.
Thus, strictly speaking, the scale at which the
broken GUT symmetry begins to leave particle remnants in the string spectrum
is actually $\mu_0 \equiv R^{-1}$, not $M'_{\rm GUT}$.
These states thus appear as GUT ``precursors''.

It is easy to see that 
these precursor states do not upset gauge coupling unification.
The zero-mode states consist, as before, of the MSSM spectrum.
By themselves, these states are well-known to lead to gauge coupling
unification.  Starting at the first excited Kaluza-Klein level,
however, all additional Kaluza-Klein states
appear in complete GUT multiplets.
As is well-known, this cannot disturb an already-present unification.
Thus, even when we take into account the Kaluza-Klein towers
corresponding to the $X$ and $Y$ bosons and Higgs triplets,
we see that rapid gauge coupling unification is preserved.
Note that this property holds regardless of the GUT group
in question, whether a minimal $SU(5)$ or $SO(10)$, or
a larger group such as $E_6$.

\subsection{Interpreting the mass scales}

Another issue that we face, upon embedding our scenario into string
theory, is the origin and interpretation of the different mass scales.
How, in particular, might such a small mass scale
$\mu_0=R^{-1}$ arise in string theory,
and how can it be implemented when constructing
a realistic string model?
There are actually two classes of possibilities.

The more traditional class of possibilities is the perturbative
one:  we simply construct a weakly coupled heterotic string
with a large radius of compactification.  In such a scenario,
the fundamental string scale is tied to the Planck scale,
and remains at the perturbative heterotic string value
$5\times 10^{17}$ GeV.
This is essentially the approach taken
in Ref.~\cite{antoniadis} (although the gauge couplings
of the specific string
models of Ref.~\cite{antoniadis} do not feel the extra dimensions,
and consequently do not evolve with power-law behavior or experience
any intermediate-scale gauge coupling unification).

However, if we now attempt to join this perturbative framework with 
our intermediate-scale grand unification,
we cannot explain why the
scale of gauge coupling unification
should be so much lower than the Planck scale.
Indeed, it is well-known~\cite{review} that weakly coupled heterotic strings
naturally lead to gauge coupling unification near the Planck scale.
Moreover, there also remains the question of the dynamical origin
of this large scale.

Given our recent understanding of the strong-coupling dynamics
of string theory, however, several more attractive possibilities arise
for interpreting the reduced unification scale.
Specifically, we might attempt to interpret the unification scale as the
 {\it string scale itself}\/, so that we have a direct embedding
into string theory immediately at the reduced unification scale $M'_{\rm GUT}$!
It may seem, at first, that this is impossible, for it is well-known
that for weakly coupled heterotic strings, the tree-level
string scale $M_{\rm string}$ is
irrevocably tied to the Planck scale $M_{\rm Planck}$ through a relation
of the form
\beq
            M_{\rm string}~\sim~ g_{\rm string}\, M_{\rm Planck}
\label{heterotic}
\eeq
where $g_{\rm string}$ is the string coupling at unification.
This relation holds regardless of the dimensionality of the spacetime
(\ie, regardless of how many of the ten spacetime dimensions
are compactified), and regardless of their
volume (radii) of compactification.
For heterotic strings at {\it strong}\/ coupling, however,
this behavior changes.
Specifically, it turns out that various (closed) heterotic strings at
strong coupling can be equivalently described as (open) Type~I strings
at weak coupling.
Thus, many non-perturbative features of heterotic string theory can
be studied by analyzing weakly coupled Type~I strings.  Remarkably,
Type~I string theory offers the interesting possibility~\cite{HWtwo}
of lowering the fundamental string scale, for (\ref{heterotic}) no
longer continues to apply.  Instead, we find the relation
\beq
            M_{\rm string}~\sim~ e^{\phi/2}\, g_{\rm gauge}\, M_{\rm Planck}
\label{TypeI}
\eeq
where $\phi$ represents the so-called ten-dimensional dilaton field
and $g_{\rm gauge}$ is the Type~I gauge coupling.
Thus, simply by adjusting the VEV of the ten-dimensional dilaton, one
can lower $M_{\rm string}$ relative to $M_{\rm Planck}$.  Of course,
the value of the dilaton indirectly affects the values of the gauge and
gravitational couplings, so that $g_{\rm gauge}$ and $M_{\rm Planck}$
themselves
change their apparent values when the dilaton is changed.
However, using other string relations
it is possible to eliminate this dependence algebraically,
and relate $M_{\rm string}$ and $M_{\rm Planck}$ directly to each other without
exhibiting the dependence on the dilaton.  We then find the general
relation
\beq
          M_{\rm string} ~\sim~
         \sqrt{ 1\over \alpha_{\rm gauge} M_{\rm Planck}}
                        \,V^{-1/4}
\label{newTypeI}
\eeq
where $\alpha_{\rm gauge}\equiv g_{\rm gauge}^2/(4\pi)$ and
where $(2\pi)^6 V$ is the (normalized) six-dimensional volume of
compactification.
In writing (\ref{newTypeI}), we have ignored (and will continue to ignore) all
numerical factors of order one, since our
goal will merely be to obtain order-of-magnitude estimates.
The relation (\ref{newTypeI}) thus represents the non-perturbative counterpart
of (\ref{heterotic}).

In order to embed our grand unification scenario
into Type~I string theory, we shall
attempt to identify $\alpha_{\rm gauge}$ with $\alpha'_{\rm GUT}$
and $M_{\rm string}$ with $M'_{\rm GUT}= 10$ TeV $\ll 10^{16}$ GeV.
Let us assume, as we have done throughout, that there are $\delta$ extra
dimensions of radius $R\equiv \mu_0^{-1}$.  As usual, these are the dimensions
that cause the gauge and Yukawa couplings to experience power-law corrections.
Given the relation (\ref{newTypeI}), we can now easily determine the required
common radius $r$ for the remaining $6-\delta$ compactified dimensions.
Writing the normalized compactification volume as
\beq
              V ~\sim~ R^\delta \,r^{6-\delta}~,
\eeq
we find
\beq
           {M'_{\rm GUT}\over M_{\rm Planck}} ~\sim~
            \alpha'_{\rm GUT} \, (M'_{\rm GUT} R)^{\delta/2}\,
                          (M'_{\rm GUT} r)^{3-\delta/2}~.
\label{fundrelation}
\eeq

%======================================================================
\begin{figure}[ht]
\centerline{ \epsfxsize 5.0 truein \epsfbox {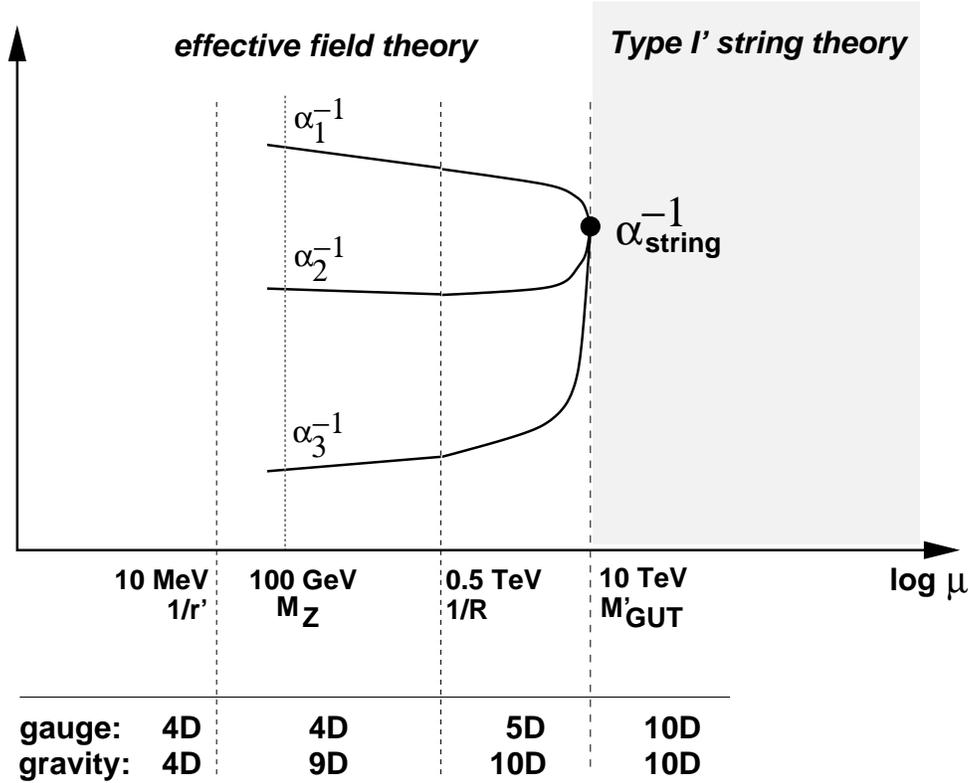}}
\caption{Sketch of the evolution of the gauge
    couplings within a Type~I$'$ realization of our scenario.
    Below the unification scale, the physics is described by
    an effective field theory,
      while the full Type~I$'$ string description is appropriate
   above this scale.
     Thus, in this realization, the string scale coincides with
       the new unification scale $M'_{\rm GUT} \approx 10$ TeV.
     This is achieved by having the gauge couplings feel a
      new dimension at $R^{-1}\equiv \mu_0 \approx 0.5$ TeV,
       while the gravitational coupling feels five new dimensions
        at $(r')^{-1}\approx 10$ MeV.
    We have also indicated the corresponding effective number
     of spacetime dimensions
    that are felt by the gauge and gravitational couplings
     at different energy scales.}
\label{firstsketch}
\end{figure}
%======================================================================

Let us begin by considering the case $\delta=1$.
Taking $M'_{\rm GUT}=10$ TeV, we see
from Figs.~\ref{gaugetwo} and \ref{gaugethree}
that $M'_{\rm GUT}R\approx 20$
and $\alpha'_{\rm GUT}\approx 1/50$.
This in turn implies that $M'_{\rm GUT} r\approx 10^{-6}$,
which implies that the radius $r$ of the five extra dimensions
must be {\it smaller}\/ than the string length scale!
Of course, this is not an inconsistency, but rather
a signal that we should pass to a slightly different description
(the so-called Type~I$'$ description) of the physics.
Technically, this procedure of passing from one description to
the other is called a $T$-duality,
and in general a Type~I theory with a compactified radius $r$
is the $T$-dual of an equivalent Type~I$'$ theory with a compactified
radius $r'\equiv(M_{\rm string}^2 r)^{-1}$:
\beq
      \hbox{$T$-duality:}~~~~~~
      M_{\rm string}r ~\leftrightarrow~ (M_{\rm string}r')^{-1}~.
\eeq
We therefore pass to a Type~I$'$ description by $T$-dualizing
our five extra dimensions.  We then find
\beq
           (r')^{-1}~\sim~ 10^{-6} \,M'_{\rm GUT} ~\sim~ 10 ~{\rm MeV}~.
\eeq

Thus, to summarize our results, we see that we can naturally
associate the scale of gauge coupling unification at $10$ TeV
with the string scale of a Type~I$'$ theory in which
one dimension has radius $R^{-1}\approx 0.5$ TeV
and the five remaining dimensions have radii $r'\sim (10\, {\rm MeV})^{-1}$.
The resulting scenario is sketched in Fig.~\ref{firstsketch}.
Note that extra dimensions of this size are not ruled out experimentally,
provided that the gauge couplings do not feel their effects~\cite{Dim,Dimtwo}. 
Likewise, a similar calculation for the $\delta=2$ case
yields the result $r'\sim (0.1\, {\rm GeV})^{-1}$
for the remaining four dimensions.
We note, however, that these results are
dependent on the chosen unification scale (string scale)
in our scenario.  For example,
taking $M'_{\rm GUT}= 10^{12}$ GeV instead yields essentially
the same result $r'\sim (10^9 \,{\rm GeV})^{-1}$ for both
the $\delta=1$ and $\delta=2$ cases.

\subsection{D-brane configurations for our scenario}

Given these results, it is natural to interpret our scenario
in terms of various configurations of the $D$-branes of
the Type~I$'$ string theory.
To see how this can be done,
let us begin by recalling that the original supersymmetric Type~I theory
contains both nine-branes and five-branes.
For the purposes of this discussion, we shall label our
ten spacetime coordinates as $\lbrace x_1,...,x_{10}\rbrace$,
with $\lbrace x_1,x_2,x_3,x_4\rbrace$ corresponding
to our observed four-dimensional world.
We shall concentrate on the $\delta=1$ case,
so that our gauge couplings feel a single extra large dimension
with radius $R$
corresponding to $x_5$, while the remaining
five dimensions $\lbrace x_6,...,x_{10}\rbrace$ have radius $r$.
The nine-branes of our theory necessarily obey Neumann boundary
conditions for all ten dimensions $\lbrace x_1,...,x_{10}\rbrace$,
but the five-branes will (by definition) obey Neumann boundary
conditions in only six dimensions and obey
Dirichlet boundary conditions in the remaining four dimensions.
The choice of which set of dimensions obeys which
set of boundary conditions amounts to a choice of
the orientation of the five-brane relative to the observed
four-dimensional spacetime and the extra large dimension.
For reasons that will become clear shortly, we shall take
our five-brane to obey Neumann boundary conditions
in the $\lbrace x_1,...,x_4,x_6,x_7\rbrace$ directions,
and to obey Dirichlet boundary conditions in the
$\lbrace x_5,x_8,x_9,x_{10}\rbrace$ directions.

As we have discussed, the fact that
the five dimensions $\lbrace x_6,...,x_{10}\rbrace$
are compactified with a radius
exceeding the Type~I string scale implies that
we must $T$-dualize these five dimensions.
Operationally, this amounts to interchanging Neumann and
Dirichlet boundary conditions in the dualized directions.
We therefore obtain a Type~I$'$ theory with the following
branes.  First, the nine-brane of the original Type~I theory
becomes a {\it four}\/-brane with Neumann boundary conditions
in the directions $\lbrace x_1,...,x_4,x_5\rbrace$ and
Dirichlet boundary conditions in the directions $\lbrace x_6,...,x_{10}\rbrace$.
By contrast, the five-brane of the original Type~I theory
becomes a {\it six}\/-brane, with Neumann boundary conditions
in directions $\lbrace x_1,...,x_4,x_8,x_9,x_{10}\rbrace$
and Dirichlet boundary conditions in
directions $\lbrace x_5,x_6,x_7\rbrace$.

\begin{figure}[ht]
\centerline{ \epsfxsize 5.0 truein \epsfbox {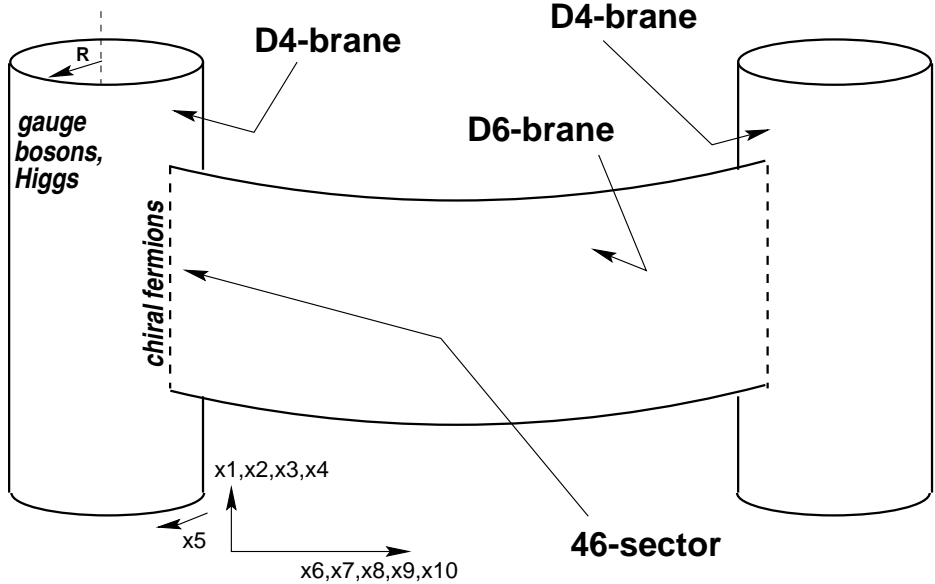}}
\caption{A $D$-brane configuration which can accommodate our scenario
   within the context of Type~I$'$ string theory.
   The observed flat four-dimensional world corresponds to the
    ``46-sector'' (\ie, the three-brane
    intersection between the (cylindrical) four-branes
     and the six-branes).  The extra (compactified) direction
   of the four-branes corresponds to our extra dimension of
    radius $R\equiv \mu_0^{-1}$.  The MSSM Higgs and gauge fields
    are presumed to lie fully on the four-branes, while in our minimal
     scenario the chiral
     MSSM fermions are restricted to the 46-sector
    and hence have no Kaluza-Klein excitations.}
\label{torah}
\end{figure}

How can we interpret this picture?
Fortunately, this Type~I$'$ brane configuration contains exactly
what we require for our scenario.
The four-brane, which extends outwards in the
directions $\lbrace x_1,...,x_4,x_5\rbrace$, can be interpreted
as the spacetime of the five-dimensional world that the gauge
couplings feel at energy scales above $R^{-1}\equiv \mu_0$:
the four directions $\lbrace x_1,...,x_4\rbrace$
are taken to be completely flat, and our fifth
dimension $\lbrace x_5\rbrace$ is compactified with radius $R$.
Thus, the MSSM Higgs fields and gauge bosons can be interpreted
as living on the four-branes.
The six-brane, by contrast, gives rise to a
separate {\it non-perturbative}\/ gauge symmetry whose
properties will not concern us here.
However, the most important feature is the presence of
a non-trivial {\it intersection}\/ between the four-brane
and the six-brane.  Note that the joint Neumann directions of this
``intersection brane'' are only
$\lbrace x_1,...,x_4\rbrace$.
Thus, particles localized on this three-brane
feel only {\it four}\/ spacetime directions, regardless of the
size of the radii $R$ and $r$.
Such states are typically said to arise in the ``46-sector''.
These states can therefore easily be interpreted as the
chiral MSSM fermions, which are required not to have Kaluza-Klein
excitations in our minimal scenario!
Note that it is crucial that our brane configuration give rise
to such a 46-sector, for
we must have a way of localizing the chiral MSSM fermions
within the context of open-string theories
so that they do not feel the extra dimensions.
This, then, explains our original choice of the orientation of the
Type~I five-brane.
The resulting brane configuration is illustrated in Fig.~\ref{torah}.

%======================================================================
%======================================================================

It is straightforward to generalize this brane configuration to
the $\delta=2$ case.
In this case we shall take $\lbrace x_5,x_6\rbrace$ as the
extra large dimensions of radius $R$.
In the original Type~I theory, we shall orient our five-brane
so that it satisfies Neumann boundary conditions
in the $\lbrace x_1,...,x_4,x_7,x_8\rbrace$ directions,
and Dirichlet boundary conditions in the
$\lbrace x_5,x_6,x_9,x_{10}\rbrace$ directions.
Upon $T$-dualizing the $\lbrace x_7,x_8,x_9,x_{10}\rbrace$
directions, we then obtain two distinct five-branes.
The five-brane that we obtain from the Type~I nine-brane
satisfies Neumann  boundary conditions in the $\lbrace
x_1,...,x_4,x_5,x_6\rbrace$ directions:  this is
the spacetime that the MSSM Higgs and gauge fields experience
at energy scales exceeding $\mu_0\equiv R^{-1}$.
The second five-brane, by contrast, has Neumann boundary
conditions in the $\lbrace x_1,...,x_4,x_9,x_{10}\rbrace$
directions.  The ``intersection'' of these two different
five-branes (\ie, the $55'$-sector) once again provides us with
an effective three-brane which
can be associated with our observed four-dimensional
world below energy scales $\mu_0\equiv R^{-1}$.  Thus, it is
this $55'$-sector which is presumed to contain
our chiral MSSM fermions.

Thus, we see that our intermediate-scale grand unification scenario
has a variety of natural interpretations and realizations
within string theory, and generalizations to higher values
of $\eta$ are obvious.
Hence we conclude
that is indeed possible to non-perturbatively
lower the string scale in such a way that
it directly coincides with our new
gauge coupling unification scale.
This opens up the exciting
possibility that our scenario can be embedded directly
into a string theory beyond the scale $M'_{\rm GUT}$.
This embedding into a finite theory such as string theory would
then be the ``cure'' for the non-renormalizability of
our effective higher-dimensional
field theory between the scales $\mu_0$ and $M'_{\rm GUT}$.

Of course, the construction of a fully realistic string
model which gives rise to these brane configurations
is a far more complicated task.  Preliminary steps towards
such model-building have been taken in Ref.~\cite{henry}.
Therefore, in order to illustrate our scenario as well
as the origin of what we have called  the ``grand unified group''
in Sect.~4,
let us briefly consider one of the explicit four-dimensional
$N=1$ supersymmetric Type~I models 
constructed in Ref.~\cite{henry}.
This model incorporates a $\IZ_3 \times \IZ_2$
orbifold defined by the actions $g$ and $R$,
where
\beqn
    g (z_1,z_2,z_3)&=& (\omega z_1, \omega z_2, \omega z_3)  \nonumber\\
    R (z_1,z_2,z_3) &=& (-z_1,-z_2,z_3) ~.
\label{emilone}
\eeqn
Here $\omega\equiv e^{2 i \pi /3}$, and 
$(z_1,z_2,z_3)$ are three complex coordinates
of the three complex planes
formed from the real coordinates
$\lbrace x_5, ..., x_{10}\rbrace$ 
of the six-dimensional compact space. 
This model contains a background 
$B$-field of rank $2$ that arises from the NS-NS sector,
and  contains both nine-branes and five-branes.  These are taken to 
be at the same fixed point. 
The gauge group from the open-string sector is
$[U(2) \times U(2) \times U(4)]^2$, with different 
massless matter representations coming
from the $99$, $55$, and $59$ sectors~\cite{henry}. 
The form of the orbifold action
(\ref{emilone}) shows that at the massive level, 
the first two complex planes $(z_1,z_2)$ give rise to an $N=4$ supersymmetric
spectrum, and the third complex plane (corresponding 
to $z_3$) has an $N=2$ supersymmetric spectrum that can provide
power-law corrections to the gauge and Yukawa couplings. 
 Based on the considerations discussed above,
we take the compact coordinates corresponding to 
$(z_1,z_2)$ to be very large 
(and therefore must $T$-dualize them, exchanging nine-branes and five-branes
in the process).  We likewise take
the inverse radius of the $z_3$ plane to be 
a factor of ten to twenty below the string scale. 
Close to the string scale, we
obtain a unified gauge group $U(8)^2$, one of whose factors to be interpreted as the
observable unified group.  This gauge group is obtained by ``undoing'' 
the $\IZ_3$ orbifold action for the massive Kaluza-Klein levels. 
The matter representations at the massive levels form $N=2$ supersymmetric
representations which are obtained by simply
decomposing the massive $U(8)$ representations with respect to a
$U(2) \times U(2) \times U(4)$ Pati-Salam subgroup. 
Given the spectrum arising from the $99$ and $95$ sectors,
the power-law corrections to 
the gauge couplings as they evolve towards lower energy scales
can then be computed as explained in Sect.~3 or Appendix~A.

%========================================================================
\setcounter{footnote}{0}
%  \vfill\eject
\section{Explaining the fermion mass hierarchy via branes}

Having explained in the previous section how our
intermediate-scale grand unification scenario can be
realized within the context of Type~I string theory
and its associated branes, we now revisit the fermion
mass hierarchy problem.
In this section, we point out that these
sorts of brane configurations
can actually provide several entirely new methods of addressing
the fermion mass hierarchy problem beyond those considered in Sect.~5.
We shall give two examples.

\subsection{A scenario with $\eta\not= 0$}

Let us begin by considering a situation in which we have two
large extra spacetime dimensions $\lbrace x_5,x_6\rbrace$ of
radius $R$.
In our original Type~I theory, we shall consider our nine-brane
along with {\it two}\/ separate five-branes.  The first five-brane
will be taken to have Neumann boundary conditions
in the directions $\lbrace x_1,...,x_4,x_5,x_7\rbrace$,
while the second will have Neumann boundary conditions
in the directions $\lbrace x_1,...,x_4,x_7,x_8\rbrace$.
All unspecified directions will be assumed to satisfy Dirichlet
boundary conditions.
As before, we must
$T$-dualize the directions $\lbrace x_7,...,x_{10}\rbrace$.
This yields a Type~I$'$ theory with the following branes.
The Type~I nine-brane becomes a Type~I five-brane with
Neumann coordinates $\lbrace x_1,...,x_4,x_5,x_6\rbrace$;
the first Type~I five-brane becomes a seven-brane with
Neumann coordinates $\lbrace x_1,...,x_4,x_5,x_8,x_9,x_{10}\rbrace$;
and
the second Type~I five-brane becomes a second Type~I$'$ five-brane
with Neumann coordinates $\lbrace x_1,...,x_4,x_9,x_{10}\rbrace$.
It is the first Type~I$'$ five-brane which is to be
interpreted as our physical spacetime at energy scales above
$R^{-1}$.

Given this brane configuration,
let us now consider the possible locations for the chiral fermion
generations of the MSSM.
One possibility is that the chiral fermions lie directly in the
first Type~I$'$ five-brane, so that they have a complete tower
of {\it two dimensions' worth}\/ of Kaluza-Klein excitations.
These are the so-called $55$ fermions, which
we shall collectively denote $\Psi_2$ (because they feel
two dimensions' worth of Kaluza-Klein excitations).
A second possibility is that the chiral MSSM fermions arise
from the 57-sector.
It is clear, given the above configuration, that the 
fermions arising in the $57$-sector
feel only {\it one dimension worth}\/ of Kaluza-Klein excitations.
We shall refer to these collectively fermions as $\Psi_1$.
Finally, the third possibility is that the chiral MSSM fermions
arise in the $55'$-sector.
Thus, these fermions have no
Kaluza-Klein excitations at all, and will be denoted $\Psi_0$.

Given these results, and given the power-law dependence that the
corresponding Yukawa couplings experience due to the extra dimensions
(as discussed in Sect.~5),
it is then natural to explain the fermion mass
hierarchy by associating one chiral MSSM generation
with each of the above sectors.
This is therefore an $\eta=2$ scenario (since two of
the chiral MSSM generations have Kaluza-Klein excitations).
As we discussed in Sect.~5, in the absence of any additional
couplings for the Higgs fields, the dominant contributions to
the evolution of the Yukawa couplings come from the diagrams
in Figs.~\ref{yukfigs}(a,b).
In the case of Fig.~\ref{yukfigs}(b), for any external fermion of
type $\Psi_i$, the internal fermion must also be of type $\Psi_i$.
However, in the case of Fig.~\ref{yukfigs}(a),
regardless of the type of the external fermion,
the internal fermion can {\it a priori}\/ be of types $\Psi_{0,1,2}$.
Each fermion carries with it an appropriate number of dimensions' worth
of Kaluza-Klein modes.
The only constraints that govern the counting of modes for each loop
are Kaluza-Klein momentum conservation at the vertices:  in general,
at any vertex, we must impose Kaluza-Klein momentum conservation
in the directions for which translational invariance is not broken.
In this way, we can generate a whole variety of powers $(\Lambda/\mu_0)^p$,
$p=0,1,2,3,4$ for the different diagrams.
Thus, depending on the fermion $\Psi_i$ in question, the corresponding
wavefunction renormalization factors $Z_{F}$ and $Z_{\overline{F}}$
can feel a different number power-law dependence, and this feature can be
used to explain the fermion mass hierarchy.

\subsection{A scenario with $\eta=0$}

Clearly, the above scenario relies on the different fermion
generations feeling different numbers of spacetime dimensions
and therefore having different numbers of Kaluza-Klein excitations.
Thus, in the language of Sects.~2 and 3, this scenario amounts to $\eta=2$
where $\eta$ is the number of MSSM fermion generations
that feel the extra dimensions.
This is therefore not the ``minimal'' scenario, which has $\eta=0$.
However, as we stressed in Sect.~4, the minimal scenario is actually
preferable for the purposes of proton decay.
It would therefore be interesting to see if there exist brane configurations
which have $\eta=0$ but which nevertheless also lead to fermion mass
hierarchies.

%======================================================================
\begin{figure}[ht]
\centerline{ \epsfxsize 5.0 truein \epsfbox {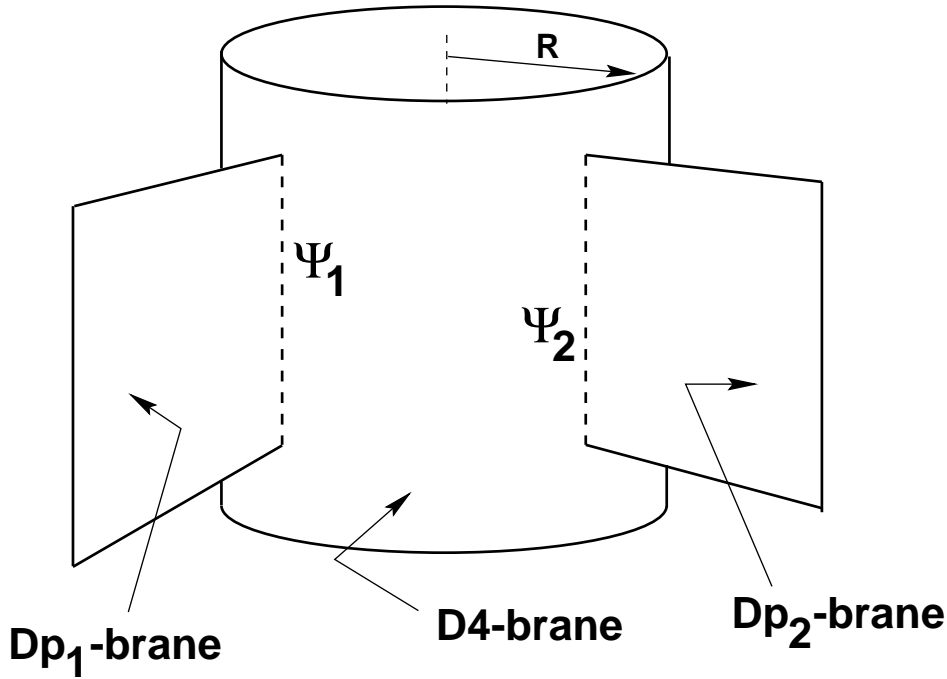}}
\caption{A $D$-brane configuration which leads to a natural fermion
     mass hierarchy for the ``minimal'' scenario with $\eta=0$
     (no Kaluza-Klein excitations for chiral MSSM fermions).}
\label{flower}
\end{figure}
%======================================================================

To do this,
let us consider the $\delta=1$ case, and
imagine a set of Type~I$'$ branes:  a four-brane (interpreted as
the observable universe at energy scales above $R^{-1}$);
and a variety of other branes of differing dimensions $p_i$.
Let us assume that each of these other branes has a four-dimensional
intersection with the four-brane, and let us place our different chiral MSSM
fermions at these different intersections, so that they arise in
the $4p_i$-sectors.
Such a situation is sketched in the case of two additional branes
in Fig.~\ref{flower}.
Because the fermions all arise in the $4p_i$ sectors,
none of them feel extra dimensions directly or have Kaluza-Klein excitations.
These are therefore minimal $\eta=0$ scenarios, as desired.
However, because each fermion $\psi_i$ arises in the $4p_i$-sector
and is restricted on the intersection of the four-brane with
the $p_i$-brane, in principle it can carry
the quantum numbers of not only the perturbative gauge group arising
from the four-brane, but also the quantum numbers of the {\it
non}\/-perturbative
gauge group $G_i$ corresponding to the $p_i$-brane.
Note that these extra quantum numbers do not affect the gauge coupling
unification that we have already observed in Sect.~3.
However, because these fermions have additional quantum numbers in different
non-perturbative gauge groups corresponding to different branes of different
dimensionalities, their corresponding wavefunction renormalization factors
$Z_F$ and $Z_{\overline{F}}$ will accrue different power-law exponents
from the analogues of Fig.~\ref{yukfigs}(b) where we replace the internal
Standard Model gauge bosons with the gauge bosons of the non-perturbative
gauge group $G_i$.
These different power-law exponents arise because these different
non-perturbative
gauge bosons each feel a different effective number of dimensions.
Thus, in such a scenario, we are able to achieve a natural fermion mass
hierarchy even while maintaining our ``minimal'' scenario with $\eta=0$.

%========================================================================
%  \vfill\eject
\section{Relations to other work}
\setcounter{footnote}{0}

Extra spacetime dimensions and their effects have been studied
in the literature from a variety of different perspectives.
For completeness, we shall briefly highlight
the novel features of our approach and compare it with
some others that have been taken.

At the field-theory level, we have seen that extra spacetime dimensions
are equivalent to the introduction of infinite towers of Kaluza-Klein
states.  It may therefore seem that our work is somehow equivalent to prior
work
in which the effects of possible extra matter beyond the MSSM are
analyzed.
However, the extra matter that is typically considered in such analyses
is vector-like and fills out complete $SU(5)$ or $SO(10)$ multiplets.
This then leads to non-perturbative (or at best semi-perturbative)
couplings at unification~\cite{strongcoupling},
and does not permit a lowering of the unification
scale.  In our scenario, by contrast, we have a
unification which remains completely perturbative, even more so than
within the MSSM itself;  moreover, our unification scale is lowered rather than
raised.   These differences essentially arise because the
prior approaches all entail shifting the one-loop beta-function coefficients
$b_i$ by a common fixed finite amount $\Delta b$:
\beq
          b_i 	~\to~ b'_i~\equiv~ b_i ~+~ \Delta b~~~~~~ {\rm for~all}~i~.
\label{them}
\eeq
In our scenario, by contrast, the $b_i$ are
not shifted but rather {\it rescaled}\/,
for at each equally-spaced Kaluza-Klein threshold we are essentially
introducing
another copy of the MSSM gauge-boson and Higgs representations.
Thus, 
our scenario can essentially be re-interpreted
in this language as one in which we continue to have logarithmic running, but
with an effective beta-function coefficient that changes with the energy
scale $\mu$ according to:
\beq
          b_i 	~\to~ b'_i(\mu)~\equiv~ (b_i-\tilde b_i) ~+~ \tilde b_i \, X_\delta\,
           \left( {\mu\over \mu_0}\right)^\delta~.
\label{us}
\eeq
Hence the $b'_i$ in (\ref{us}) are not all driven in the same direction
towards positive values, as they are in (\ref{them}).  It is this feature
that enables us
to achieve rapid gauge coupling unification at low unification
scales without encountering non-perturbative gauge couplings.
Moreover, the rescaling factor in (\ref{us}) is itself $\mu$-dependent.
This enables us to generate the {\it power-law}\/ evolution for the
gauge couplings which
is the hallmark of our scenario.
It is also this feature, for example, which enables us to address the fermion
mass hierarchy problem.

Extra spacetime dimensions and Kaluza-Klein towers of states
have also been analyzed within the context of string theory.
For example, the consequences of
extra large (TeV-scale) dimensions have been previously examined
in a notable series of papers~\cite{antoniadis}.
However, this analysis was carried out within the context of
string models that were deliberately constructed in such a way that
the effects of the extra spacetime dimensions were shielded from the evolution
of
the gauge couplings.  In other words, the Kaluza-Klein towers of states were
arranged to arise in only certain $N=4$ supersymmetric
sectors of the underlying string model
so that they had no effect on the running of the gauge couplings, giving
rise to $\tilde b_i=0$.
Thus, in these restricted scenarios, the gauge couplings can run only
in their usual logarithmic fashion, and unify only at the usual
GUT scale $M_{\rm GUT}\approx 2\times 10^{16}$ GeV.
By contrast, it is precisely the effects of these extra dimensions
on the gauge and Yukawa couplings which have been our main focus in this paper
---
rather than avoid these effects, we have exploited them!
Moreover, such effects can be expected to be the generic case in string
theory (the existence of certain specially constructed
string models notwithstanding).

More recently, extra dimensions have also played a role in
understanding the strong-coupling behavior of various string theories.
The most famous example of this phenomenon
is the ten-dimensional $E_8\times E_8$ heterotic string:
at strong coupling it has been proposed~\cite{HW} that this string
``grows'' an eleventh dimension of finite length
whose natural size is much larger
than the eleven-dimensional Planck length, and in particular
is much larger than the presumed size of the six-dimensional manifold
on which a subsequent compactification to four dimensions
takes place.  This leads
to a scenario in which our four-dimensional low-energy world should
successively
look five-dimensional and then ultimately eleven-dimensional
as the energy scale is increased.
The fundamental distinction between this scenario and our own, however,
is the effect of this fifth dimension.  In our scenario, the extra dimension(s)
are universal, affecting both gauge and gravitational couplings.
In the $E_8\times E_8$ case, by contrast, the extra fifth dimension
is felt only by the gravitational couplings, and the gauge couplings are
again immune to its effects.

Of course, our analysis should be directly applicable to
string theories which have generic, large-radius compactifications.
Such theories have recently been discussed in a number
of theoretical and phenomenological contexts~\cite{largeradius}.

Finally, there recently appeared a proposal~\cite{Dim,Dimtwo} for
solving the {\it gauge}\/ hierarchy problem through the appearance
of new {\it millimeter-scale}\/ extra dimensions!
These new dimensions are presumed to affect the gravitational
interaction only, and have no effect on the gauge couplings.
This proposal is therefore, in some sense, the gravitational
counterpart of our proposal, effectively reducing the {\it Planck}\/ scale.
Of course, this proposal differs from ours in essentially the
same way that previous proposals have differed:
it does not address gauge coupling unification;
the Standard Model particles, unlike the graviton,
are presumed to be essentially
trapped on a four-dimensional submanifold relative to these
extra dimensions;
and the Standard Model particles are therefore once again largely immune
to the effects of these extra dimensions.
Nevertheless, it would be very interesting to combine our scenario
(in which the {\it gauge}\/ part of observable low-energy world
feels new extra dimensions as large as a TeV) with the scenario
of Ref.~\cite{Dim} (in which the {\it gravitational}\/ part of the
observable low-energy world feels extra dimensions as large
as a millimeter).  Such a synthesis could proceed
along the lines sketched in Sect.~7,
and might well lead to a unified
picture of gauge {\it and gravitational}\/
unification, all occurring at around a TeV.
Preliminary steps in realizing this 
possibility within the context
of Type~I string theory have already been taken in Ref.~\cite{henry}.
Earlier discussions of such ``TeV-scale superstrings'' can also
be found in Ref.~\cite{lykken}.
Likewise, recent advances in understanding ``the universe
as brane''~\cite{sundrum}
are likely to prove crucial in developing these scenarios
at both the string-theoretic and field-theoretic levels.
Note that the gauge hierarchy problem has also
been addressed within the context of a higher-dimensional
field theory in Ref.~\cite{Hatanaka}.

%========================================================================
%  \vfill\eject
\section{Collider signals and cosmological implications}
\setcounter{footnote}{0}

As might be expected, the appearance of extra
large spacetime dimensions can give rise to many
interesting signals for collider experiments (see, \eg,
Refs.~\cite{chunk,ABQ}).
They can also have profound implications for cosmology~\cite{kolb}.
In this section we will give a short sketch of some of these
connections.  

If the scale $\mu_0\equiv R^{-1}$ of the new dimensions
is close to the electroweak scale, then future colliders will be
able to probe the new dimensions directly.
Let us first consider our ``minimal'' scenario with $\eta=0$.
In this scenario, only the non-chiral MSSM states will have an
infinite tower of Kaluza-Klein excitations, and these will be
separated by an energy scale $\mu_0$.
The importance of such Kaluza-Klein excitations for collider phenomenology or
cosmology depends crucially on their transformation properties under
the $\IZ_2$ orbifold action $y_i\rightarrow -y_i$, where
$y_i$ are the coordinates of the new compactified dimensions.
Let us first consider the Kaluza-Klein states whose wavefunctions
are even with respect to this action.
Such states can directly
couple to the Standard Model fermions because these
fermions are located either at the orbifold fixed points
or on three-branes;  in either case we cannot impose Kaluza-Klein
momentum conservation on the vertices because
translational invariance in the compactified directions is broken.
Therefore, such even Kaluza-Klein states
can be directly produced at future colliders if
$\mu_0\simeq {\cal O}$(TeV), with the lowest-lying even Kaluza-Klein mode
directly decaying into (s)fermions.
Note, on the other hand, that
because the gauge bosons and Higgs particles
feel the extra dimensions,
Kaluza-Klein momentum conservation continues to apply
at their coupling vertices.
This feature then prevents
the lowest-lying even Kaluza-Klein modes from directly decaying
into the zero-mode gauge bosons and/or Higgs particles.
Of course, the Kaluza-Klein modes corresponding to the gauge bosons
will be more important than the Kaluza-Klein modes corresponding to the
Higgs fields because the Higgs field couplings
is usually suppressed by small Yukawa factors.

Given these observations, we see that
Kaluza-Klein states corresponding to the gauge bosons
might be observable via Drell-Yan production
in proton-(anti) proton collisions;  one would seek to
identify charged leptons in the final state.
The analysis for the lowest-lying neutral gauge boson
Kaluza-Klein state is exactly analogous to that for $Z'$ bosons in $E_6$
superstring-inspired models \cite{dAQZ,ABQ}. Typically, the branching ratio
into fermion pairs is reduced
by the presence of other supersymmetric channels \cite{gkk}. Thus, bounds
on $\mu_0$ tend to be model-dependent.  However, using
the results in Ref.~\cite{ABQ}, recent Fermilab data suggests
the simple estimate
$\mu_0\equiv R^{-1} \gsim 500$ GeV for
the lower bound on the scale of the extra dimensions.

Alternatively, if the scale of extra dimensions is much larger than
$\cal O$(TeV), the effects of the infinite Kaluza-Klein tower can be seen at
low
energies via an effective contact interaction.
Such an effective contact interaction
can arise, for example,
from the tree-level exchange of massive gauge-boson Kaluza-Klein modes.
Let us assume, for simplicity, that only one extra dimension exists.
Then the amplitude for the scattering process
          $ l^+ l^- \to l^+ l^-$
receives a contribution
\beq
         g^2 R^2 \, \sum_{n=1}^{\infty} \,{1\over q^2 R^2 + n^2}
\eeq
from the infinite tower of even Kaluza-Klein states.
At low energy scales $q^2\ll \mu_0^2$,
we find that the above expression is approximately $(\pi^2/6) g^2 R^2$.
This then gives rise to a four-fermion contact interaction
of the form
\beq
    {\cal L} ~\approx~  {\pi^2\over 6}\, g^2 R^2
                          \, (\bar\Psi\gamma_\mu\Psi)^2
\eeq
where $g$ is a gauge coupling and $\Psi$ schematically
denotes either quarks or leptons.
Recent bounds on four-fermion contact interactions then imply
a lower bound $\mu_0 \gsim 300$ GeV.

On the other hand, Kaluza-Klein states that are odd under the $\IZ_2$ orbifold
action do not couple to chiral fermions.
Consequently, they can be probed at collider experiments only via
higher-loop processes.
This implies there are no significant bounds from collider
phenomenology arising from these states.
However, since the lowest-lying odd Kaluza-Klein states
are stable due to Kaluza-Klein momentum conservation, such states can have
important cosmological implications.
For example,
if the annihilation processes
are sufficiently strong~\cite{ks},
such states might serve as ideal dark-matter candidates.

Our ``non-minimal'' scenarios with $\eta >0$
(\ie, with Kaluza-Klein excitations for chiral MSSM fermions)
can also lead to interesting collider phenomenology, provided
that proton decay is not a problem.
This issue was discussed in Sect.~4.
Massive Kaluza-Klein states for the chiral fermions will appear as
heavier versions of the usual zero-mode fermions.
If there is at least one generation of chiral fermions with an infinite
tower of Kaluza-Klein excitations,
then Kaluza-Klein momentum conservation will prevent
the first-excited gauge-boson Kaluza-Klein state from
decaying into the low-energy fermions associated with the MSSM
 generation that experiences the extra dimensions.
This would therefore be a unique experimental signature
of the fact that not all fermion generations have a Kaluza-Klein tower
(\ie, that $\eta <3$).
Thus, by probing such signatures, one has the possibility of
 {\it experimentally}\/ choosing between viable TeV-scale string
models or grand unification scenarios!
However, if all three fermion
generations experience the extra dimensions,
then the first-excited even gauge-boson Kaluza-Klein states
are now stable and can also serve as suitable dark matter
candidates.

One might worry about the fact that in a more general scenario
in which gravity experiences extra dimensions, the thermal regeneration
of unstable gravitinos could cause a problem during nucleosynthesis~\cite{as}.
However, there may be solutions to this difficulty when the effects
of extra dimensions are taken into account in analyzing the
dynamics of the early universe~\cite{kolb}.
 {\it A priori}\/, there are many effects that come into play in the context of
a
higher-dimensional cosmology.
In addition to the issue surrounding higher-dimensional inflation,
additional issues include
the effects of extra dimensions on adiabatic density perturbations,
on topological defects, and on cosmological phase transitions.
Indeed,
being slightly bolder,
one might even imagine developing a possible explanation for the
dimensionality
of spacetime (\ie, the number
of large dimensions) along the lines of
the approach followed in Ref.~\cite{bv}.
All of these issues can be expected to have a profound effect on
our understanding of dynamics of the early universe,
and are worthy of further study.

%========================================================================
%  \vfill\eject
\section{Conclusions and future prospects}
\setcounter{footnote}{0}

In this paper, we have proposed a new framework in which
the physics of conventional grand unification might be brought
down to accessible energy scales, perhaps even as low as a TeV.
Our fundamental idea involves the appearance of extra large spacetime
dimensions.  The appearance of extra spacetime dimensions is a natural feature
in string theory, and their radii are generally unfixed by string dynamics.
Therefore, by postulating the appearance of relatively large
extra dimensions, we have shown that the physics of conventional
grand unification can be addressed in an entirely new context.
Specifically, we have shown that gauge coupling unification is
preserved by extra dimensions, but that the unification scale is
significantly lowered.  This leads to the exciting possibility of
intermediate-scale grand unification.
We found that proton decay can also be
avoided --- even with the smaller unification scale --- thanks
to various new symmetry properties pertaining to the extra
spacetime dimensions.  Furthermore, we showed that extra dimensions
also provide a natural setting in which to address the fermion
mass hierarchy problem, for such extra dimensions
tend to significantly amplify
the effects of relatively small flavor-dependent couplings.
It is also possible to consider the effects of extra spacetime
dimensions on the running of the soft SUSY-breaking masses.

Although we primarily applied our scenario to the Minimal Supersymmetric
Standard Model (MSSM), the effects of extra dimensions are completely
general and can just as easily be applied in a number of other different
contexts.  To illustrate this, we also considered the role of
extra dimensions in the {\it non-supersymmetric}\/ Standard Model,
and found that once again they can lead to gauge coupling unification
at very low energy scales.  This would then be a non-supersymmetric
``solution'' to the gauge hierarchy problem.
We also considered the embedding of
our scenario into string theory, and found that there exist several
very natural string and D-brane settings in which our scenario
can be realized.  Moreover,
within this context, we also proposed a new
method for addressing the fermion mass hierarchy which can be
realized in non-perturbative open-string theories and which does
not require the {\it ad hoc}\/ introduction of low-energy
flavor-dependent couplings.

Overall, however, we stress that the most exciting aspect of
this approach to grand unification is that it permits
the predictions of GUT physics (and indeed even of string theory itself)
to be brought down to accessible energy scales!  Thus,
if this framework is correct,  we can expect to witness strong and unmistakable
signals in the next round of accelerator experiments.
Such signals were discussed in Sect.~10, and in fact
experimental evidence for extra large spacetime dimensions
has ``already'' been found~\cite{kane} in the year 2011.
Moreover, our scenario should also have important and dramatic implications
for cosmology.
Taken together, therefore,
our results suggest an entirely new approach towards probing
--- both theoretically and experimentally ---
the physics of grand unification as well as the
phenomenology of large-radius string compactifications.

%=============================================================================
\bigskip
\medskip
\leftline{\large\bf Acknowledgments}
\medskip

We wish to thank G.~Kane, J.~March-Russell,
L.~Randall, R.~Rattazzi, S.-H.H.~Tye, and C.~Wagner
for useful discussions.
The LPTHE is laboratoire associ\'e au CNRS-URA-D0063.
%=============================================================================

%=============================================================================
%  \vfill\eject

%=============================================================================
\setcounter{section}{0}   %  starts Appendix lettering at "A"
\Appendix{Relation between extra spacetime dimensions and Kaluza-Klein modes}

In this Appendix, we shall give the precise relation between extra
spacetime dimensions and Kaluza-Klein modes.  Specifically,
we shall exactly calculate the effects of the
infinite towers of Kaluza-Klein states on the ``running''
of the gauge couplings.  This will also enable us to determine
the extent to which the results of such a calculation can be
approximated by the expression given in (\ref{pizero}).

Before beginning our calculation, let us discuss the main
idea.  As discussed in Sect.~3,
the result (\ref{pizero}) can be easily obtained by treating the
non-chiral sector of the MSSM as
effectively being in $D$ flat spacetime dimensions, where
$D=4+\delta$.
Of course, geometrically speaking,
a spacetime consisting of four flat dimensions and $\delta$
circles of fixed radius $R=\mu_0^{-1}$ is {\it never}\/
equivalent to a flat $(4+\delta)$-dimensional spacetime.
However, we expect that as the energy scale
$\mu$ increases, the effective length scale decreases,
and consequently the fixed radius $R$ ``appears'' to
become large.
Thus, as we shall see, there are essentially two equivalent
pictures that can be used to describe the same physics.

\begin{figure}[th]
\centerline{ \epsfxsize 3.0 truein \epsfbox {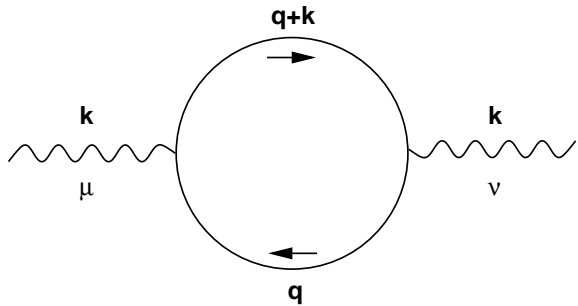}}
\caption{The vacuum polarization diagram.  We include the effects
   of extra Kaluza-Klein modes in the loops.}
\label{bubblefig}
\end{figure}

Our procedure will be to
adopt the strict four-dimensional point
of view, and to
evaluate the vacuum polarization
diagram shown in Fig.~\ref{bubblefig} where we include the effects
of the MSSM particles as well as the
appropriate Kaluza-Klein excitations in the loops.
Note that parts of our calculation are similar
to a calculation in Ref.~\cite{TV}.
For simplicity, we shall begin by performing our calculation in the case
of a single Dirac fermion and its
corresponding Kaluza-Klein excitations.  Since the
effects of these Kaluza-Klein excitations
will essentially be the same for each particle that has Kaluza-Klein
excitations, and since
these effects are likewise universal for all theories (whether
QED or the MSSM), we can generalize our results to the full
MSSM in the final step.

%======================================================================
%======================================================================

For a single Dirac fermion with Kaluza-Klein excitations,
the vacuum polarization
diagram in Fig.~\ref{bubblefig} is given by
\beq
   \Pi_{\mu\nu}(k)~=~ -\,
     \sum_{n_i=-\infty}^\infty\, g^2 \,
    \int_0^\infty {d^4 q\over (2\pi)^4} \, {\rm Tr}\,\left(
     \gamma_\mu {1\over \not q - m_n } \gamma_\nu {1\over \not k + \not q -
m_n}
      \right)
\label{firsteq}
\eeq
where we have used the notation
\beq
      \sum_{n_i= -\infty}^\infty ~\equiv ~
     \sum_{n_1= -\infty}^\infty
     \sum_{n_2= -\infty}^\infty
     \cdots
     \sum_{n_\delta= -\infty}^\infty
\eeq
to represent a summation over all
corresponding Kaluza-Klein excitations with masses
$m_n^2$ given in (\ref{KKmasses}).
In (\ref{KKmasses}),
$m_0$ is the energy of the ground state, which we will henceforth
take to be zero for simplicity.
Restricting our summation to only the $n_i=0$ term therefore amounts
to considering only the original fermionic state without its Kaluza-Klein
excitations,
and opposite values of $n_i$ correspond to Kaluza-Klein states whose
$i^{\rm th}$ circle momenta are in opposite directions.
The overall sign in (\ref{firsteq}) arises due the fermion loop.

Here and throughout we
shall assume the presence of a suitable ultraviolet regulator
with cutoff $\Lambda$
in order to justify our subsequent manipulations.
We shall discuss our regulator explicitly when its form becomes crucial
for our analysis.

Our initial steps are completely standard.
Using gauge invariance to define $\Pi(k^2)$ via
(\ref{gauginv}),
we contract the Lorentz indices and evaluate the trace to obtain
\beq
     \Pi(k^2) ~=~ -{8 g^2\over 3k^2} \,
     \sum_{n_i=-\infty}^\infty\,
    \int_0^\infty {d^4 q\over (2\pi)^4} \,
           \left\lbrace { -(k+q)\cdot q + 2 m_n^2  \over
           (q^2-m_n^2) \,\lbrack (k+q)^2 - m_n^2 \rbrack }
           \right\rbrace~.
\eeq
Passing to Euclidean momenta, introducing the Feynman $x$-parameter
to combine the propagators, and keeping only terms in the integrand
that are even in $q$ then yields
\beq
     \Pi(k^2) ~=~ -{8 g^2\over 3k^2} \,
     \sum_{n_i=-\infty}^\infty\,
    \int_0^1 dx \int_0^\infty {d^4 q\over (2\pi)^4} \,
           \left\lbrace {q^2 - x(1-x) k^2 + 2 m_n^2 \over
          \lbrack q^2 + x(1-x) k^2 + m_n^2 \rbrack^2 }
           \right\rbrace~.
\eeq

Our next step is to rewrite this expression in terms of a Schwinger
proper-time parameter $t$ using the identity
\beq
          {1\over A^2} = \int_0^\infty dt\, t\, e^{-At}~.
\eeq
This yields
\beqn
      \Pi(k^2) &=& -{8 g^2\over 3k^2} \,
      \sum_{n_i=-\infty}^\infty\,
     \int_0^1 dx \int_0^\infty dt\, t\,\int_0^\infty {d^4 q\over (2\pi)^4} \,
           \lbrack q^2 - x(1-x) k^2 + 2 m_n^2 \rbrack ~\times\nonumber\\
         && ~~~~~~~\times~ \exp\left\lbrace
            -t\lbrack q^2 + x(1-x) k^2 + m_n^2 \rbrack \right\rbrace~,
\eeqn
and performing the momentum integrations via the identities
\beq
     \int_0^\infty {d^4 q\over (2\pi)^4}
          \,e^{-t q^2} = {1\over 16\pi^2 t^2} ~,~~~~~~~~~
     \int_0^\infty {d^4 q\over (2\pi)^4}
          \,q^2\, e^{-t q^2} = {1\over 8\pi^2 t^3} ~,
\eeq
we obtain
\beqn
      \Pi(k^2) &=& -{ g^2\over 6\pi^2 k^2} \,
      \sum_{n_i=-\infty}^\infty\,
     \int_0^1 dx \int_0^\infty {dt\over t} \,
           \left\lbrack {2\over t} - x(1-x) k^2 + 2 m_n^2
\right\rbrack\,\times\nonumber\\
        && ~~~~~~~\times ~\exp\left\lbrace
            -t\lbrack x(1-x) k^2 + m_n^2 \rbrack \right\rbrace~.
\eeqn
Integrating the first term by parts then yields
\beq
      \Pi(k^2) ~=~ { g^2\over 2\pi^2 } \,
      \sum_{n_i=-\infty}^\infty\,
     \int_0^1 dx \,x(1-x)\, \int_0^\infty {dt\over t} \,
        \exp\left\lbrace
            -t\lbrack x(1-x) k^2 + m_n^2 \rbrack \right\rbrace~.
\label{regulatorintroduced}
\eeq

Let us now perform the summation over the Kaluza-Klein states.
In order to do this,
we recall the definition of the Jacobi $\vartheta_3$ function:
\beq
      \vartheta_3(\tau) ~\equiv~ \sum_{n_i=-\infty}^\infty \,\exp(\pi i \tau
n^2)~
\eeq
where $\tau$ is a complex number.
This function has the remarkable property that
\beq
     \vartheta_3(-1/\tau) ~=~ \sqrt{-i\tau} \, \vartheta_3(\tau)~
\label{Strans}
\eeq
where one chooses the branch of the square root
with non-negative real part.
We can thus rewrite our result (\ref{regulatorintroduced})
in terms of this function as
\beq
      \Pi(k^2) ~=~ { g^2\over 2\pi^2 } \,
     \int_0^1 dx \,x(1-x)\, \int_0^\infty {dt\over t} \,
            e^{-t x(1-x) k^2}\,
          \left\lbrace \vartheta_3\left( {it\over \pi R^2}
\right)\right\rbrace^\delta~,
\eeq
whereupon we find that $\Pi(0)$ is
\beq
      \Pi(0) ~=~ { g^2\over 12\pi^2 } \,
             \int_0^\infty {dt\over t} \,
          \left\lbrace \vartheta_3\left( {it\over \pi R^2}
\right)\right\rbrace^\delta~.
\label{intermed}
\eeq

At this step, we must introduce our infrared and ultraviolet regulators,
along with their corresponding cutoffs.
Let us first recall that the ultraviolet and infrared divergences
in this expression arise from the $t\to 0$ and $t\to \infty$ limits of
integration respectively.  Therefore, it is simplest to render this expression
finite in both limits by introducing upper and lower cutoffs on the
$t$-integration:
\beq
       \int_0^\infty \, dt ~\longrightarrow~
       \int_{r\Lambda^{-2}}^{r\mu_0^{-2}}  \, dt~.
\label{cutoff}
\eeq
Here $\Lambda$ is our ultraviolet cutoff, $\mu_0$ is our infrared cutoff,
and the numerical coefficient $r$ (which ultimately relates these cutoff
parameters
to underlying physical {\it mass scales}) is defined as
\beq
           r~\equiv~ \pi \,(X_\delta)^{-2/\delta}
\eeq
where $X_\delta$ is defined in (\ref{Xdef}).
As we discussed in Sect.~3, such a precise value of $X_\delta$ or $r$ cannot be
deduced purely on the basis of such a non-renormalizable theory alone, and
instead
requires outside information.  However, in Appendix~B we shall see
that this is the correct relative normalization factor that relates the cutoff
parameters $r\Lambda^{-2}$ and $r\mu_0^{-2}$ in (\ref{cutoff}) to
the underlying physical mass scales $\Lambda$ and $\mu_0$ used in Sect.~3.

Looking at (\ref{intermed}),
we see that the limit of the usual four-dimensional theory
without Kaluza-Klein modes can be obtained
by setting $\vartheta_3=1$.
This is equivalent to setting $R\to 0$, which essentially
makes all Kaluza-Klein modes infinitely massive.
For all values of $\delta$, this then produces the expected result
\beq
       \Pi(0)  ~=~
      {g^2 \over 6\pi^2}\, \ln {\Lambda\over \mu_0}
      ~=~ {g^2 b\over 8\pi^2}\, \ln {\Lambda\over \mu_0}
\eeq
where we have identified $b=4/3$ as the
beta-function coefficient of our single Dirac fermion.

Let us now generalize this result to the present case of the full MSSM.
As we discussed in Sect.~3, not all of the MSSM states have Kaluza-Klein
excitations.  Indeed, while the zero-mode states with $n_i=0$ correspond
to the full MSSM spectrum (for which the corresponding beta-function
coefficients
are denoted $b_i$), only the gauge bosons and Higgs fields in the MSSM
will have Kaluza-Klein excitations.
The beta-function coefficients $\tilde b_i$ corresponding to these
Kaluza-Klein modes at each non-zero mass level $\lbrace n_i\rbrace$
are given in (\ref{btilde}).
Thus, generalizing (\ref{intermed}) to the case of the full MSSM with
Kaluza-Klein excitations from gauge bosons and Higgs fields only,
we find that
\beqn
      \Pi(0) &=&
      {g_i^2 b_i\over 8\pi^2}\, \ln {\Lambda\over \mu_0}
       ~+~
      { g_i^2 \tilde b_i \over 16\pi^2 } \,
             \int_{r\Lambda^{-2}}^{r\mu_0^{-2}} {dt\over t} \,
          \left\lbrace
           \left\lbrack
          \vartheta_3\left( {it\over \pi R^2} \right) \right\rbrack^\delta -1
          \right\rbrace
       \nonumber\\
       &=&
      {g_i^2 (b_i-\tilde b_i)\over 8\pi^2}\, \ln {\Lambda\over \mu_0}
       ~+~
      { g_i^2 \tilde b_i \over 16\pi^2 } \,
             \int_{r\Lambda^{-2}}^{r\mu_0^{-2}} {dt\over t} \,
          \left\lbrace
          \vartheta_3\left( {it\over \pi R^2} \right) \right\rbrace^\delta~.
\label{anequation}
\eeqn
In the first line, we have explicitly separated the
zero-mode contributions (which yield the first term)
from the higher-mode contributions (which yield the second term).
Since the $\vartheta_3$ functions implicitly include the contributions from
the zero-modes, we have explicitly subtracted these
contributions from the integrand of the second term
by writing $\vartheta_3^\delta -1$.
Thus, passing to the second line of (\ref{anequation}), we see
that the second term represents the contributions
from the complete Kaluza-Klein towers that would have existed if the
zero-mode states had matched the excited states in our theory,
while the first term represents the compensating adjustment
that arises because the zero-modes and excited states are actually different.
Taken together, (\ref{anequation}) then implies that
\beq
       \alpha_i^{-1}(\Lambda) ~=~ \alpha_i^{-1}(\mu_0) ~-~
            {b_i-\tilde b_i\over 2\pi}\,\ln{\Lambda\over \mu_0}
          ~-~ {\tilde b_i\over 4\pi}\,
             \int_{r\Lambda^{-2}}^{r\mu_0^{-2}} {dt\over t} \,
     \left\lbrace \vartheta_3\left( {it\over \pi R^2} \right)
\right\rbrace^\delta~.
\label{KKresult}
\eeq

The result (\ref{KKresult}) gives the exact running of the MSSM
gauge couplings in the presence of an infinite tower of Higgs and gauge-boson
Kaluza-Klein states associated with $\delta$ extra dimensions
compactified on circles of radius $R$.  Indeed, the effect of the Kaluza-Klein
modes
is completely incorporated within the $\vartheta_3$ function.
Note that this result is true for {\it any}\/ mass
scales $\Lambda$ and $\mu_0$ --- in particular, we need not identify $\mu_0$
with $R^{-1}$.

However, we expect that it is a valid description
of the physics to treat the non-chiral sector of the MSSM as being
in $D$ flat dimensions
for energy scales much larger than $R^{-1}$.
We shall now demonstrate in what sense this is true.

Let us suppose, for the moment, that $\mu_0$ and $\Lambda$ are both
much larger than $R^{-1}$.
(Of course, strictly speaking, we will ultimately want to identify $\mu_0$ with
$R^{-1}$, but we will assume $\mu_0 \gg R^{-1}$ for now and defer a discussion
of the errors this introduces until later.)
In this case, we have $t/R^2 \ll 1$,
and we can approximate the $\vartheta_3$ function
using (\ref{Strans}), obtaining
\beq
      \vartheta_3 \left( {it \over \pi R^2} \right)
            \approx~    R\sqrt{\pi\over t}~.
\label{approxed}
\eeq
Inserting this approximation back into
(\ref{KKresult}) and evaluating the integral, we then obtain
\beq
       \alpha_i^{-1}(\Lambda) ~=~ \alpha_i^{-1}(\mu_0) ~-~
            {b_i-\tilde b_i\over 2\pi}\,\ln{\Lambda\over \mu_0}
    ~-~ {\tilde b_i X_\delta \over 2\pi\delta}\,  \, R^\delta \,
         (\Lambda^\delta-\mu_0^\delta)~.
\label{almostdone}
\eeq
If we now identify $R^{-1}$ with $\mu_0$,
we find
\beq
       \alpha_i^{-1}(\Lambda) ~=~ \alpha_i^{-1}(\mu_0) ~-~
            {b_i-\tilde b_i\over 2\pi}\,\ln{\Lambda\over \mu_0}
    ~-~ {\tilde b_i X_\delta \over 2\pi\delta}\,  \,
         \left\lbrack \left({\Lambda\over \mu_0}\right)^\delta-1\right\rbrack~,
\eeq
in agreement with (\ref{geffd}).
Thus, for sufficiently high energy scales,
we see that our explicit Kaluza-Klein calculation
reproduces the
gauge coupling relations used in Sect.~3.

Finally, we must discuss the validity of the approximation $\mu_0,\Lambda\gg
R^{-1}$
that was used in obtaining (\ref{approxed}),
especially in light of the fact that we ultimately wish to identify
$\mu_0=R^{-1}$ and $\Lambda=M'_{\rm GUT}$.
To what extent does this disturb the validity of the above calculation?
Due to the difficulty of analytically integrating the $\vartheta$-function,
this question is best answered numerically.
However, as we stated in Sect.~3,
one cannot discern any difference between
Fig.~\ref{unifII} (in which the exact results (\ref{KKresult}) are plotted)
and the approximate results based on (\ref{geffd}).
Thus, we conclude that the assumption
of $D$ flat spacetime dimensions for the non-chiral sector of the MSSM
at energy scales above $\mu_0$
provides an excellent approximation
to the full Kaluza-Klein theory.

%=============================================================================
% \setcounter{section}{0}   %  starts Appendix lettering at "A"
\Appendix{The truncated Kaluza-Klein theory}

In this Appendix, we shall show that the truncated Kaluza-Klein theory
is also an excellent approximation to the full Kaluza-Klein
theory for the purposes of calculating the scale-dependence
of the gauge couplings and Yukawa couplings.
This will ultimately enable us to calculate the
exact value of $X_\delta$ given in (\ref{Xdef}).

To do this, let us for the moment
consider a simple five-dimensional
$U(1)$ gauge theory
along with an arbitrary tower of
Kaluza-Klein excitations with beta-function coefficient $\tilde b_1$.
If the extra dimension has radius $R$, then
this Kaluza-Klein tower of excited states will consist
of two Kaluza-Klein states
of mass $\approx \mu_0\equiv R^{-1}$, two additional Kaluza-Klein states
of mass $\approx 2 \mu_0$,
two more of mass $\approx 3\mu_0$, and so forth.
There are two Kaluza-Klein states
at each mass level because there are two possible directions
for the Kaluza-Klein momentum in the fifth direction.

Let us now consider the running of our $U(1)$ gauge coupling in the presence
of this Kaluza-Klein spectrum.
We shall first ignore the effects of the zero-modes (which by themselves always
give logarithmic running), and concentrate
solely on the contributions of the non-zero Kaluza-Klein excitations.
Rather than consider all of these Kaluza-Klein states running in the
loops at once (as in Appendix~A), our fundamental idea in this Appendix
will be to introduce these states only at their
thresholds, two each at every mass threshold $m_n=n\mu_0$.
If our Kaluza-Klein tower has beta-function coefficient $\tilde b_1$,
then each time we cross a threshold
the effective beta-function coefficient
increases by $2\tilde b_1$ because each threshold
produces two extra massive copies of the same Kaluza-Klein states.
Thus, after $n$ thresholds (\ie, for energy scales $n\mu_0\leq \mu \leq
(n+1)\mu_0$),
our beta-function coefficient has grown to $(2n+1)\tilde b_1$.
Adding these incremental contributions together (and restoring the contribution
from the zero-modes), we thus find the result
\beq
      \alpha_1^{-1}(\mu) ~=~ \alpha_1^{-1}(M_Z) ~-~
     {b_1-\tilde b_1\over 2\pi} \ln{\mu\over M_Z}
         ~-~ {\tilde b_1\over 2\pi}\,
           \left(
                2n \ln {\mu\over\mu_0}
                    - 2 \ln n! \right)~.
\label{discreteresult}
\eeq
In other words, for any value of $\mu$, our gauge couplings will be given
by (\ref{discreteresult})  where we identify $n\equiv \lbrack \mu/
\mu_0\rbrack$
where $\lbrack r\rbrack$ signifies the greatest integer not exceeding $r$.

Since we are interested in physics only below the scale $M'_{\rm GUT}$, in this
approach we are free to disregard Kaluza-Klein states of masses exceeding
$M'_{\rm GUT}$.
Thus, at every step in the evolution of our $U(1)$ gauge coupling,
we have a {\it completely renormalizable field theory}\/.  Indeed,
the only difference relative to the usual MSSM
is a finite set of extra states
whose masses are regularly spaced in multiples of $\mu_0$.
Note that we have chosen
to limit our attention to an {\it abelian}\/ gauge group.
This is because our Kaluza-Klein states will necessarily
include massive copies of our low-energy gauge bosons, and it is immediately
clear that abelian massive gauge bosons are consistent with renormalizability.
However, it turns out that such a truncated Kaluza-Klein theory is
renormalizable even in the case of non-abelian gauge groups;  this
will be discussed in Appendix~C.

Given that this theory is completely renormalizable, there are no ambiguities
regarding the interpretation of cutoffs and mass scales.  Indeed,
(\ref{discreteresult})
may be regarded as a true renormalization group equation.  We can therefore
compare this prediction with that based on our non-renormalizable
higher-dimensional theory in order to resolve any numerical cutoff or
mass-scale
ambiguities.

To do this, let us consider the prediction (\ref{newsoln}) from our
non-renormalizable field theory, restricted to the $U(1)$ case.
For simplicity, we may formally
take a derivative of (\ref{newsoln}) to write
\beq
        {d\over d\ln \mu} \alpha_1^{-1}(\mu) ~=~  -
          {b_1-\tilde b_1\over 2\pi} ~-~ {\tilde b_1 X_\delta \over 2\pi }\,
               \left({\mu\over \mu_0}\right)^\delta
\label{diffeqnewsoln}
\eeq
where we have rewritten $\Lambda\to\mu$ for notational convenience.
Indeed, we may regard (\ref{diffeqnewsoln}) as providing a {\it definition}\/
for $X_\delta$.
Note that this result should hold, in principle, for {\it any}\/ value of
$\mu$, no matter how large.
Let us now compare this differential equation with the
prediction of our above discrete approach for $\delta=1$.
It is clear that our discrete approach gives the corresponding RGE
\beq
        {d\over d\ln \mu} \alpha_1^{-1}(\mu) ~=~  - {b_1-\tilde b_1 \over 2\pi
}
         - {\tilde b_1 \over \pi }\,
               \left\lbrack {\mu\over \mu_0}\right\rbrack ~.
\label{res}
\eeq
Once again, this result should hold for any $\mu$, no matter how large.
As $\mu/\mu_0\to\infty$,
we can approximate $\lbrack \mu/\mu_0\rbrack \approx \mu/\mu_0$ in (\ref{res}).
We thus immediately find that $X_1=2$.

\begin{figure}[ht]
\centerline{ \epsfxsize 3.5 truein \epsfbox {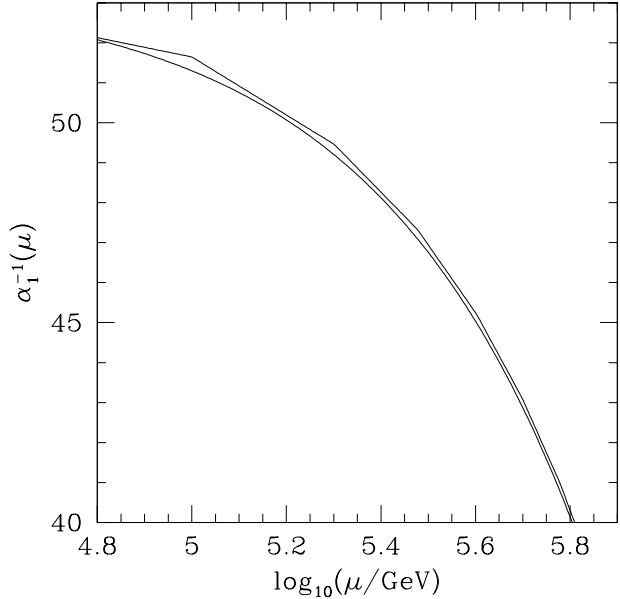}}
\caption{
   Comparison between the renormalizable discrete threshold
   approach (upper curve) and the full non-renormalizable
   approach (lower curve).  We have taken $\mu_0=10^5$ GeV and
   $\delta=1$, and for the clarity of this figure
   we have artificially {\it magnified}\/ the difference between the
   two curves by a factor of 11 (see text).
   We see that the agreement between the two curves
   remains excellent, even in this extreme case.}
\label{finalplot}
\end{figure}

We may also check this result numerically.
Taking $X_1=2$, it is straightforward to verify that
that (\ref{discreteresult}) and (\ref{newsoln}) give closely matching
curves, even for relatively small values of $\mu$.
In Fig.~\ref{finalplot}, we show an extreme case:  we take $\mu_0=10^5$
GeV, and compare the discrete result (\ref{discreteresult}) against
the {\it full}\/ analytical result (\ref{KKresult}).
Moreover, for this figure
we have artificially inflated the value of $\tilde b_1$ (taking $\tilde
b_1=b_1=33/5$
rather than its true value $\tilde b_1=3/5$) in order to magnify the
differences
between the two curves and render these differences
visible in Fig.~\ref{finalplot}.
Even with this magnification,
we see that the agreement is excellent.
Thus, we see that we are free to interpret $\Lambda$ as the physical mass scale
provided we take $X_1=2$.

%======================================================================
%======================================================================

Finally, let us consider the situation in higher dimensions.
Once again, our discrete threshold approach yields the RGE
\beq
        {d\over d\ln \mu} \alpha_1^{-1}(\mu) ~=~  -
           {b_1-\tilde b_1 \over 2\pi }~-~
           {\tilde b_1 \over 2\pi }\, N(\mu,\mu_0)
\eeq
where $N(\mu,\mu_0)$ is the number of Kaluza-Klein states with
masses less than $\mu$.
Note that by definition, $N(\mu,\mu_0)$ is the number of solutions
to the equation
\beq
        \sum_{i=1}^\delta n_i^2 ~\leq~ \left({\mu\over \mu_0}\right)^2~,~~~~~
n_i\in\IZ~.
\eeq
For large $\mu/\mu_0$, this is well-approximated
as the volume of
a $\delta$-dimensional sphere of radius $\mu/\mu_0$:
\beq
           N(\mu,\mu_0) ~=~ {\pi^{\delta/2} \over \Gamma(1+\delta/2) }
                \, \left( {\mu\over \mu_0}\right)^\delta~.
\eeq
Comparing with (\ref{diffeqnewsoln}) then yields the result for
$X_\delta$ quoted in (\ref{Xdef}).
Thus, once again, we find that we may interpret the cutoff $\Lambda$
in Sect.~3 as a physical mass scale provided we take the appropriate
value for $X_\delta$.  Although our analysis in this section is restricted
to the case of a $U(1)$ gauge group, these values for $X_\delta$ are
universal for all gauge groups because they reflect nothing more than
the universal enhancement factors due to the appearance of Kaluza-Klein
states and/or extra spacetime dimensions.
Indeed, as we shall demonstrate in Appendix~C, the
truncated Kaluza-Klein theory is renormalizable
even in the non-abelian case.

Finally, let us briefly discuss the significance
of the fact that we can model the
scale-dependence
(or cutoff-dependence) of the gauge couplings as
resulting from an effective non-renormalizable theory.\footnote{
    We thank R.~Rattazzi for discussions on this point.}
In general, since we are evolving the physics
from the infrared to the ultraviolet within the context
of a non-renormalizable field theory, there is always
the danger that there will exist additional relevant
operators at higher scales whose effects we are not including.
In general, this should limit the validity of our approach.
However, for the purposes of examining the evolution of
gauge and Yukawa couplings (which are ultimately wavefunction
renormalization calculations), such operators will have no effect.
This then explains why the truncated Kaluza-Klein theory
succeeds so well in modelling the evolution of these couplings,
which has been our main focus in this paper.
Nevertheless, it is possible
that there will exist physical processes for which such
operators will play an important role, and for which
a renormalizable truncated Kaluza-Klein theory
will not be appropriate.

%  \vfill\eject
%=============================================================================
% \setcounter{section}{0}   %  starts Appendix lettering at "A"
\Appendix{Renormalizability of truncated Kaluza-Klein theories}

In this Appendix, we shall provide a technical background discussion
concerning the renormalizability and gauge invariance of
truncated Kaluza-Klein theories.
Specifically, we shall demonstrate the renormalizability
of the non-abelian truncated
Kaluza-Klein theory on the $\IZ_2$ orbifold discussed in Sect.~2 by
demonstrating the close analogy
between this theory and
the ordinary Higgs mechanism of four-dimensional gauge
theories (which we know preserves renormalizability).

For simplicity, we shall restrict ourselves to the
pure gauge part,
and begin the discussion by considering abelian gauge fields.
For an abelian gauge theory in five dimensions,
the pure gauge Lagrangian is given by
\beq
            {\cal L} ~=~ -{1\over 4}\, F_{ab}F^{ab}
\eeq
where $a,b=1,...,5$ and where $F_{ab}\equiv \partial_a A_b - \partial_b A_a$.
Let us now compactify the fifth dimension on a circle of radius $R$
and rescale our Kaluza-Klein modes:
\beq
          A_{\mu}^{(n)}, A_5^{(n)} ~\rightarrow~ \sqrt{2} (A_{\mu}^{(n)},
A_5^{(n)})~.
\eeq
We then obtain
\beq
        {\cal L}~=~ -{1 \over 4} \sum_{n=0}^{\infty}
        F_{\mu \nu}^{(n)2} ~+~
          {1 \over 4} \sum_{n=1}^{\infty} (\partial_{\mu}
           A_5^{(n)}+ {n \over R} A_{\mu}^{(n)})^2 ~.
\eeq
Note that the resulting Lagrangian is gauge-invariant for
any $A_{\mu}^{(n)}$, with nonlinear gauge transformations
\beqn
     A_{\mu}^{(n)} &\rightarrow& A_{\mu}^{(n)} + \partial_{\mu} \theta^{(n)}
{}~,\nonumber\\
     A_5^{(n)} &\rightarrow& A_5^{(n)} - {n \over R} \theta^{(n)}
\eeqn
where $\theta^{(n)}$ is the gauge transformation parameter.
It is straightforward to compare this result
with the usual $U(1)$ abelian Higgs model.
If we define the massive gauge field
\beq
       B_{\mu}^{(n)} ~=~ A_{\mu}^{(n)} + {R\over n} \partial_{\mu}
         A_5^{(n)}~,
\eeq
we see that $B_{\mu}^{(n)}$ has mass $n/R$.
We thus construct the analogy with the abelian Higgs model
by associating
\beq
          e ~\Longleftrightarrow ~n ~,~~~~~ v ~\Longleftrightarrow~ 1/R
\eeq
where $e$ is the electric charge of the abelian gauge field
in the abelian Higgs model
and where $v$ is the VEV of the Higgs field.
The Goldstone boson of the spontaneously broken
$U(1)$ associated with $A_{\mu}^{(n)}$ is therefore $A_5^{(n)}$.

In five dimensions, Lorentz invariance allows us to add a gauge-fixing
term, and the full pure gauge Lagrangian reads
\beq
     {\cal L} ~=~ -{1 \over 4} F_{ab}^2 \, -\,
           {\lambda \over 2} (\partial^a A_a )^2 ~.
\label{eqone}
\eeq
Note that when reduced to four dimensions, the gauge-fixing term
becomes
\beq
      -{\lambda \over 2} (\partial^a A_a)^2 ~=~
     -{\lambda \over 2} \sum_{n=0}^{\infty}
     { \left[ \partial^{\mu} A_{\mu}^{(n)} - {n \over R} A_5^{(n)} \right] }^2
{}~.
\eeq
Thus, for $\lambda=1$, we see that we obtain
a four-dimensional 't Hooft renormalizable gauge.
Hence the dimensional reduction of the Lagrangian in (\ref{eqone})
gives the four-dimensional result
\beqn
       {\cal L} &=& -{1 \over 4} \sum_{n=0}^{\infty} F_{\mu \nu}^{(n)2}
    -{1 \over 2} \sum_{n=0}^{\infty} \left[ (\partial^{\mu} A_{\mu}^{(n)})^2
      -{n^2 \over R^2} (A_{\mu}^{(n)})^2  \right] \nonumber\\
      && ~~~+~ {1 \over 2} \sum_{n=1}^{\infty}
    \left[ (\partial_{\mu} A_5^{(n)})^2-{n^2 \over R^2} (A_5^{(n)})^2 \right] \
{}.
\eeqn
For $n\geq 1$,
this Lagrangian then leads to the
propagators
$\Delta_{\mu \nu}^{(n)}(k)$
and $\Delta^{(n)}(k)$
for the gauge fields and Goldstone bosons respectively,
where
\beqn
     \Delta_{\mu \nu}^{(n)} (k) &=&
          {-i g_{\mu \nu} \over k^2-n^2/R^2 +i \epsilon} \nonumber\\
       \Delta^{(n)} (k) &=& {i \over k^2 - n^2/R^2 + i \epsilon} ~.
\eeqn
These propagators are well-behaved as $k\to \infty$.
Thus, we conclude that the whole theory is renormalizable
for any (finite) number of Kaluza-Klein
states.  Of course, the theory becomes non-renormalizable
if consider the full infinite tower of Kaluza-Klein states.

Having explicitly demonstrated that the truncated abelian Kaluza-Klein
theory is renormalizable, we can now easily repeat the above
steps for the non-abelian case.  Every step carries through as before.
Thus, we conclude that
even the {\it non-abelian}\/ truncated Kaluza-Klein theory is renormalizable.
For completeness, we give the non-linear realization of the gauge symmetry
for the massive Kaluza-Klein levels:
\beqn
  \delta A_{\mu}^{a(n)} &=& \partial_{\mu} \theta^{a(n)}-
  {1 \over 2} f^{abc} \sum_m \left[ A_{\mu}^{b(n-m)}+ A_{\mu}^{b(n+m)}\right]
         \theta^{c(m)} ~\nonumber \\
   \delta A_5^{a(n)} &=& -{n \over R} \theta^{a(n)}-
   {1 \over 2} f^{abc} \sum_m \left[ A_5^{b(n-m)}+ A_5^{b(n+m)}\right]
           \theta^{c(m)}
\eeqn
where $f^{abc}$ are the structure constants of the non-abelian gauge group
and where $\theta^{a(n)}$ are the gauge transformation parameters.

%=============================================================================

\vfill\eject
\bigskip
\medskip

\bibliographystyle{unsrt}

\end{document}